\documentclass[12pt]{article}
\usepackage{amsmath,amssymb}
\usepackage{bm}
\usepackage{color}
\usepackage{cite}
\usepackage{mathtools}
\usepackage{here}

\usepackage[pdftex]{graphicx}

\usepackage{multirow}

\begin{document}

\title{
\begin{flushright}
\ \\*[-80pt]
\begin{minipage}{0.2\linewidth}
\normalsize
%arXiv:YYMM.NNNN \\
EPHOU-23-004\\*[50pt]
\end{minipage}
\end{flushright}
% Title
{\Large \bf
Quark mass hierarchies and CP violation \\ in $A_4\times A_4\times A_4$ modular symmetric flavor models 
\\*[20pt]}}
% /Title

\author{
Shota Kikuchi,
%\footnote{A's mail}
~Tatsuo Kobayashi,
%\footnote{B's mail}
~Kaito Nasu,
%\footnote{C's mail}
\\~Shohei Takada, and
%\footnote{D's mail}
~Hikaru Uchida
%\footnote{E's mail}
\\*[20pt]
\centerline{
\begin{minipage}{\linewidth}
\begin{center}
{\it \normalsize
Department of Physics, Hokkaido University, Sapporo 060-0810, Japan} \\*[5pt]
\end{center}
\end{minipage}}
\\*[50pt]}

\date{
\centerline{\small \bf Abstract}
\begin{minipage}{0.9\linewidth}
\medskip
\medskip
\small
We study $A_4 \times A_4 \times A_4$  modular symmetric flavor models  to realize 
quark mass hierarchies and mixing angles without fine-tuning.
Mass matrices are written in terms of modular forms. At modular fixed points $\tau = i\infty$ and $\omega$, $A_4$ is broken to $Z_3$ residual symmetry. 
When the modulus $\tau$ is deviated from the fixed points, modular forms show hierarchies depending on their residual charges. Thus, we obtain hierarchical structures in mass matrices. Since we begin with $A_4\times A_4 \times A_4$, the residual symmetry is $Z_3 \times Z_3 \times Z_3$ which can generate sufficient hierarchies to realize quark mass ratios and absolute values of the CKM matrix $|V_{\textrm{CKM}}|$ without fine-tuning. Furthermore, CP violation is studied. We present necessary conditions for CP violation caused by the value of $\tau$. We also show possibilities to realize observed values of the Jarlskog invariant $J_{\textrm{CP}}$, quark mass ratios and CKM matrix $|V_{\textrm{CKM}}|$ simultaneously,
if $\mathcal{O}(10)$ adjustments in coefficients of Yukawa couplings are allowed or moduli values are non-universal.  
\end{minipage}
}
\begin{titlepage}
\maketitle
\thispagestyle{empty}
\end{titlepage}

\newpage

% ------------------------------------------------------ %
% ------------------------------------------------------ %
% ------------------------------------------------------ %
% ------------------------------------------------------ %

\section{Introduction}
\label{Intro}

The origin of quark and lepton flavor structures such as hierarchical masses and mixing angles are one of the biggest mysteries in current particle physics.
Recently as one of the approaches to the flavor structures, the modular invariant flavor models have been widely studied.
In these models three generations of quarks and leptons are regarded as three-dimensional (reducible or irreducible) representations of the finite modular groups.
Their mass matrices are written in terms of the modular forms for the finite groups, which are holomorphic functions of the modulus $\tau$ \cite{Feruglio:2017spp} \footnote{The modular flavor symmetry was also studied from the top-down approach such as stringtheory \cite{Ferrara:1989bc,Ferrara:1989qb,Lerche:1989cs,Lauer:1989ax,Lauer:1990tm,Kobayashi:2018rad,Kobayashi:2018bff,Ohki:2020bpo,Kikuchi:2020frp,Kikuchi:2020nxn,
Kikuchi:2021ogn,Almumin:2021fbk,Baur:2019iai,Nilles:2020kgo,Baur:2020jwc,Nilles:2020gvu}.}.
Interestingly, the finite modular groups $\Gamma_N$ for $N=2,3,4,5$ are isomorphic to the non-Abelian discrete groups $S_3$, $A_4$, $S_4$ and $A_5$, respectively \cite{deAdelhartToorop:2011re}.
These non-Abelian groups have been used in the flavor models for quarks and leptons \cite{Altarelli:2010gt,Ishimori:2010au,Ishimori:2012zz,Kobayashi:2022moq,Hernandez:2012ra,King:2013eh,King:2014nza,Tanimoto:2015nfa,King:2017guk,Petcov:2017ggy,Feruglio:2019ktm}.
Motivated by this, the modular symmetric lepton flavor models have been proposed in $\Gamma_2\simeq S_3$ \cite{Kobayashi:2018vbk}, $\Gamma_3\simeq A_4$ \cite{Feruglio:2017spp}, $\Gamma_4\simeq S_4$ \cite{Penedo:2018nmg} and $\Gamma_5\simeq A_5$ \cite{Novichkov:2018nkm,Ding:2019xna}.
Also modular symmetries including higher levels and covering groups were studied
\cite{Li:2021buv,Ding:2020msi,Kobayashi:2018bff,Liu:2019khw,Novichkov:2020eep,Liu:2020akv,Liu:2020msy}.

Phenomenological studies using modular forms have been implemented in many works \cite{
%Feruglio:2017spp,Kobayashi:2018vbk,Penedo:2018nmg,Novichkov:2018nkm,
Criado:2018thu,
Kobayashi:2018scp,
Ding:2019zxk,
Novichkov:2018ovf,
Kobayashi:2019mna,Wang:2019ovr,%Ding:2019xna,
%Liu:2019khw,
Chen:2020udk,%Novichkov:2020eep,
%Liu:2020akv,
deMedeirosVarzielas:2019cyj,
  	Asaka:2019vev,%Ding:2020msi,
Asaka:2020tmo,deAnda:2018ecu,Kobayashi:2019rzp,Novichkov:2018yse,Kobayashi:2018wkl,Okada:2018yrn,Okada:2019uoy,Nomura:2019jxj, Okada:2019xqk,
  	Nomura:2019yft,Nomura:2019lnr,Criado:2019tzk,
  	King:2019vhv,Gui-JunDing:2019wap,deMedeirosVarzielas:2020kji,Zhang:2019ngf,Nomura:2019xsb,Kobayashi:2019gtp,Lu:2019vgm,Wang:2019xbo,King:2020qaj,Abbas:2020qzc,Okada:2020oxh,Okada:2020dmb,Ding:2020yen,Okada:2020rjb,Okada:2020ukr,Nagao:2020azf,Wang:2020lxk,
  	Okada:2020brs,Yao:2020qyy}.
Nevertheless it is a difficult issue to describe flavor structures by fewer parameters.
Especially it seems that some kind of fine-tuning is necessary to realize the large hierarchies of fermions masses. 
Indeed many works need to fine-tune coefficients of modular forms in Yukawa couplings for hierarchical masses, in particular quark mass hierarchies.

One way to describe hierarchical fermion masses without fine-tuning is use of the residual symmetry of the modular symmetry.
At three modular fixed points of the modulus, $\tau=i$, $\omega$ $(=e^{2\pi i/3})$ and $i\infty$, the modular symmetry breaks into residual $Z_2$, $Z_3$ and $Z_N$ symmetries, respectively, where $N$ is the level of the finite modular group \cite{Novichkov:2018ovf}.
Due to these residual symmetries, values of the modular forms become hierarchical in the vicinity of the modular fixed points depending on their residual charges.
Thus, deviation of the modulus $\tau$ from the modular fixed points can generate hierarchical structures of fermion masses.
Along in this way, the lepton flavor structure was successfully described without fine-tuning in modular invariant models  Ref.~\cite{Feruglio:2021dte,Novichkov:2021evw}.
Also in Ref.~\cite{Petcov:2022fjf}, the quark flavor structure was described by $\Gamma_3\simeq A_4$ modular symmetry in the vicinity of $\tau=\omega$.
They combined relaxation of quark masses by ${\cal O}(10)$ coefficient in Yukawa couplings to reproduce large quark mass hierarchies.
Realization of  the quark flavor structure which originates solely from the deviation of the modulus was studied in $\Gamma_6\simeq A_4\times S_3$ \cite{Kikuchi:2023cap} and $S'_4\times S_3$ \cite{Abe:2023ilq}.

Higher dimensional theories such as superstring theory can be an origin of the modular symmetry. 
For example, the torus compactification $T^2_1\times T^2_2\times T^2_3$ of extra six-dimensions in the superstring theory has the modular symmetry as a geometrical symmetry.
Actually some modular forms are derived from the torus compactification of the low-energy effective theory of the superstring theory with magnetic flux background \cite{Kobayashi:2018rad,Kobayashi:2018bff,Ohki:2020bpo,Kikuchi:2020frp,Kikuchi:2020nxn,Kikuchi:2021ogn, Cremades:2004wa}.
Therefore it may be expected that the modular invariant models with $\Gamma_{N_1}\times \Gamma_{N_2}\times \Gamma_{N_3}$ originate from $T^2_1\times T^2_2\times T^2_3$.
Indeed, quark flavors in $\Gamma_6\simeq A_4\times S_3$ \cite{Kikuchi:2023cap} and $S'_4\times S_3$ \cite{Abe:2023ilq} may be derived from the torus compactification with the moduli stabilization $\tau_1=\tau_2\equiv \tau$.
Also multi modular symmetries were studied in Refs.~\cite{King:2021fhl,Du:2022lij,Abbas:2022slb}.
Inspired by this point, we study the quark flavor structure in the modular invariant models with the  $A_4\times A_4\times A_4$ 
symmetry.
For simplicity, we focus on the case that all of the moduli values are same, i.e.  $\tau_1=\tau_2=\tau_3\equiv\tau$.
We discuss two modular fixed points $\tau=\omega$ and $i\infty$ where $A_4\times A_4\times A_4$ breaks into $Z_3\times Z_3\times Z_3$.
Hence, Yukawa couplings can possess $Z_3\times Z_3\times Z_3$ charges 0 to 6 at $\tau=\omega$ and $i\infty$.
Thus, we can obtain hierarchical values of Yukawa couplings such as 1, $\varepsilon$, $\varepsilon^2$, $\varepsilon^3$, $\varepsilon^4$, $\varepsilon^5$ and $\varepsilon^6$ in the vicinity of $\tau=\omega$ and $i\infty$, where $\varepsilon$ denotes the deviation of the modulus from the modular fixed points.

This paper is organized as follows.
In section 2, we study general aspects of $A_4 \times A_4 \times A_4$ modular symmetric quark flavor models leading to the desirable hierarchical structures without fine-tuning. 
In section 3, we perform more concrete analysis of $A_4 \times A_4 \times A_4$ quark flavor models with numerical examples.
In section 4, CP violation is discussed. Section 5 is our conclusion.
We give brief reviews of the group theoretical aspects of $A_4$ and modular forms of $A_4$ in appendix \ref{app:A4group} and \ref{app:A4modularforms}.
We classify the phase factors and hierarchical structures of mass matrices in phenomenologically viable models obtained by our studies in appendix \ref{app:viable_models}.

%----------------------------------------------------------------------
%----------------------------------------------------------------------
%----------------------------------------------------------------------

\section{Quark mass hierarchy without fine-tuning}
\label{sec:quark_mass}

The two generators of the modular group are denoted by $S$ and $T$, 
which are represented by the $2 \times 2$ matrices as,
\begin{align}
T=
\begin{pmatrix}
1 & 1 \\ 0 & 1
\end{pmatrix},\qquad  
S =
\begin{pmatrix}
0 & 1 \\ -1 & 0
\end{pmatrix}.
\end{align} 
They act on the modulus $\tau$ as 
\begin{align}
T~:~\tau \xrightarrow{T} \tau +1, \qquad S~:~\tau \xrightarrow{S} -1/\tau.
\end{align}
There are three fixed points, $\tau=i, \omega$, and $i \infty$.
The $Z_3$ symmetry remains at $\tau = \omega$ and $i \infty$, while 
the $Z_2$ symmetry remains at $\tau=i$.

Here, we present general aspects of modular $A_4 \times A_4 \times A_4$ quark flavor models without fine-tuning. 
Firstly, we assign modular weights to supermultiplets. 
In general, a superfield may have different modular weights among the first, second, and third $A_4$'s.
However, we consider the simplest case that each superfield has the same weights of three $A_4$'s. 
Thus, assignments of weights corresponding to one of the $A_4$ are shown below.
\begin{itemize}
    \item{quark doublets $Q = (Q^1, Q^2, Q^3)$ are three-dimensional representation (reducible or irreducible) of $A_4$ with modular weight $-k_Q$.} 
    \item{up sector quark singlets $u_R = (u_R^1 , u_R^2, u_R^3)$ are three-dimensional representation  (reducible or irreducible) of $A_4$ with modular weight $-k_u$.}
    \item{down sector quark singlets  $d_R = (d_R^1, d_R^2, d_R^3)$ are three-dimensional representation  (reducible or irreducible) of $A_4$ with weight $-k_d$.} 
    \item{both up and down sector Higgs fields $H_{u,d}$ are one-dimensional representations of $A_4$ with modular weight $-k_{H_{u,d}}$.}
\end{itemize}

Secondly, we write down the general form of $A_4$ invariant superpotential for the up sector as
\begin{equation}
\label{eq: up_superpotential}
    W_u = \sum_{{\bf{r}}}
    \left[ Y_{{\bf{r}}}^{(k_{Y_u})} (Q^1\ Q^2\ Q^3) 
    \begin{pmatrix}
    \alpha_{{\bf{r}}}^{11} & \alpha_{{\bf{r}}}^{12} & \alpha_{{\bf{r}}}^{13} \\
    \alpha_{{\bf{r}}}^{21} & \alpha_{{\bf{r}}}^{22} & \alpha_{{\bf{r}}}^{23} \\
    \alpha_{{\bf{r}}}^{31} & \alpha_{{\bf{r}}}^{32} & \alpha_{{\bf{r}}}^{33}
    \end{pmatrix}
    \begin{pmatrix}
    u_R^1 \\ u_R^2 \\ u_R^3
    \end{pmatrix}
    H_u
    \right]_{\bf{1}},
\end{equation}
where some of coupling constants $\alpha^{ij}$ can be related each other depending on the representations of $Q$ and $u_R$.
In above $Y_{{\bf{r}}}^{(k_{{Y}_u})}$ denote modular forms which transform as  irreducible representations ${\bf{r}}$ of $A_4$ with modular weight $k_{Y_u}=k_{Q}+k_{u}+k_{H_u}$.
This ensures the cancellation of modular weights with quark and Higgs fields. Thus, modular invariant superpotential is obtained once the trivial singlet terms are picked up from above combinations, which is represented by writing $\bf{1}$.  

Extension to $A_4 \times A_4 \times A_4$ is straightforward.
The superpotential term for the up sector can be written by
\begin{equation}
\label{eq: up_superpotential_(A4)^3}
    W_u = \sum_{{\bf{r}}_1, {\bf{r}}_2, {\bf{r}}_3}
    \left[ Y_{{\bf{r}}_1}^{(k_{Y_u})}
    Y_{{\bf{r}}_2}^{(k_{Y_u})}
    Y_{{\bf{r}}_3}^{(k_{Y_u})}
    (Q^1\ Q^2\ Q^3) 
    \begin{pmatrix}
    \alpha_{{\bf{r}}_1{\bf{r}}_2{\bf{r}}_3}^{11} &  \alpha_{{\bf{r}}_1{\bf{r}}_2{\bf{r}}_3}^{12} &  \alpha_{{\bf{r}}_1{\bf{r}}_2{\bf{r}}_3}^{13} \\
    \alpha_{{\bf{r}}_1{\bf{r}}_2{\bf{r}}_3}^{21} &  \alpha_{{\bf{r}}_1{\bf{r}}_2{\bf{r}}_3}^{22} &  \alpha_{{\bf{r}}_1{\bf{r}}_2{\bf{r}}_3}^{23} \\
     \alpha_{{\bf{r}}_1{\bf{r}}_2{\bf{r}}_3}^{31} &  \alpha_{{\bf{r}}_1{\bf{r}}_2{\bf{r}}_3}^{32} &  \alpha_{{\bf{r}}_1{\bf{r}}_2{\bf{r}}_3}^{33}
    \end{pmatrix}
    \begin{pmatrix}
    u_R^1 \\ u_R^2 \\ u_R^3
    \end{pmatrix}
    H_u
    \right]_{\bf{1}},
\end{equation}
where 
$Y_{{\bf{r}}_n}^{k_{(Y_u)}}, (n=1,2,3)$ denote the modular forms which transform as irreducible representations of ${\bf{r}}_n$ with respect to $n$-th $A_4$. By taking  products of them as in Eq.(\ref{eq: up_superpotential_(A4)^3}) the modular weights are cancelled for each $A_4$.\footnote{Also in magnetized $T^2 \times T^2 \times T^2$ compactification, Yukawa couplings are given by the product of three modular forms corresponding to the contribution of each torus\cite{Cremades:2004wa}.}  

Similarly, superpotential for the down sector is written as
\begin{equation}
\label{eq: down_superpotential}
     W_d = \sum_{{\bf{r}}_1, {\bf{r}}_2, {\bf{r}}_3}
    \left[ Y_{{\bf{r}}_1}^{(k_{Y_d})}
    Y_{{\bf{r}}_2}^{(k_{Y_d})}
    Y_{{\bf{r}}_3}^{(k_{Y_d})}
    (Q^1\ Q^2\ Q^3) 
    \begin{pmatrix}
    \beta_{{\bf{r}}_1{\bf{r}}_2{\bf{r}}_3}^{11} &  \beta_{{\bf{r}}_1{\bf{r}}_2{\bf{r}}_3}^{12} &  \beta_{{\bf{r}}_1{\bf{r}}_2{\bf{r}}_3}^{13} \\
    \beta_{{\bf{r}}_1{\bf{r}}_2{\bf{r}}_3}^{21} &  \beta_{{\bf{r}}_1{\bf{r}}_2{\bf{r}}_3}^{22} &  \beta_{{\bf{r}}_1{\bf{r}}_2{\bf{r}}_3}^{23} \\
    \beta_{{\bf{r}}_1{\bf{r}}_2{\bf{r}}_3}^{31} &  \beta_{{\bf{r}}_1{\bf{r}}_2{\bf{r}}_3}^{32} &  \beta_{{\bf{r}}_1{\bf{r}}_2{\bf{r}}_3}^{33}
    \end{pmatrix}
    \begin{pmatrix}
    d_R^1 \\ d_R^2 \\ d_R^3
    \end{pmatrix}
    H_d
    \right]_{\bf{1}},
\end{equation}
where $k_{Y_d}=k_Q + k_d + k_{H_d}$.
Mass terms are obtained when the Higgs fields acquire non-zero vacuum expectation values as
\begin{align}
\begin{aligned}
\label{eq: up_mass_(A4)^3}
    (Q^1\ Q^2\ & Q^3) M_u
    \begin{pmatrix}
      u_R^1 \\ u_R^2 \\ u_R^3
    \end{pmatrix}  \\
    &= \sum_{{\bf{r}}_1, {\bf{r}}_2, {\bf{r}}_3}
    \left[ \prod_{n=1}^3 Y_{{\bf{r}}_n}^{(k_{Y_u})}
    (Q^1\ Q^2\ Q^3) 
    \begin{pmatrix}
    \alpha_{{\bf{r}}_1{\bf{r}}_2{\bf{r}}_3}^{11} &  \alpha_{{\bf{r}}_1{\bf{r}}_2{\bf{r}}_3}^{12} &  \alpha_{{\bf{r}}_1{\bf{r}}_2{\bf{r}}_3}^{13} \\
    \alpha_{{\bf{r}}_1{\bf{r}}_2{\bf{r}}_3}^{21} &  \alpha_{{\bf{r}}_1{\bf{r}}_2{\bf{r}}_3}^{22} &  \alpha_{{\bf{r}}_1{\bf{r}}_2{\bf{r}}_3}^{23} \\
     \alpha_{{\bf{r}}_1{\bf{r}}_2{\bf{r}}_3}^{31} &  \alpha_{{\bf{r}}_1{\bf{r}}_2{\bf{r}}_3}^{32} &  \alpha_{{\bf{r}}_1{\bf{r}}_2{\bf{r}}_3}^{33}
    \end{pmatrix}
    \begin{pmatrix}
    u_R^1 \\ u_R^2 \\ u_R^3
    \end{pmatrix}
   \langle H_u \rangle
    \right]_{\bf{1}},
    \end{aligned}
\end{align}

\begin{align}
\begin{aligned}
\label{eq: down_mass_(A4)^3}
    (Q^1\ Q^2\ & Q^3) M_d
    \begin{pmatrix}
      d_R^1 \\ d_R^2 \\ d_R^3
    \end{pmatrix} \\
    &= \sum_{{\bf{r}}_1, {\bf{r}}_2, {\bf{r}}_3}
    \left[ \prod_{n=1}^3 Y_{{\bf{r}}_n}^{(k_{Y_d})}
    (Q^1\ Q^2\ Q^3) 
 \begin{pmatrix}
    \beta_{{\bf{r}}_1{\bf{r}}_2{\bf{r}}_3}^{11} &  \beta_{{\bf{r}}_1{\bf{r}}_2{\bf{r}}_3}^{12} &  \beta_{{\bf{r}}_1{\bf{r}}_2{\bf{r}}_3}^{13} \\
    \beta_{{\bf{r}}_1{\bf{r}}_2{\bf{r}}_3}^{21} &  \beta_{{\bf{r}}_1{\bf{r}}_2{\bf{r}}_3}^{22} &  \beta_{{\bf{r}}_1{\bf{r}}_2{\bf{r}}_3}^{23} \\
    \beta_{{\bf{r}}_1{\bf{r}}_2{\bf{r}}_3}^{31} &  \beta_{{\bf{r}}_1{\bf{r}}_2{\bf{r}}_3}^{32} &  \beta_{{\bf{r}}_1{\bf{r}}_2{\bf{r}}_3}^{33}
    \end{pmatrix}
    \begin{pmatrix}
    d_R^1 \\ d_R^2 \\ d_R^3
    \end{pmatrix}
   \langle H_d \rangle
    \right]_{\bf{1}}.
\end{aligned}
\end{align}

 We only use the complex structure modulus $\tau$ as a continuous free-parameter. 
In order to realize quark masses and mixing angles, we do not consider fine-tuning of coupling constants 
 $\alpha^{ij}$ and $\beta^{ij}$, but  we expect that they are typically $\mathcal{O}(1)$.
In order to make our point clear, 
we restrict them to either $+1$ or $-1$, i.e.
\begin{equation}
    \begin{pmatrix}
    \alpha^{11} & \alpha^{12} & \alpha^{13} \\
    \alpha^{21} & \alpha^{22} & \alpha^{23} \\
    \alpha^{31} & \alpha^{32} & \alpha^{33}
    \end{pmatrix} =
    \begin{pmatrix}
    +1 & +1 & +1 \\
    +1 & \pm1 & \pm1 \\
    +1 & \pm1 & \pm1
    \end{pmatrix},\quad 
        \begin{pmatrix}
    \beta^{11} & \beta^{12} & \beta^{13} \\
    \beta^{21} & \beta^{22} & \beta^{23} \\
    \beta^{31} & \beta^{32} & \beta^{33}
    \end{pmatrix} =
    \begin{pmatrix}
    +1 & +1 & +1 \\
    \pm1 & \pm1 & \pm1 \\
    \pm1 & \pm1 & \pm1
    \end{pmatrix}.
\end{equation}
By using these values of $\alpha^{ij}$ and $\beta^{ij}$, we try to realize the order of quark mass ratios and mixing angles.
Note that we may fix the signs of $(1,1), (1,2), (1,3), (2,1)$ and $(3,1)$ components of $\alpha^{ij}$ to $+1$ by redefinition of fields $Q$ and $u_R$. Similarly, the signs of $(1,1), (1,2)$ and $(1,3)$ components of $\beta^{ij}$ are fixed to $+1$ by the redefinition of field $d_R$.

Thirdly, to reproduce hierarchies in quark mass ratios without fine-tuning, modular forms must be the source of hierarchical structures. This can be achieved when the complex structure modulus $\tau$ takes its value in the vicinity of modular fixed points, $\tau = i, \omega$ and $i \infty$. At the fixed points, $Z_N$ residual symmetries exist.
For example, $A_4$ is broken to $Z_3$ when $\tau = i \infty$ and $\omega$.
In addition, we have $Z_2$ residual symmetry at $\tau =i$.
Since $Z_3$ residual symmetry is more attractive to produce large hierarchies of quarks,  we  study $\tau = i \infty $ 
and $\omega$ in this paper.

For illustration, we begin with considering a single $A_4$ symmetry. 
Suppose that quark doublets $Q$, up sector quark singlets $u_R$ and up-type Higgs field $H_u$ with the following $Z_3$ residual charges,
\begin{equation}
    Q:(1,1,0),\ u_R: (0,1,0),\ H_u:0.
\end{equation}
Then the modular invariance of the superpotential $W_u$ fixes the $Z_3$ residual charges of up sector mass matrix $M_u$ as 
\begin{equation}
    M_u^{ij}: 
    \begin{pmatrix}
      2 & 1 & 2 \\
      2 & 1 & 2\\
      0 & 2  & 0
    \end{pmatrix}.
\end{equation}
When $\tau$ is in the vicinity of the fixed point, modular form $f(\tau)$ with $Z_3$ residual charge $r$ can be expanded by powers of the deviation from symmetric point as\cite{Novichkov:2021evw},
\begin{itemize}
    \item{$\tau \sim i \infty:\ f(\tau) \sim \varepsilon^r,\quad \varepsilon \propto  q = e^{2\pi i \tau/3}$,}
    \item{$\tau \sim \omega:\  f(\tau) \sim \varepsilon^r,\quad  \varepsilon \propto u = \frac{\tau - \omega}{\tau - \omega^2}$.} 
\end{itemize}
Thus, the following hierarchical structure is generated in $M_u^{ij}$, 
\begin{equation}
    M_u^{ij} \sim 
    \begin{pmatrix}
      \varepsilon^2 & \varepsilon & \varepsilon^2 \\
      \varepsilon^2 & \varepsilon & \varepsilon^2 \\
      1 & \varepsilon^2 & 1
   \end{pmatrix}.
\end{equation}
In this way, hierarchies in mass matrices can be generated by the values of modular forms.

However, to realize the quark mass ratios in both up and down sectors, $\varepsilon$ up to the power of $2$ seems not enough.
Hence, we consider the direct product $A_4 \times A_4 \times A_4$ which would yield $\varepsilon$ up to the power of $6$. This is possible because three modular forms corresponding to each $A_4$ are multiplied in the superpotential as in Eqs.(\ref{eq: up_superpotential_(A4)^3}) and (\ref{eq: down_superpotential}). 
Then, we expect to obtain phenomenologically viable quark flavor models as we study in the next section.

%----------------------------------------------------------------------
%----------------------------------------------------------------------
%----------------------------------------------------------------------

\section{The models with $A_4\times A_4\times A_4$}
\label{sec:models_A4x3}

In this section, we study concrete models with $A_4 \times A_4 \times A_4$ modular symmetry when $\tau$ is in the vicinity of $i \infty$ and $\omega$.

Here, we only use singlet modular forms of $A_4$.
There are three singlets, $\bf{1}, \bf{1}'$ and $\bf{1}''$ in $A_4$ as reviewed in appendix \ref{app:A4group}.
They represent the generators $S$ and $T$ as
\begin{equation}
\label{eq: A4_singlet_rep}
    S({\bf1'}) = 1,\quad S({\bf1''}) = 1,\quad T({\bf1'}) = \omega,\quad T({\bf1''}) = \omega^2.
\end{equation}
We have modular forms corresponding to each singlet at modular weight $8$.
As shown in appendix \ref{app:A4modularforms}, when the weight is less than $8$, there is a lack of modular forms.
Thus, our assignments of modular weights are 
\begin{equation}
    k_{Y_{u}}=k_{Y_d}=8,\quad k_Q = k_u = k_d = 4,\quad k_{H_u}=k_{H_d} = 0.
\end{equation}
The Higgs fields are always assigned to the trivial singlet.

First, we look at the case when $\tau \sim i \infty$. Under the $T$-transformation, $\tau = i\infty$ is invariant.
%\footnote{$T$- and $S$-transformations are given by
%$T = 
%\begin{pmatrix}
%1 & 1 \\ 0 & 1
%\end{pmatrix},\ 
%S =
%\begin{pmatrix}
%0 & 1 \\ -1 & 1
%\end{pmatrix}$. They act on $\tau$ as $\tau \xrightarrow{T} \tau +1$ and $\tau \xrightarrow{S} -1/\tau$.} 
This means the residual charge is the $T$-charge. Therefore, three singlet modular forms of $A_4$ with weight 8 show corresponding dependence on $\varepsilon$ as shown in Table \ref{tab: A4_T-charge}.
\begin{table}[H]
  \begin{center}
    \renewcommand{\arraystretch}{1.3}
    $\begin{array}{c|c c c} \hline
     \textrm{singlet modular form } & Y_{\bf{1}}^{(8)} & Y_{\bf{1}'}^{(8)} & Y_{\bf{1}''}^{(8)} \\ \hline
    \textrm{$T$-charge} & 0 & 1 & 2 \\ \hline
    \textrm{order} & 1 & \varepsilon & \varepsilon^2 \\ \hline
    \end{array}$
  \end{center}
  \caption{$T$-charges of three $A_4$ singlets and their orders in the vicinity of $\tau = i\infty$.}
\label{tab: A4_T-charge}
\end{table}

Next, consider the case when $\tau \sim \omega$. Under the $ST$-transformation, $\tau = \omega$ is invariant. This means the residual charge is related to the $ST$-charge taking into account the effect of automorphy factor. Let us briefly explain it based on the discussion in Ref.\cite{Petcov:2022fjf}.
Under the $ST$-transformation, we have
\begin{equation}
    Y_{\bf{r}}^{(8)}(\tau) \xrightarrow{ST} Y_{\bf{r}}^{(8)}(-1/(\tau + 1)) = (-1-\tau)^{8} \rho_{\bf r}(ST) Y_{\bf{r}}^{(8)}(\tau), 
\end{equation}
where ${\bf{r}} \in \{\bf{1}, \bf{1}', \bf{1}'' \}$. From Eq. (\ref{eq: A4_singlet_rep}), we have $\rho_{\bf{1}}(ST)=1,\ \rho_{\bf{1}'}(ST)=\omega$ and ${\rho}_{{\bf{1}}''}(ST) = \omega^2$. For convenience, let us define $\tilde{\rho}_{\bf r} \equiv \omega^{-8} \rho_{\bf r}$. Then, we obtain
\begin{equation}
 Y_{\bf{r}}^{(8)}(-1/(\tau + 1)) = [-\omega (1+\tau)]^{8} \tilde{\rho}_{\bf r}(ST) Y_{\bf{r}}^{(8)}(\tau).
\end{equation}
A convenient parameter for the deviation of $\tau$ from $\omega$ is \cite{Novichkov:2021evw}
\begin{equation}
    u \equiv \frac{\tau - \omega}{\tau - \omega^2}.
\end{equation}
By noting $u \xrightarrow{ST} \omega^2 u$, we find
\begin{equation}
     Y_{\bf{r}}^{(8)}(\omega^2 u) = \left( \frac{1-\omega^2 u}{1-u}
     \right)^{8} \tilde{\rho}_{\bf r}(ST) Y_{\bf{r}}^{(8)}(u),
\end{equation}
where we regard $Y_{\bf r}^{(8)}$ as functions of $u$. If we define $\tilde{Y}_{\bf r}^{(8)}(u) \equiv (1-u)^{-8} Y^{(8)}_{\bf r}(u)$, we get
\begin{equation}
    {\tilde{Y}_{\bf{r}}}^{(8)}(\omega^2 u) = \tilde{\rho}_{\bf{r}}(ST) {\tilde{Y}_{\bf{r}}}^{(8)}(u).
\end{equation}
Expansion with respect to $u,\ (|u| \ll 1)$ yields
\begin{equation}
    (\omega^{2l}-\tilde{\rho}_{\bf{r}}(ST)) \frac{d^{l}{\tilde{Y}_{\bf{r}}}^{(8)}(0)}{du^{l}} =0.
\end{equation}
This relation shows when $l(=0,1,2)$ satisfies $\omega^{2l}-\tilde{\rho}_{\bf r}(ST)=0$, the modular forms behave as ${\tilde{Y}_{\bf r}}^{(8)} \sim  Y_{\bf r}^{(8)} \sim \mathcal{O}(|u|^l)$. We call such $l$ as $ST$-charge, namely $( \textrm{$ST$-charge}) \equiv 2-(\textrm{$T$-charge}) \pmod{3}$. Three singlet modular forms show the behaviors as in Table
\ref{tab: A4_Z3-charge}.
\begin{table}[H]
  \begin{center}
    \renewcommand{\arraystretch}{1.3}
    $\begin{array}{c|c c c} \hline
     \textrm{singlet modular form } & Y_{\bf{1}}^{(8)} & Y_{\bf{1}'}^{(8)} & Y_{\bf{1}''}^{(8)} \\ \hline
    \textrm{$ST$-charge} & 2 & 1 & 0 \\ \hline
    \textrm{order} & \varepsilon^2 & \varepsilon & 1 \\ \hline
    \end{array}$
  \end{center}
      \caption{$ST$-charges of three $A_4$ singlets and their orders in the vicinity of $\tau = \omega$.}
\label{tab: A4_Z3-charge}
\end{table}

The reference values of up and down quark mass ratios are shown in Table \ref{tab:quark_masses}.
Values at a high scale energy include renormalization group effects, which depend on the scenario.
We use the values of Refs.~\cite{Antusch:2013jca,Bjorkeroth:2015ora} at the GUT scale  in the minimal supersymmetric standard model with 
$\tan \beta =5$.
\begin{table}[H]
    \begin{center}
    \renewcommand{\arraystretch}{1.3}
    \begin{tabular}{c|cccc} \hline
      & $\frac{m_u}{m_t}\times10^{6}$ & $\frac{m_c}{m_t}\times10^3$ & $\frac{m_d}{m_b}\times10^4$ & $\frac{m_s}{m_b}\times10^2$ \\ \hline
      GUT scale values & $5.39$ & $2.80$ & $9.21$ & $1.82$ \\
      $1\sigma$ errors & $\pm 1.68$ & $\pm 0.12$ & $\pm 1.02$ & $\pm 0.10$ \\ \hline
    \end{tabular}
  \end{center}
      \caption{Quark mass ratios at GUT scale $2\times 10^{16}$ GeV with $\tan \beta=5$ \cite{Antusch:2013jca,Bjorkeroth:2015ora}.}
\label{tab:quark_masses}
\end{table}
To realize these hierarchical structures of quarks, let us consider the mass matrices of the form,
\begin{align}
\label{eq: hierarchy_mass}
    M_u &\propto 
    \begin{pmatrix}
    \mathcal{O}(\varepsilon^6) & * & * \\
    * & \mathcal{O}(\varepsilon^3) & * \\
    * & * & \mathcal{O}(1)
    \end{pmatrix},\quad
     M_d \propto 
    \begin{pmatrix}
    \mathcal{O}(\varepsilon^4) & * & * \\
    * & \mathcal{O}(\varepsilon^2) & * \\
    * & * &  \mathcal{O}(1)
    \end{pmatrix},
\end{align}
where we assume $\varepsilon \sim 0.15$ and the order is unfixed for elements with $*$ at this stage.

%----------------------------------------------------------------------
%----------------------------------------------------------------------

\subsection{Types}

We concentrate on $A_4 \times A_4 \times A_4$ models which lead to mass matrices of the form shown in Eq. (\ref{eq: hierarchy_mass}). We find a number of possibilities in generating $\varepsilon^n$ depending on how much each of the $A_4$ contributes to the power.
Thus, let us distinguish contributions from each $A_4$ in producing $\varepsilon^n,\ n\in \{1,\cdots, 6\}$. We denote the contribution of $i$-th $A_4$ by $\varepsilon_i$ where $i=1,2,3$. 

We have $2$ possibilities in $M_u$,
\begin{align}
   \text{Type $123$}:\  M_u &\propto
    \begin{pmatrix}
    \mathcal{O}({\varepsilon_1}^2{\varepsilon_2}^2{\varepsilon_3}^2) & & \\
    & \mathcal{O}({\varepsilon_1}{\varepsilon_2}{\varepsilon_3}) & \\
    & & \mathcal{O}(1)
    \end{pmatrix},\\
    \label{eq: 1^22}
    \text{Type $1^22$}:\  M_u &\propto
    \begin{pmatrix}
    \mathcal{O}({\varepsilon_1}^2{\varepsilon_2}^2{\varepsilon_3}^2) & & \\
    & \mathcal{O}({\varepsilon_1}^2{\varepsilon_2}) & \\
    & & \mathcal{O}(1)
    \end{pmatrix}.
\end{align}
Type $123$ has a symmetry under the permutation of three $A_4$'s. On the other hand, we
do not have such symmetry in type $1^22$.
Instead we do not need to consider other types $1^23$, $12^2$, $2^23$, $13^2$ and $23^2$ which are equivalent to Eq.(\ref{eq: 1^22}) up to the permutation of three $A_4$'s.
For example, up quark mass matrix $M_u$ of type $1^23$,
\begin{align}
\text{Type $1^23$}:\ M_u \propto 
\begin{pmatrix}
\mathcal{O}({\varepsilon_1}^2{\varepsilon_2}^2{\varepsilon_3}^2) & & \\
& \mathcal{O}({\varepsilon_1}^2\varepsilon_3) & \\
& & \mathcal{O}(1) \\
\end{pmatrix},
\end{align}
is equivalent to one of type $1^22$ up to the permutation of $\varepsilon_2$ and $\varepsilon_3$, that is, the permutation of the second and third $A_4$.
Similarly, it can be shown that other types are equivalent to type $1^22$ up to the permutation.

%----------------------------------------------------------------------

\subsubsection{Type $123$}

We have $8$ patterns of down quark mass matrix $M_d$ when $M_u$ is in type 123. The $\mathcal{O}(\varepsilon^4)$ element has $2$ patterns,
\begin{align}
\begin{aligned}
    \text{$1^22^2$}:\ M_d &\propto
    \begin{pmatrix}
    \mathcal{O}({\varepsilon_1}^2{\varepsilon_2}^2) & & \\
    & \mathcal{O}(\varepsilon^2) & \\
    & & \mathcal{O}(1)
    \end{pmatrix}, \\
    \text{$123^2$}:\ M_d &\propto
    \begin{pmatrix}
    \mathcal{O}(\varepsilon_1\varepsilon_2{\varepsilon_3}^2) & & \\
    & \mathcal{O}(\varepsilon^2) & \\
    & & \mathcal{O}(1)
    \end{pmatrix}.
\end{aligned}
\end{align}
In both cases, the permutation symmetry is partially broken. We still have a symmetry under the exchange of first and second $A_4$'s. 
Thus, we only need to treat 4 patterns of $\mathcal{O}(\varepsilon^2)$ element given by
\begin{align}
\begin{aligned}
\label{eq: O(e^2)}
\text{12}:\ \mathcal{O}(\varepsilon_1 \varepsilon_2),\quad
\text{23}:\ \mathcal{O}(\varepsilon_2 \varepsilon_3), \quad
\text{$1^2$}:\ \mathcal{O}({\varepsilon_1}^2),\quad
\text{$3^2$}:\ \mathcal{O}({\varepsilon_3}^2).
\end{aligned}
\end{align}

%----------------------------------------------------------------------

\subsubsection{Type $1^22$}

We have 36 patterns of down quark mass matrix $M_d$ when $M_u$ is in type $1^22$. The $\mathcal{O}(\varepsilon^4)$ element of $M_d$ in Eq. (\ref{eq: hierarchy_mass}) is given by,
\begin{align}
    \begin{aligned}
\text{$1^22^2$}:\ &  \mathcal{O}({\varepsilon_1}^2 {\varepsilon_2}^2),\quad 
\text{$2^23^2$}:\ 
\mathcal{O}({\varepsilon_2}^2{\varepsilon_3}^2),\quad 
\text{$1^23^2$}:\  \mathcal{O}({\varepsilon_1}^2{\varepsilon_3}^2), \\
\text{$1^223$}:\ &
\mathcal{O}({\varepsilon_1}^2{\varepsilon_2}{\varepsilon_3}),\quad
\text{$12^23$}:\  \mathcal{O}({\varepsilon_1}{\varepsilon_2}^2{\varepsilon_3}),\quad 
\text{$123^2$}:\  \mathcal{O}({\varepsilon_1} {\varepsilon_2}{\varepsilon_3}^2).
    \end{aligned}
\end{align}

 The $\mathcal{O}(\varepsilon^2)$ element of $M_d$ in Eq. (\ref{eq: hierarchy_mass}) is given by,
\begin{align}
\begin{aligned}
\text{$1^2$}:\ &  \mathcal{O}({\varepsilon_1}^2),
\quad 
\text{$2^2$}:\ 
\mathcal{O}({\varepsilon_2}^2),
\quad 
\text{$3^2$}:\  \mathcal{O}({\varepsilon_3}^2),\\
\text{$12$}:\ &  \mathcal{O}({\varepsilon_1} {\varepsilon_2}),\quad \text{$23$}:\ 
\mathcal{O}({\varepsilon_2}{\varepsilon_3}),\quad 
\text{$13$}:\  \mathcal{O}({\varepsilon_1}{\varepsilon_3}).
\end{aligned}
\end{align}

%\subsubsection{Type $1^22$}
%We have $36$ patterns of down quark mass matrix $M_d$ when $M_u$ is in type $1^22$. The $\mathcal{O}(\varepsilon^4)$ element has $6$ patterns,
%\begin{align}
%\begin{aligned}
%    1^22^2:&\  \mathcal{O}(\varepsilon_1^2 \varepsilon_2^2),\quad
%    23:\ \mathcal{O}(\varepsilon_1^2 \varepsilon_3^2),\quad
%    2^23^3:\  \mathcal{O}(\varepsilon_2^2 \varepsilon_3^2),\\
%    123^2:\  \mathcal{O}&(\varepsilon_1 \varepsilon_2 \varepsilon_3^2),\quad
%    12^23:\  \mathcal{O}(\varepsilon_1 \varepsilon_2^2 \varepsilon_3),\quad
%    1^223:\  \mathcal{O}(\varepsilon_1^2 \varepsilon_2 \varepsilon_3).
%\end{aligned}
%\end{align}
%The $\mathcal{O}(\varepsilon^2)$ element has $6$ patterns,
%\begin{align}
%\begin{aligned}
%    1^2:\  \mathcal{O}(\varepsilon_1^2),\quad
 %   3^2:\ \mathcal{O}(\varepsilon_3^2),\quad
  %  2^2:\  \mathcal{O}(\varepsilon_2^2),\\
   % 12:\  \mathcal{O}(\varepsilon_1 \varepsilon_2), \quad
 %13:\  \mathcal{O}(\varepsilon_1\varepsilon_3),\quad
%    23:\  
%    \mathcal{O}(\varepsilon_2 \varepsilon_3).
%\end{aligned}
%\end{align}

%----------------------------------------------------------------------
%----------------------------------------------------------------------

\subsection{Favorable models}

Here we investigate phenomenologically viable models in types.
In the vicinity of $\tau=i\infty$ and $\omega$, we choose two benchmark points $\tau=2.1i$ and $\tau=\omega+0.051i$ where $Y_{\bm{1'}}^{(8)}/Y_{\bm{1''}}^{(8)}\sim\varepsilon\sim0.15$.
We enumerate the models for each choice of the signs in $\alpha$ and $\beta$ for each type.
Our purpose is to find models to realize the order of quark mass ratios and mixing angles without fine-tuning.
Thus, we require the following conditions:
\begin{align}
\begin{aligned}
\label{eq: mass_ratio_order}
    &1/3< \frac{(m_u/m_t)_{\textrm{model}}}{(m_u/m_t)_{\textrm{GUT}}} < 3,\quad  
    1/3 < \frac{(m_c/m_t)_{\textrm{model}}}{(m_c/m_t)_{\textrm{GUT}}} < 3, \\
    &1/3< \frac{(m_d/m_b)_{\textrm{model}}}{(m_d/m_b)_{\textrm{GUT}}} < 3,\quad  
    1/3 < \frac{(m_s/m_b)_{\textrm{model}}}{(m_s/m_b)_{\textrm{GUT}}} < 3, \\
    &2/3 < \frac{|V_{\textrm{CKM}}^x|_{\textrm{model}}}{|V_{\textrm{CKM}}^x|_{\textrm{GUT}}} < 3/2, \quad (x\in\{us,cb,ub\}).
\end{aligned}
\end{align}
Then, we find 1,584 number of models satisfying these conditions at both benchmark points $\tau = 2.1i$ and $\tau=\omega+0.051i$.
Results at $\tau=2.1i$ are shown in Table \ref{tab:chi<0.01atinfinite} and ones at $\tau=\omega+0.051i$ are in Table \ref{tab:chi<0.01atomega}.
%We also classify the mass matrix structures of the models in Tables \ref{tab:chi<0.01atinfinite} and \ref{tab:chi<0.01atomega} in appendix \ref{app:viable_models}.
\begin{table}[H]
  \centering
  \begin{tabular}{cc|cc}
    \hline
Type & Number of models  & Type & Number of models  \\
    \hline \hline
$123\textrm{-}1^22^2\textrm{-}12$ & 64 & $1^22\textrm{-}1^23^2\textrm{-}3^2$ & 64 \\
$123\textrm{-}1^22^2\textrm{-}23$ & 64 & $1^22\textrm{-}1^23^2\textrm{-}12$ & 32 \\
$123\textrm{-}1^22^2\textrm{-}1^2$ & 96 & $1^22\textrm{-}1^23^2\textrm{-}23$ & 32 \\
$123\textrm{-}1^22^2\textrm{-}3^2$ & 96 & $1^22\textrm{-}1^23^2\textrm{-}13$ & 64 \\
$123\textrm{-}123^2\textrm{-}12$ & 32 & $1^22\textrm{-}1^223\textrm{-}1^2$ & 16 \\
$123\textrm{-}123^2\textrm{-}23$ & 16 & $1^22\textrm{-}1^223\textrm{-}2^2$ & 48 \\
$123\textrm{-}123^2\textrm{-}1^2$ & 32 & $1^22\textrm{-}1^223\textrm{-}3^2$ & 48 \\
$123\textrm{-}123^2\textrm{-}3^2$ & 32 & $1^22\textrm{-}1^223\textrm{-}12$ & 32 \\
$1^22\textrm{-}1^22^2\textrm{-}1^2$ & 32 & $1^22\textrm{-}1^223\textrm{-}23$ & 16 \\
$1^22\textrm{-}1^22^2\textrm{-}2^2$ & 64 & $1^22\textrm{-}1^223\textrm{-}13$ & 48 \\
$1^22\textrm{-}1^22^2\textrm{-}3^2$ & 64 & $1^22\textrm{-}12^23\textrm{-}1^2$ & 16 \\
$1^22\textrm{-}1^22^2\textrm{-}12$ & 32 & $1^22\textrm{-}12^23\textrm{-}2^2$ & 32 \\
$1^22\textrm{-}1^22^2\textrm{-}23$ & 32 & $1^22\textrm{-}12^23\textrm{-}3^2$ & 32 \\
$1^22\textrm{-}1^22^2\textrm{-}13$ & 64 & $1^22\textrm{-}12^23\textrm{-}12$ & 16 \\
$1^22\textrm{-}2^23^2\textrm{-}1^2$ & 32 & $1^22\textrm{-}12^23\textrm{-}23$ & 16 \\
$1^22\textrm{-}2^23^2\textrm{-}2^2$ & 32 & $1^22\textrm{-}12^23\textrm{-}13$ & 32 \\
$1^22\textrm{-}2^23^2\textrm{-}3^2$ & 32 & $1^22\textrm{-}123^2\textrm{-}1^2$ & 0 \\
$1^22\textrm{-}2^23^2\textrm{-}12$ & 0 & $1^22\textrm{-}123^2\textrm{-}2^2$ & 16 \\
$1^22\textrm{-}2^23^2\textrm{-}23$ & 32 & $1^22\textrm{-}123^2\textrm{-}3^2$ & 16 \\
$1^22\textrm{-}2^23^2\textrm{-}13$ & 32 & $1^22\textrm{-}123^2\textrm{-}12$ & 16 \\
$1^22\textrm{-}1^23^2\textrm{-}1^2$ & 32 & $1^22\textrm{-}123^2\textrm{-}23$ & 0 \\
$1^22\textrm{-}1^23^2\textrm{-}2^2$ & 64 & $1^22\textrm{-}123^2\textrm{-}13$ & 16 \\
    \hline
  \end{tabular}
    \caption{Number of models satisfying hierarchy conditions in Eq. (\ref{eq: mass_ratio_order}) at the benchmark point $\tau = 2.1i$.}
    \label{tab:chi<0.01atinfinite}
\end{table}

%We find common $N\sigma$ values in different types for the following reason.
%Equality such as $Y_{\bf{1}''} = (Y_{\bf{1}'})^2$ holds in 3 significant digits at $\tau = 2.1i$. It then follows $\mathcal{O}(\varepsilon_1^2)$ and $\mathcal{O}(\varepsilon_1 \varepsilon_2)$ are numerically indistinguishable in this precision. Thus, we find models producing numerically the same mass matrices and $N\sigma$ value although $T$-charges of $A_4\times A_4 \times A_4$ are different.

\begin{table}[H]
  \centering
  \begin{tabular}{cc|cc}
    \hline
Type & Number of models  & Type & Number of models  \\
    \hline \hline
$123\textrm{-}1^22^2\textrm{-}12$ & 64 & $1^22\textrm{-}1^23^2\textrm{-}3^2$ & 64 \\
$123\textrm{-}1^22^2\textrm{-}23$ & 64 & $1^22\textrm{-}1^23^2\textrm{-}12$ & 32 \\
$123\textrm{-}1^22^2\textrm{-}1^2$ & 96 & $1^22\textrm{-}1^23^2\textrm{-}23$ & 32 \\
$123\textrm{-}1^22^2\textrm{-}3^2$ & 96 & $1^22\textrm{-}1^23^2\textrm{-}13$ & 64 \\
$123\textrm{-}123^2\textrm{-}12$ & 32 & $1^22\textrm{-}1^223\textrm{-}1^2$ & 16 \\
$123\textrm{-}123^2\textrm{-}23$ & 16 & $1^22\textrm{-}1^223\textrm{-}2^2$ & 48 \\
$123\textrm{-}123^2\textrm{-}1^2$ & 32 & $1^22\textrm{-}1^223\textrm{-}3^2$ & 48 \\
$123\textrm{-}123^2\textrm{-}3^2$ & 32 & $1^22\textrm{-}1^223\textrm{-}12$ & 32 \\
$1^22\textrm{-}1^22^2\textrm{-}1^2$ & 32 & $1^22\textrm{-}1^223\textrm{-}23$ & 16 \\
$1^22\textrm{-}1^22^2\textrm{-}2^2$ & 64 & $1^22\textrm{-}1^223\textrm{-}13$ & 48 \\
$1^22\textrm{-}1^22^2\textrm{-}3^2$ & 64 & $1^22\textrm{-}12^23\textrm{-}1^2$ & 16 \\
$1^22\textrm{-}1^22^2\textrm{-}12$ & 32 & $1^22\textrm{-}12^23\textrm{-}2^2$ & 32 \\
$1^22\textrm{-}1^22^2\textrm{-}23$ & 32 & $1^22\textrm{-}12^23\textrm{-}3^2$ & 32 \\
$1^22\textrm{-}1^22^2\textrm{-}13$ & 64 & $1^22\textrm{-}12^23\textrm{-}12$ & 16 \\
$1^22\textrm{-}2^23^2\textrm{-}1^2$ & 32 & $1^22\textrm{-}12^23\textrm{-}23$ & 16 \\
$1^22\textrm{-}2^23^2\textrm{-}2^2$ & 32 & $1^22\textrm{-}12^23\textrm{-}13$ & 32 \\
$1^22\textrm{-}2^23^2\textrm{-}3^2$ & 32 & $1^22\textrm{-}123^2\textrm{-}1^2$ & 0 \\
$1^22\textrm{-}2^23^2\textrm{-}12$ & 0 & $1^22\textrm{-}123^2\textrm{-}2^2$ & 16 \\
$1^22\textrm{-}2^23^2\textrm{-}23$ & 32 & $1^22\textrm{-}123^2\textrm{-}3^2$ & 16 \\
$1^22\textrm{-}2^23^2\textrm{-}13$ & 32 & $1^22\textrm{-}123^2\textrm{-}12$ & 16 \\
$1^22\textrm{-}1^23^2\textrm{-}1^2$ & 32 & $1^22\textrm{-}123^2\textrm{-}23$ & 0 \\
$1^22\textrm{-}1^23^2\textrm{-}2^2$ & 64 & $1^22\textrm{-}123^2\textrm{-}13$ & 16 \\
\hline
  \end{tabular}
  \caption{Number of models satisfying hierarchy conditions in Eq. (\ref{eq: mass_ratio_order}) at the benchmark point $\tau = \omega+0.051i$.}
  \label{tab:chi<0.01atomega}
\end{table}

We comment on why the number of models is zero for certain types.
As shown in Tables \ref{tab:chi<0.01atinfinite} and \ref{tab:chi<0.01atomega}, we cannot find the models satisfying hierarchy conditions in Eq. (\ref{eq: mass_ratio_order}) for types $1^22\textrm{-}2^23^2\textrm{-}12$, $1^22\textrm{-}123^2\textrm{-}1^2$ and $1^22\textrm{-}123^2\textrm{-}23$.
We find that all models in these types lead to not favorable structures of the CKM matrix or ${\cal O}(0.1)$ size of strange quark mass compared to the GUT scale value.
For later cases, it may be possible to obtain realistic values when we vary the coefficients $\alpha^{ij},\beta^{ij}={\cal O}(10)$.

%----------------------------------------------------------------------
%----------------------------------------------------------------------

\subsection{Numerical examples}
\label{subsec:numerical}

Here we show some numerical examples of the models satisfying hierarchy conditions in Eq. (\ref{eq: mass_ratio_order}).

%----------------------------------------------------------------------

\subsubsection{$\tau\sim i\infty$}
We choose $\tau = 2.1i$ as a benchmark point of the modulus. Then, modular forms become hierarchical
\begin{equation}
    Y^{(8)}_{{\bf{1}}}/ Y^{(8)}_{{\bf{1}}} = 1 \rightarrow 1,\quad 
    Y^{(8)}_{{\bf{1}}'}/ Y^{(8)}_{{\bf{1}}} = -0.148 \rightarrow \varepsilon,\quad
    Y^{(8)}_{{\bf{1}}''}/ Y^{(8)}_{{\bf{1}}} = 0.0218 \rightarrow \varepsilon^2.
\end{equation}

%----------------------------------------------------------------------

\paragraph{Example 1. Type $1^22\textrm{-}1^22^2\textrm{-}1^2$}{}\ \\
In type $1^22\textrm{-}1^22^2\textrm{-}1^2$, possible assignments of the $T$-charges to quark fields are 
\begin{align}
&\{Q^1,Q^2,Q^3\}:~\{(a_1,a_2,a_3),(b_1,b_2,b_3),(0,0,0)\}, \\
&\{u_R^1,u_R^2,u_R^3\}:~\{(1-a_1,1-a_2,1-a_3)_{\textrm{mod~3}},(1-b_1,2-b_2,-b_3)_{\textrm{mod~3}},(0,0,0)\}, \\
&\{d_R^1,d_R^2,d_R^3\}:~\{(1-a_1,1-a_2,-a_3)_{\textrm{mod~3}},(1-b_1,-b_2,-b_3)_{\textrm{mod~3}},(0,0,0)\},
\end{align}
where $a_i \in \{0,1,2 \}$ and $b_i \in \{0,1,2 \}$ are $T$-charges of the $i$-th $A_4$ for $Q^1$ and $Q^2$ respectively.
The mass matrices of the best-fit model are given by
\begin{align}
&M_u = \begin{pmatrix}
Y^{(8)}_{\bm{1''}}Y^{(8)}_{\bm{1''}}Y^{(8)}_{\bm{1''}} & Y^{(8)}_{\bm{1}}Y^{(8)}_{\bm{1''}}Y^{(8)}_{\bm{1''}} & Y^{(8)}_{\bm{1''}}Y^{(8)}_{\bm{1''}}Y^{(8)}_{\bm{1''}} \\
Y^{(8)}_{\bm{1'}}Y^{(8)}_{\bm{1'}}Y^{(8)}_{\bm{1}} & Y^{(8)}_{\bm{1''}}Y^{(8)}_{\bm{1'}}Y^{(8)}_{\bm{1}} & -Y^{(8)}_{\bm{1'}}Y^{(8)}_{\bm{1'}}Y^{(8)}_{\bm{1}} \\
Y^{(8)}_{\bm{1}}Y^{(8)}_{\bm{1}}Y^{(8)}_{\bm{1}} & -Y^{(8)}_{\bm{1'}}Y^{(8)}_{\bm{1}}Y^{(8)}_{\bm{1}} & -Y^{(8)}_{\bm{1}}Y^{(8)}_{\bm{1}}Y^{(8)}_{\bm{1}} \\
\end{pmatrix}, \\
&M_d = \begin{pmatrix}
Y^{(8)}_{\bm{1''}}Y^{(8)}_{\bm{1''}}Y^{(8)}_{\bm{1}} & Y^{(8)}_{\bm{1}}Y^{(8)}_{\bm{1'}}Y^{(8)}_{\bm{1''}} & Y^{(8)}_{\bm{1''}}Y^{(8)}_{\bm{1''}}Y^{(8)}_{\bm{1''}} \\
Y^{(8)}_{\bm{1'}}Y^{(8)}_{\bm{1'}}Y^{(8)}_{\bm{1'}} & -Y^{(8)}_{\bm{1''}}Y^{(8)}_{\bm{1}}Y^{(8)}_{\bm{1}} & -Y^{(8)}_{\bm{1'}}Y^{(8)}_{\bm{1'}}Y^{(8)}_{\bm{1}} \\
Y^{(8)}_{\bm{1}}Y^{(8)}_{\bm{1}}Y^{(8)}_{\bm{1'}} & Y^{(8)}_{\bm{1'}}Y^{(8)}_{\bm{1''}}Y^{(8)}_{\bm{1}} & Y^{(8)}_{\bm{1}}Y^{(8)}_{\bm{1}}Y^{(8)}_{\bm{1}} \\
\end{pmatrix}.
\end{align}

They correspond to the following assignments of representations of $A_4 \times A_4 \times A_4$ to quark fields,
\begin{align}
&(Q^1,Q^2,Q^3) = (\bm{1'}_1\otimes \bm{1'}_2\otimes \bm{1'}_3,\bm{1''}_1\otimes \bm{1''}_2\otimes \bm{1}_3,\bm{1}_1\otimes \bm{1}_2\otimes \bm{1}_3), \\
&(u_R^1,u_R^2,u_R^3) = (\bm{1}_1\otimes \bm{1}_2\otimes \bm{1}_3,\bm{1''}_1\otimes \bm{1}_2\otimes \bm{1}_3,\bm{1}_1\otimes \bm{1}_2\otimes \bm{1}_3), \\
&(d_R^1,d_R^2,d_R^3) = (\bm{1}_1\otimes \bm{1}_2\otimes \bm{1''}_3,\bm{1''}_1\otimes \bm{1'}_2\otimes \bm{1}_3,\bm{1}_1\otimes \bm{1}_2\otimes \bm{1}_3),
\end{align}
where $a_1=1$, $a_2=1$, $a_3=1$, $b_1=2$, $b_2=2$, and $b_3=0$.
The coupling coefficients $\alpha^{ij}$ and $\beta^{ij}$ are chosen as 
\begin{equation}
\begin{pmatrix}
\alpha^{11} & \alpha^{12} & \alpha^{13} \\
\alpha^{21} & \alpha^{22} & \alpha^{23} \\
\alpha^{31} & \alpha^{32} & \alpha^{33} \\
\end{pmatrix}
=
\begin{pmatrix}
1 & 1 & 1 \\
1 & 1 & -1 \\
1 & -1 & -1 \\
\end{pmatrix}, \quad\begin{pmatrix}
\beta^{11} & \beta^{12} & \beta^{13} \\
\beta^{21} & \beta^{22} & \beta^{23} \\
\beta^{31} & \beta^{32} & \beta^{33} \\
\end{pmatrix}
=
\begin{pmatrix}
1 & 1 & 1 \\
1 & -1 & -1 \\
1 & 1 & 1 \\
\end{pmatrix}.
\end{equation}
The hierarchical structures of the mass matrices are numerically obtained as  
\begin{align}
|M_u/M_u^{33}| &=
\begin{pmatrix}
1.03\times 10^{-5} & 4.74\times 10^{-4} & 1.03\times 10^{-5} \\
2.18\times 10^{-2} & 3.21\times 10^{-3} & 2.18\times 10^{-2} \\
1.00 & 1.48\times 10^{-1} & 1.00 \\
\end{pmatrix} \\
&\sim
\begin{pmatrix}
{\cal O}(\varepsilon^6) & {\cal O}(\varepsilon^4) & {\cal O}(\varepsilon^6) \\
{\cal O}(\varepsilon^2) & {\cal O}(\varepsilon^3) & {\cal O}(\varepsilon^2) \\
{\cal O}(1) & {\cal O}(\varepsilon) & {\cal O}(1) \\
\end{pmatrix}, \\
|M_d/M_d^{33}| &=
\begin{pmatrix}
4.74\times 10^{-4} & 3.21\times 10^{-3} & 1.03\times 10^{-5} \\
3.21\times 10^{-3} & 2.18\times 10^{-2} & 2.18\times 10^{-2} \\
1.48\times 10^{-1} & 3.21\times 10^{-3} & 1.00 \\
\end{pmatrix} \\
&\sim
\begin{pmatrix}
{\cal O}(\varepsilon^4) & {\cal O}(\varepsilon^3) & {\cal O}(\varepsilon^6) \\
{\cal O}(\varepsilon^3) & {\cal O}(\varepsilon^2) & {\cal O}(\varepsilon^2) \\
{\cal O}(\varepsilon) & {\cal O}(\varepsilon^3) & {\cal O}(1) \\
\end{pmatrix}.
\end{align}
Here, we show the orders in $\varepsilon$ where $\varepsilon = \varepsilon_i, (i=1,2,3)$.

Results are summarized in Table \ref{tab: 1^22-123^2-13}.
Recall that our purpose is to realize the order of quark mass rations and mixing angles without fine-tuning.
For this purpose, we have fixed the coefficients, $\alpha^{ij}, \beta^{ij}=\pm 1$ to make our point clear.
We could obtain more realistic values when we vary  $\alpha^{ij}, \beta^{ij}={\cal O}(1)$.
Also other models in this type could be realistic when we vary  $\alpha^{ij}, \beta^{ij}={\cal O}(1)$.
In addition, we have a remark on normalization of modular forms.
The normalization of modular forms has ambiguity, but we expect naturally that 
such normalization would not lead to a large hierarchy.
Our models may originate from compactification of higher dimensional field theory or superstring theory.
In that case, values in our models appear in high energy scale such as the GUT scale.
Renormalization group effects change values by some factors, although those effects depend on the scenario.
For example, renormalization group effects in the minimal supersymmetric scenario were studied in Refs.~\cite{Antusch:2013jca,Bjorkeroth:2015ora}.
Table \ref{tab:quark_masses} shows those values at the GUT scale for $\tan\beta=5$ as reference values.
\begin{table}[H]
  \begin{center}
    \renewcommand{\arraystretch}{1.3}
    \begin{tabular}{c|ccccccc} \hline
      & $\frac{m_u}{m_t}\times10^{6}$ & $\frac{m_c}{m_t}\times10^3$ & $\frac{m_d}{m_b}\times10^4$ & $\frac{m_s}{m_b}\times10^2$ & $|V_{\textrm{CKM}}^{us}|$ & $|V_{\textrm{CKM}}^{cb}|$ & $|V_{\textrm{CKM}}^{ub}|$ \\ \hline
      obtained values & 10.22 & 4.50 & 13.22 & 2.27 & 0.202 & 0.0419 & 0.00318 \\ \hline
      GUT scale values & 5.39 & 2.80 & 9.21 & 1.82 & 0.225 & 0.0400 & 0.00353 \\
      $1\sigma$ errors & $\pm 1.68$ & $\pm 0.12$ & $\pm 1.02$ & $\pm 0.10$ & $\pm 0.0007$ & $\pm 0.0008$ & $\pm 0.00013$ \\ \hline
    \end{tabular}
  \end{center}
  \caption{The mass ratios of the quarks and the absolute values of the CKM matrix elements at the benchmark point $\tau=2.1i$.
GUT scale values at $2\times 10^{16}$ GeV with $\tan \beta=5$ \cite{Antusch:2013jca,Bjorkeroth:2015ora} and $1\sigma$ errors are shown.}
\label{tab: 1^22-123^2-13}
\end{table}

As mentioned above, when we vary $\alpha^{ij},\beta^{ij}={\cal O}(1)$, we can obtain more realistic values.
For example, we set
\begin{equation}
\begin{pmatrix}
\alpha^{11} & \alpha^{12} & \alpha^{13} \\
\alpha^{21} & \alpha^{22} & \alpha^{23} \\
\alpha^{31} & \alpha^{32} & \alpha^{33} \\
\end{pmatrix}
=
\begin{pmatrix}
2.71 & 1.94 & 2.67 \\
2.53 & 1.99 & -2.23 \\
2.82 & -1.39 & -2.44 \\
\end{pmatrix}, \quad\begin{pmatrix}
\beta^{11} & \beta^{12} & \beta^{13} \\
\beta^{21} & \beta^{22} & \beta^{23} \\
\beta^{31} & \beta^{32} & \beta^{33} \\
\end{pmatrix}
=
\begin{pmatrix}
1.24 & 1.96 & 3.00 \\
2.45 & -1.88 & -2.26 \\
1.00 &  1.20 & 2.35 \\
\end{pmatrix}.
\end{equation}
Then, we obtain the following quark mass ratios,
\begin{align}
&(m_u,m_c,m_t)/m_t = (5.39\times 10^{-6}, 2.80\times 10^{-3}, 1), \\
&(m_d,m_s,m_b)/m_b = (9.21\times 10^{-4}, 1.82\times 10^{-2}, 1),
\end{align}
and the absolute values of the CKM matrix elements,
\begin{align}
|V_{\textrm{CKM}}| =
\begin{pmatrix}
0.974 & 0.225 & 0.00353 \\
0.225 & 0.974 & 0.0400 \\
0.00556 & 0.0397 & 0.999 \\
\end{pmatrix}.
\end{align}
Results are shown in Table \ref{tab: orderone}.
\begin{table}[H]
  \begin{center}
    \renewcommand{\arraystretch}{1.3}
    \begin{tabular}{c|ccccccc} \hline
      & $\frac{m_u}{m_t}\times10^{6}$ & $\frac{m_c}{m_t}\times10^3$ & $\frac{m_d}{m_b}\times10^4$ & $\frac{m_s}{m_b}\times10^2$ & $|V_{\textrm{CKM}}^{us}|$ & $|V_{\textrm{CKM}}^{cb}|$ & $|V_{\textrm{CKM}}^{ub}|$ \\ \hline
      obtained values & 5.39 & 2.80 & 9.21 & 1.82 & 0.225 & 0.0400 & 0.00353 \\ \hline
      GUT scale values & 5.39 & 2.80 & 9.21 & 1.82 & 0.225 & 0.0400 & 0.00353 \\
      $1\sigma$ errors & $\pm 1.68$ & $\pm 0.12$ & $\pm 1.02$ & $\pm 0.10$ & $\pm 0.0007$ & $\pm 0.0008$ & $\pm 0.00013$ \\ \hline
    \end{tabular}
  \end{center}
  \caption{The mass ratios of the quarks and the absolute values of the CKM matrix elements at the benchmark point $\tau=2.1i$.
GUT scale values at $2\times 10^{16}$ GeV with $\tan \beta=5$ \cite{Antusch:2013jca,Bjorkeroth:2015ora} and $1\sigma$ errors are shown.}
\label{tab: orderone}
\end{table}

%----------------------------------------------------------------------
\paragraph{Example 2. Type $123\textrm{-}1^22^2\textrm{-}12$} {}\ \\
In type $123\textrm{-}1^22^2\textrm{-}12$, possible assignments of the $T$-charges to quark fields are
\begin{align}
&\{Q^1,Q^2,Q^3\}:~\{(a_1,a_2,a_3),(b_1,b_2,b_3),(0,0,0)\}, \\
&\{u_R^1,u_R^2,u_R^3\}:~\{(1-a_1,1-a_2,1-a_3)_{\textrm{mod~3}},(2-b_1,2-b_2,2-b_3)_{\textrm{mod~3}},(0,0,0)\}, \\
&\{d_R^1,d_R^2,d_R^3\}:~\{(1-a_1,1-a_2,-a_3)_{\textrm{mod~3}},(2-b_1,2-b_2,-b_3)_{\textrm{mod~3}},(0,0,0)\},
\end{align}
where $a_i \in \{0,1,2 \}$ and $b_i \in \{0,1,2 \}$ are $T$-charges of the $i$-th $A_4$ for $Q^1$ and $Q^2$ respectively.
The mass matrices of the best-fit model are given by
\begin{align}
&M_u = \begin{pmatrix}
Y^{(8)}_{\bm{1''}}Y^{(8)}_{\bm{1''}}Y^{(8)}_{\bm{1''}} & Y^{(8)}_{\bm{1''}}Y^{(8)}_{\bm{1}}Y^{(8)}_{\bm{1''}} & Y^{(8)}_{\bm{1''}}Y^{(8)}_{\bm{1''}}Y^{(8)}_{\bm{1''}} \\
Y^{(8)}_{\bm{1'}}Y^{(8)}_{\bm{1}}Y^{(8)}_{\bm{1'}} & Y^{(8)}_{\bm{1'}}Y^{(8)}_{\bm{1'}}Y^{(8)}_{\bm{1'}} & -Y^{(8)}_{\bm{1'}}Y^{(8)}_{\bm{1}}Y^{(8)}_{\bm{1'}} \\
Y^{(8)}_{\bm{1}}Y^{(8)}_{\bm{1}}Y^{(8)}_{\bm{1}} & -Y^{(8)}_{\bm{1}}Y^{(8)}_{\bm{1'}}Y^{(8)}_{\bm{1}} & -Y^{(8)}_{\bm{1}}Y^{(8)}_{\bm{1}}Y^{(8)}_{\bm{1}} \\
\end{pmatrix}, \\
&M_d = \begin{pmatrix}
Y^{(8)}_{\bm{1''}}Y^{(8)}_{\bm{1''}}Y^{(8)}_{\bm{1}} & Y^{(8)}_{\bm{1''}}Y^{(8)}_{\bm{1}}Y^{(8)}_{\bm{1'}} & Y^{(8)}_{\bm{1''}}Y^{(8)}_{\bm{1''}}Y^{(8)}_{\bm{1''}} \\
-Y^{(8)}_{\bm{1'}}Y^{(8)}_{\bm{1}}Y^{(8)}_{\bm{1''}} & -Y^{(8)}_{\bm{1'}}Y^{(8)}_{\bm{1'}}Y^{(8)}_{\bm{1}} & Y^{(8)}_{\bm{1'}}Y^{(8)}_{\bm{1}}Y^{(8)}_{\bm{1'}} \\
Y^{(8)}_{\bm{1}}Y^{(8)}_{\bm{1}}Y^{(8)}_{\bm{1'}} & -Y^{(8)}_{\bm{1}}Y^{(8)}_{\bm{1'}}Y^{(8)}_{\bm{1''}} & -Y^{(8)}_{\bm{1}}Y^{(8)}_{\bm{1}}Y^{(8)}_{\bm{1}} \\
\end{pmatrix}.
\end{align}
They correspond to the following assignments of representations of $A_4 \times A_4 \times A_4$ to quark fields,
\begin{align}
&(Q^1,Q^2,Q^3) = (\bm{1'}_1\otimes \bm{1'}_2\otimes \bm{1'}_3,\bm{1''}_1\otimes \bm{1}_2\otimes \bm{1''}_3,\bm{1}_1\otimes \bm{1}_2\otimes \bm{1}_3), \\
&(u_R^1,u_R^2,u_R^3) = (\bm{1}_1\otimes \bm{1}_2\otimes \bm{1}_3,\bm{1}_1\otimes \bm{1''}_2\otimes \bm{1}_3,\bm{1}_1\otimes \bm{1}_2\otimes \bm{1}_3), \\
&(d_R^1,d_R^2,d_R^3) = (\bm{1}_1\otimes \bm{1}_2\otimes \bm{1''}_3,\bm{1}_1\otimes \bm{1''}_2\otimes \bm{1'}_3,\bm{1}_1\otimes \bm{1}_2\otimes \bm{1}_3),
\end{align}
where $a_1=1$, $a_2=1$, $a_3=1$, $b_1=2$, $b_2=0$, and $b_3=2$.
The coupling coefficients $\alpha^{ij}$ and $\beta^{ij}$ are chosen as 
\begin{equation}
\begin{pmatrix}
\alpha^{11} & \alpha^{12} & \alpha^{13} \\
\alpha^{21} & \alpha^{22} & \alpha^{23} \\
\alpha^{31} & \alpha^{32} & \alpha^{33} \\
\end{pmatrix}
=
\begin{pmatrix}
1 & 1 & 1 \\
1 & 1 & -1 \\
1 & -1 & -1 \\
\end{pmatrix}, \quad\begin{pmatrix}
\beta^{11} & \beta^{12} & \beta^{13} \\
\beta^{21} & \beta^{22} & \beta^{23} \\
\beta^{31} & \beta^{32} & \beta^{33} \\
\end{pmatrix}
=
\begin{pmatrix}
1 & 1 & 1 \\
-1 & -1 & 1 \\
1 & -1 & -1 \\
\end{pmatrix}.
\end{equation}
The hierarchical structures of the mass matrices are numerically obtained as  
\begin{align}
|M_u/M_u^{33}| &=
\begin{pmatrix}
1.03\times 10^{-5} & 4.74\times 10^{-4} & 1.03\times 10^{-5} \\
2.18\times 10^{-2} & 3.21\times 10^{-3} & 2.18\times 10^{-2} \\
1.00 & 1.48\times 10^{-1} & 1.00 \\
\end{pmatrix} \\
&\sim
\begin{pmatrix}
{\cal O}(\varepsilon^6) & {\cal O}(\varepsilon^4) & {\cal O}(\varepsilon^6) \\
{\cal O}(\varepsilon^2) & {\cal O}(\varepsilon^3) & {\cal O}(\varepsilon^2) \\
{\cal O}(1) & {\cal O}(\varepsilon) & {\cal O}(1) \\
\end{pmatrix}, \\
|M_d/M_d^{33}| &=
\begin{pmatrix}
4.74\times 10^{-4} & 3.21\times 10^{-3} & 1.03\times 10^{-5} \\
3.21\times 10^{-3} & 2.18\times 10^{-2} & 2.18\times 10^{-2} \\
1.48\times 10^{-1} & 3.21\times 10^{-3} & 1.00 \\
\end{pmatrix} \\
&\sim
\begin{pmatrix}
{\cal O}(\varepsilon^4) & {\cal O}(\varepsilon^3) & {\cal O}(\varepsilon^6) \\
{\cal O}(\varepsilon^3) & {\cal O}(\varepsilon^2) & {\cal O}(\varepsilon^2) \\
{\cal O}(\varepsilon) & {\cal O}(\varepsilon^3) & {\cal O}(1) \\
\end{pmatrix}.
\end{align}
Here, we show the orders in $\varepsilon$ where  $\varepsilon = \varepsilon_i, (i=1,2,3)$.

Results are summarized in Table \ref{tab: 123-1^22^2-12}.

\begin{table}[H]
  \begin{center}
    \renewcommand{\arraystretch}{1.3}
    \begin{tabular}{c|ccccccc} \hline
      & $\frac{m_u}{m_t}\times10^{6}$ & $\frac{m_c}{m_t}\times10^3$ & $\frac{m_d}{m_b}\times10^4$ & $\frac{m_s}{m_b}\times10^2$ & $|V_{\textrm{CKM}}^{us}|$ & $|V_{\textrm{CKM}}^{cb}|$ & $|V_{\textrm{CKM}}^{ub}|$ \\ \hline
      obtained values & 10.22 & 4.50 & 4.57 & 2.17 & 0.219 & 0.0430 & 0.00330 \\ \hline
      GUT scale values & 5.39 & 2.80 & 9.21 & 1.82 & 0.225 & 0.0400 & 0.00353 \\
      $1\sigma$ errors & $\pm 1.68$ & $\pm 0.12$ & $\pm 1.02$ & $\pm 0.10$ & $\pm 0.0007$ & $\pm 0.0008$ & $\pm 0.00013$ \\ \hline
    \end{tabular}
  \end{center}
  \caption{The mass ratios of the quarks and the absolute values of the CKM matrix elements at the benchmark point $\tau=2.1i$.
  GUT scale values at $2\times 10^{16}$ GeV with $\tan \beta=5$ \cite{Antusch:2013jca,Bjorkeroth:2015ora} and $1\sigma$ errors are shown.}
\label{tab: 123-1^22^2-12}
\end{table}

%----------------------------------------------------------------------

\subsubsection{$\tau\sim\omega$}
We choose $\tau = \omega + 0.051i$ as a benchmark point of the modulus. Then, modular forms become hierarchical
\begin{equation}
   | Y^{(8)}_{{\bf{1}}''}/ Y^{(8)}_{{\bf{1}}''}| = 1 \rightarrow 1,\quad 
    |Y^{(8)}_{{\bf{1}}'}/ Y^{(8)}_{{\bf{1}}''} | = 0.148 \rightarrow \varepsilon,\quad
    |Y^{(8)}_{{\bf{1}}}/ Y^{(8)}_{{\bf{1}}''}| = 0.0218  \rightarrow \varepsilon^2.
\end{equation}

\paragraph{Example 1. $123\textrm{-}1^22^2\textrm{-}12$}\ {} \\
In type $123\textrm{-}1^22^2\textrm{-}12$, possible  assignments of the $ST$-charges to quark fields are 
\begin{align}
&\{Q^1,Q^2,Q^3\}:~\{(a_1,a_2,a_3),(b_1,b_2,b_3),(0,0,0)\}, \\
&\{u_R^1,u_R^2,u_R^3\}:~\{(1-a_1,1-a_2,1-a_3)_{\textrm{mod~3}},(2-b_1,2-b_2,2-b_3)_{\textrm{mod~3}},(0,0,0)\}, \\
&\{d_R^1,d_R^2,d_R^3\}:~\{(1-a_1,1-a_2,-a_3)_{\textrm{mod~3}},(2-b_1,2-b_2,-b_3)_{\textrm{mod~3}},(0,0,0)\},
\end{align}
where $a_i \in \{0,1,2 \}$ and $b_i \in \{0,1,2 \}$ are $ST$-charges of the $i$-th $A_4$ for $Q^1$ and $Q^2$ respectively.
The mass matrices of the best-fit model are given by
\begin{align}
&M_u = \begin{pmatrix}
Y^{(8)}_{\bm{1}}Y^{(8)}_{\bm{1}}Y^{(8)}_{\bm{1}} & Y^{(8)}_{\bm{1}}Y^{(8)}_{\bm{1''}}Y^{(8)}_{\bm{1}} & Y^{(8)}_{\bm{1}}Y^{(8)}_{\bm{1}}Y^{(8)}_{\bm{1}} \\
Y^{(8)}_{\bm{1'}}Y^{(8)}_{\bm{1''}}Y^{(8)}_{\bm{1'}} & Y^{(8)}_{\bm{1'}}Y^{(8)}_{\bm{1'}}Y^{(8)}_{\bm{1'}} & -Y^{(8)}_{\bm{1'}}Y^{(8)}_{\bm{1''}}Y^{(8)}_{\bm{1'}} \\
Y^{(8)}_{\bm{1''}}Y^{(8)}_{\bm{1''}}Y^{(8)}_{\bm{1''}} & -Y^{(8)}_{\bm{1''}}Y^{(8)}_{\bm{1'}}Y^{(8)}_{\bm{1''}} & -Y^{(8)}_{\bm{1''}}Y^{(8)}_{\bm{1''}}Y^{(8)}_{\bm{1''}} \\
\end{pmatrix}, \\
&M_d = \begin{pmatrix}
Y^{(8)}_{\bm{1}}Y^{(8)}_{\bm{1}}Y^{(8)}_{\bm{1''}} & Y^{(8)}_{\bm{1}}Y^{(8)}_{\bm{1''}}Y^{(8)}_{\bm{1'}} & Y^{(8)}_{\bm{1}}Y^{(8)}_{\bm{1}}Y^{(8)}_{\bm{1}} \\
Y^{(8)}_{\bm{1'}}Y^{(8)}_{\bm{1''}}Y^{(8)}_{\bm{1}} & -Y^{(8)}_{\bm{1'}}Y^{(8)}_{\bm{1'}}Y^{(8)}_{\bm{1''}} & Y^{(8)}_{\bm{1'}}Y^{(8)}_{\bm{1''}}Y^{(8)}_{\bm{1'}} \\
Y^{(8)}_{\bm{1''}}Y^{(8)}_{\bm{1''}}Y^{(8)}_{\bm{1'}} & -Y^{(8)}_{\bm{1''}}Y^{(8)}_{\bm{1'}}Y^{(8)}_{\bm{1}} & -Y^{(8)}_{\bm{1''}}Y^{(8)}_{\bm{1''}}Y^{(8)}_{\bm{1''}} \\
\end{pmatrix}.
\end{align}
They correspond to the following assignments of representations of $A_4 \times A_4 \times A_4$ to quark fields,
\begin{align}
&(Q^1,Q^2,Q^3) = (\bm{1'}_1\otimes \bm{1'}_2\otimes \bm{1'}_3,\bm{1}_1\otimes \bm{1''}_2\otimes \bm{1}_3,\bm{1''}_1\otimes \bm{1''}_2\otimes \bm{1''}_3), \\
&(u_R^1,u_R^2,u_R^3) = (\bm{1''}_1\otimes \bm{1''}_2\otimes \bm{1''}_3,\bm{1''}_1\otimes \bm{1}_2\otimes \bm{1''}_3,\bm{1''}_1\otimes \bm{1''}_2\otimes \bm{1''}_3), \\
&(d_R^1,d_R^2,d_R^3) = (\bm{1''}_1\otimes \bm{1''}_2\otimes \bm{1}_3,\bm{1''}_1\otimes \bm{1}_2\otimes \bm{1'}_3,\bm{1''}_1\otimes \bm{1''}_2\otimes \bm{1''}_3),
\end{align}
where $a_1=1$, $a_2=1$, $a_3=1$, $b_1=2$, $b_2=0$, and $b_3=2$.
The coupling coefficients $\alpha^{ij}$ and $\beta^{ij}$ are chosen as 
\begin{equation}
\begin{pmatrix}
\alpha^{11} & \alpha^{12} & \alpha^{13} \\
\alpha^{21} & \alpha^{22} & \alpha^{23} \\
\alpha^{31} & \alpha^{32} & \alpha^{33} \\
\end{pmatrix}
=
\begin{pmatrix}
1 & 1 & 1 \\
1 & 1 & -1 \\
1 & -1 & -1 \\
\end{pmatrix}, \quad\begin{pmatrix}
\beta^{11} & \beta^{12} & \beta^{13} \\
\beta^{21} & \beta^{22} & \beta^{23} \\
\beta^{31} & \beta^{32} & \beta^{33} \\
\end{pmatrix}
=
\begin{pmatrix}
1 & 1 & 1 \\
1 & -1 & 1 \\
1 & -1 & -1 \\
\end{pmatrix}.
\end{equation}
The hierarchical structures of the mass matrices are numerically obtained as  
\begin{align}
|M_u/M_u^{33}| &=
\begin{pmatrix}
1.04\times 10^{-5} & 4.76\times 10^{-4} & 1.04\times 10^{-5} \\
2.18\times 10^{-2} & 3.22\times 10^{-3} & 2.18\times 10^{-2} \\
1.00 & 1.48\times 10^{-1} & 1.00 \\
\end{pmatrix} \\
&\sim
\begin{pmatrix}
{\cal O}(\varepsilon^6) & {\cal O}(\varepsilon^4) & {\cal O}(\varepsilon^6) \\
{\cal O}(\varepsilon^2) & {\cal O}(\varepsilon^3) & {\cal O}(\varepsilon^2) \\
{\cal O}(1) & {\cal O}(\varepsilon) & {\cal O}(1) \\
\end{pmatrix}, \\
|M_d/M_d^{33}| &=
\begin{pmatrix}
4.76\times 10^{-4} & 3.22\times 10^{-3} & 1.04\times 10^{-5} \\
3.22\times 10^{-3} & 2.18\times 10^{-2} & 2.18\times 10^{-2} \\
1.48\times 10^{-1} & 3.22\times 10^{-3} & 1.00 \\
\end{pmatrix} \\
&\sim
\begin{pmatrix}
{\cal O}(\varepsilon^4) & {\cal O}(\varepsilon^3) & {\cal O}(\varepsilon^6) \\
{\cal O}(\varepsilon^3) & {\cal O}(\varepsilon^2) & {\cal O}(\varepsilon^2) \\
{\cal O}(\varepsilon) & {\cal O}(\varepsilon^3) & {\cal O}(1) \\
\end{pmatrix}.
\end{align}
Here, we show the orders in $\varepsilon$ where  $\varepsilon = \varepsilon_i, (i=1,2,3)$.

Results are summarized in Table \ref{tab: 123-123^2-12_omega}.
\begin{table}[H]
  \begin{center}
    \renewcommand{\arraystretch}{1.3}
    \begin{tabular}{c|ccccccc} \hline
      & $\frac{m_u}{m_t}\times10^{6}$ & $\frac{m_c}{m_t}\times10^3$ & $\frac{m_d}{m_b}\times10^4$ & $\frac{m_s}{m_b}\times10^2$ & $|V_{\textrm{CKM}}^{us}|$ & $|V_{\textrm{CKM}}^{cb}|$ & $|V_{\textrm{CKM}}^{ub}|$ \\ \hline
      obtained values & 10.3 & 4.52 & 13.29 & 2.27 & 0.202 & 0.0420 & 0.00319 \\ \hline
      GUT scale values & 5.39 & 2.80 & 9.21 & 1.82 & 0.225 & 0.0400 & 0.00353 \\ 
      $1\sigma$ errors & $\pm 1.68$ & $\pm 0.12$ & $\pm 1.02$ & $\pm 0.10$ & $\pm 0.0007$ & $\pm 0.0008$ & $\pm 0.00013$ \\ \hline
    \end{tabular}
  \end{center}
  \caption{The mass ratios of the quarks and the absolute values of the CKM matrix elements at the benchmark point $\tau=\omega + 0.051i$.
GUT scale values at $2\times 10^{16}$ GeV with $\tan \beta=5$ \cite{Antusch:2013jca,Bjorkeroth:2015ora} and $1\sigma$ errors are shown.}
\label{tab: 123-123^2-12_omega}
\end{table}

\paragraph{Example 2. Type $1^22\textrm{-}1^22^2\textrm{-}1^2$}\ {} \\
In type $1^22\textrm{-}1^22^2\textrm{-}1^2$, possible  assignments of the $ST$-charges to quark fields are 
\begin{align}
&\{Q^1,Q^2,Q^3\}:~\{(a_1,a_2,a_3),(b_1,b_2,b_3),(0,0,0)\}, \\
&\{u_R^1,u_R^2,u_R^3\}:~\{(1-a_1,1-a_2,1-a_3)_{\textrm{mod~3}},(1-b_1,2-b_2,-b_3)_{\textrm{mod~3}},(0,0,0)\}, \\
&\{d_R^1,d_R^2,d_R^3\}:~\{(1-a_1,1-a_2,-a_3)_{\textrm{mod~3}},(1-b_1,-b_2,-b_3)_{\textrm{mod~3}},(0,0,0)\},
\end{align}
where $a_i \in \{0,1,2 \}$ and $b_i \in \{0,1,2 \}$ are $ST$-charges of the $i$-th $A_4$ for $Q_1$ and $Q_2$ respectively.
The mass matrices of the best-fit model are given by
\begin{align}
&M_u = \begin{pmatrix}
Y^{(8)}_{\bm{1}}Y^{(8)}_{\bm{1}}Y^{(8)}_{\bm{1}} & Y^{(8)}_{\bm{1''}}Y^{(8)}_{\bm{1}}Y^{(8)}_{\bm{1}} & Y^{(8)}_{\bm{1}}Y^{(8)}_{\bm{1}}Y^{(8)}_{\bm{1}} \\
Y^{(8)}_{\bm{1'}}Y^{(8)}_{\bm{1'}}Y^{(8)}_{\bm{1''}} & Y^{(8)}_{\bm{1}}Y^{(8)}_{\bm{1'}}Y^{(8)}_{\bm{1''}} & -Y^{(8)}_{\bm{1'}}Y^{(8)}_{\bm{1'}}Y^{(8)}_{\bm{1''}} \\
Y^{(8)}_{\bm{1''}}Y^{(8)}_{\bm{1''}}Y^{(8)}_{\bm{1''}} & -Y^{(8)}_{\bm{1'}}Y^{(8)}_{\bm{1''}}Y^{(8)}_{\bm{1''}} & -Y^{(8)}_{\bm{1''}}Y^{(8)}_{\bm{1''}}Y^{(8)}_{\bm{1''}} \\
\end{pmatrix}, \\
&M_d = \begin{pmatrix}
Y^{(8)}_{\bm{1}}Y^{(8)}_{\bm{1}}Y^{(8)}_{\bm{1''}} & Y^{(8)}_{\bm{1''}}Y^{(8)}_{\bm{1'}}Y^{(8)}_{\bm{1}} & Y^{(8)}_{\bm{1}}Y^{(8)}_{\bm{1}}Y^{(8)}_{\bm{1}} \\
-Y^{(8)}_{\bm{1'}}Y^{(8)}_{\bm{1'}}Y^{(8)}_{\bm{1'}} & -Y^{(8)}_{\bm{1}}Y^{(8)}_{\bm{1''}}Y^{(8)}_{\bm{1''}} & Y^{(8)}_{\bm{1'}}Y^{(8)}_{\bm{1'}}Y^{(8)}_{\bm{1''}} \\
Y^{(8)}_{\bm{1''}}Y^{(8)}_{\bm{1''}}Y^{(8)}_{\bm{1'}} & Y^{(8)}_{\bm{1'}}Y^{(8)}_{\bm{1}}Y^{(8)}_{\bm{1''}} & -Y^{(8)}_{\bm{1''}}Y^{(8)}_{\bm{1''}}Y^{(8)}_{\bm{1''}} \\
\end{pmatrix}.
\end{align}
They correspond to the following assignments of representations of $A_4 \times A_4 \times A_4$ to quark fields,
\begin{align}
&(Q^1,Q^2,Q^3) = (\bm{1'}_1\otimes \bm{1'}_2\otimes \bm{1'}_3,\bm{1}_1\otimes \bm{1}_2\otimes \bm{1''}_3,\bm{1''}_1\otimes \bm{1''}_2\otimes \bm{1''}_3), \\
&(u_R^1,u_R^2,u_R^3) = (\bm{1''}_1\otimes \bm{1''}_2\otimes \bm{1''}_3,\bm{1}_1\otimes \bm{1''}_2\otimes \bm{1''}_3,\bm{1''}_1\otimes \bm{1''}_2\otimes \bm{1''}_3), \\
&(d_R^1,d_R^2,d_R^3) = (\bm{1''}_1\otimes \bm{1''}_2\otimes \bm{1}_3,\bm{1}_1\otimes \bm{1'}_2\otimes \bm{1''}_3,\bm{1''}_1\otimes \bm{1''}_2\otimes \bm{1''}_3),
\end{align}
where $a_1=1$, $a_2=1$, $a_3=1$, $b_1=2$, $b_2=2$, and $b_3=0$.
The coupling coefficients $\alpha^{ij}$ and $\beta^{ij}$ are chosen as

\begin{equation}
\begin{pmatrix}
\alpha^{11} & \alpha^{12} & \alpha^{13} \\
\alpha^{21} & \alpha^{22} & \alpha^{23} \\
\alpha^{31} & \alpha^{32} & \alpha^{33} \\
\end{pmatrix}
=
\begin{pmatrix}
1 & 1 & 1 \\
1 & 1 & -1 \\
1 & -1 & -1 \\
\end{pmatrix}, \quad\begin{pmatrix}
\beta^{11} & \beta^{12} & \beta^{13} \\
\beta^{21} & \beta^{22} & \beta^{23} \\
\beta^{31} & \beta^{32} & \beta^{33} \\
\end{pmatrix}
=
\begin{pmatrix}
1 & 1 & 1 \\
-1 & -1 & 1 \\
1 & 1 & -1 \\
\end{pmatrix}.
\end{equation}
The hierarchical structures of the mass matrices are numerically obtained as  
\begin{align}
|M_u/M_u^{33}| &=
\begin{pmatrix}
1.04\times 10^{-5} & 4.76\times 10^{-4} & 1.04\times 10^{-5} \\
2.18\times 10^{-2} & 3.22\times 10^{-3} & 2.18\times 10^{-2} \\
1.00 & 1.48\times 10^{-1} & 1.00 \\
\end{pmatrix} \\
&\sim
\begin{pmatrix}
{\cal O}(\varepsilon^6) & {\cal O}(\varepsilon^4) & {\cal O}(\varepsilon^6) \\
{\cal O}(\varepsilon^2) & {\cal O}(\varepsilon^3) & {\cal O}(\varepsilon^2) \\
{\cal O}(1) & {\cal O}(\varepsilon) & {\cal O}(1) \\
\end{pmatrix}, \\
|M_d/M_d^{33}| &=
\begin{pmatrix}
4.76\times 10^{-4} & 3.22\times 10^{-3} & 1.04\times 10^{-5} \\
3.22\times 10^{-3} & 2.18\times 10^{-2} & 2.18\times 10^{-2} \\
1.48\times 10^{-1} & 3.22\times 10^{-3} & 1.00 \\
\end{pmatrix} \\
&\sim
\begin{pmatrix}
{\cal O}(\varepsilon^4) & {\cal O}(\varepsilon^3) & {\cal O}(\varepsilon^6) \\
{\cal O}(\varepsilon^3) & {\cal O}(\varepsilon^2) & {\cal O}(\varepsilon^2) \\
{\cal O}(\varepsilon) & {\cal O}(\varepsilon^3) & {\cal O}(1) \\
\end{pmatrix}.
\end{align}
Here, we show the orders in $\varepsilon$ where  $\varepsilon = \varepsilon_i, (i=1,2,3)$.

Results are summarized in Table \ref{tab: 1^22-1^223-23_omega}.
\begin{table}[H]
  \begin{center}
    \renewcommand{\arraystretch}{1.3}
    \begin{tabular}{c|ccccccc} \hline
      & $\frac{m_u}{m_t}\times10^{6}$ & $\frac{m_c}{m_t}\times10^3$ & $\frac{m_d}{m_b}\times10^4$ & $\frac{m_s}{m_b}\times10^2$ & $|V_{\textrm{CKM}}^{us}|$ & $|V_{\textrm{CKM}}^{cb}|$ & $|V_{\textrm{CKM}}^{ub}|$ \\ \hline
      obtained values & 10.3 & 4.52 & 4.62 & 2.17 & 0.219 & 0.0431 & 0.00329 \\ \hline
      GUT scale values & 5.39 & 2.80 & 9.21 & 1.82 & 0.225 & 0.0400 & 0.00353 \\
      $1\sigma$ errors & $\pm 1.68$ & $\pm 0.12$ & $\pm 1.02$ & $\pm 0.10$ & $\pm 0.0007$ & $\pm 0.0008$ & $\pm 0.00013$ \\ \hline
    \end{tabular}
  \end{center}
  \caption{The mass ratios of the quarks and the absolute values of the CKM matrix elements at the benchmark point $\tau=\omega + 0.051i$.
GUT scale values at $2\times 10^{16}$ GeV with $\tan \beta=5$ \cite{Antusch:2013jca,Bjorkeroth:2015ora} and $1\sigma$ errors are shown.}
\label{tab: 1^22-1^223-23_omega}
\end{table}

%----------------------------------------------------------------------
%----------------------------------------------------------------------

\subsection{Comment on the models using $A_4$ triplet}

We comment on the models using $A_4$ triplet.
When we assign $A_4$ triplet to either up or down quarks, some of coefficients $\alpha$ ($\beta$) are related each other.
Under such a restriction, we can find some models leading to realistic quark mass hierarchies, but they cannot realize mixing angles, which are small compared with experimental values.
It is challenging to derive both quark mass hierarchies and mixing angles in models with $A_4$ triplet.
We would study it elsewhere.

%----------------------------------------------------------------------
%----------------------------------------------------------------------
%----------------------------------------------------------------------

\section{CP violation}
\label{sec:CP}
\begin{figure}[H]
  \centering
  \includegraphics[width=5cm]{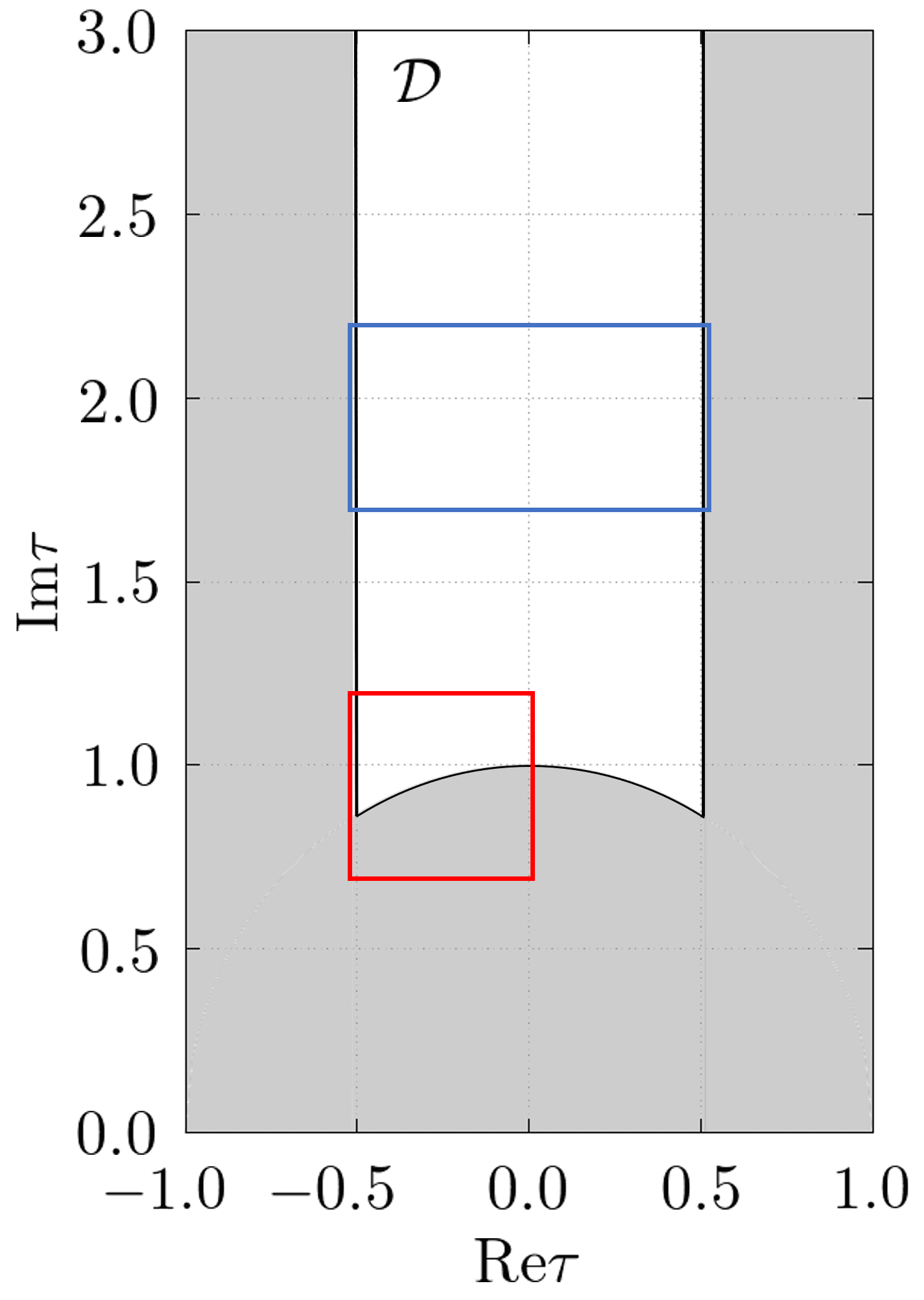}
  \caption{Fundamental region ${\cal D}$ of the modulus $\tau$.
  White corresponds to fundamental region and gray is out of the region.
  The region squared by red shows the region focused in Figure \ref{fig:chi_J_omega}.
  One squared by blue shows the region focused in Figure \ref{fig:chi_J_infinite}.}
  \label{fig:fnd}
\end{figure}

Here we study CP violation on quark flavor models in $A_4\times A_4\times A_4$ modular symmetry.
We consider CP violation induced by the vacuum expectation value (VEV) of the modulus $\tau$.
Figure \ref{fig:fnd} shows the fundamental region ${\cal D}$ of the modulus $\tau$.
The modulus $\tau$ transforms 
\begin{align}
\tau \to -\tau^*,
\end{align}
under the CP transformation \cite{Baur:2019kwi,Novichkov:2019sqv,Baur:2019iai}.
Obviously, the CP symmetry is not violated at ${\rm Re }\tau=0$.
On the other hand, the line ${\rm Re}\tau = -1/2$ transforms as 
\begin{align}
\tau = -\frac12 + i{\rm Im}\tau \to  -\tau^*=\frac12 + i{\rm Im}\tau,
\end{align}
under the above CP transformation.
However, these transform each other by the $T$-transformation.
Thus, the CP symmetry is not violated along  ${\rm Re}\tau = \pm1/2$
because of the modular symmetry.
Similarly,  CP violation does not occur at the arc of the fundamental region.
Actually, our numerical examples in section \ref{subsec:numerical} are results at such modulus and therefore CP phase of those vanishes.
In this section, we find necessary conditions for CP violation and give numerical studies in $A_4\times A_4\times A_4$ modular symmetry.

%----------------------------------------------------------------------
%----------------------------------------------------------------------

\subsection{Necessary conditions for CP violation}
\label{subsec:necessary}

%The modular forms of level 3 can be expanded by cube root of $q=e^{2\pi i\tau}$ ($q$-expansion).
%Since $q^{1/3}$ has $T$-charge 1, the modular forms with $T$-charge 0, 1 and 2 can be expanded by $q^{n}$, $q^{1/3+n}$ and $q^{2/3+n}$ for $n\in\{0,1,2,...\}$, respectively.
%As an example, we show expansions of singlet modular forms of level 3 and weight 8, $Y_{\bm{1}}^{(8)}$, $Y_{\bm{1'}}^{(8)}$ and $Y_{\bm{1''}}^{(8)}$:
%\begin{align}
%  &Y_{\bm{1}}^{(8)}(\tau) = 1+480q+61920q^2+\cdots, \\
%  &Y_{\bm{1'}}^{(8)}(\tau) = -12q^{1/3}(1+232q+260q^2+\cdots), \\
%  &Y_{\bm{1''}}^{(8)}(\tau) = 144q^{2/3}(1-16q+104q^2+\cdots).
%\end{align}

As we will see soon, there are two types of mass matrices; CP violation does not occur in one type while it occurs in another type.
First of all, let us consider the vicinity of the cusp, $\tau=i\infty$ and $A_4$ modular symmetry instead of $A_4\times A_4\times A_4$.
As we have mentioned in section \ref{sec:quark_mass}, the mass matrix elements in the vicinity of $\tau=i\infty$ can be written in terms of $\varepsilon \propto q^{1/3} =e^{2\pi i\tau/3}$ ($q$-expansion).
Then powers of $\varepsilon$ in mass matrix elements are determined by its $T$-charge \footnote{$\varepsilon=q^{1/3}$ has $T$-charge 1; therefore mass matrix elements with $T$-charge 0, 1 and 2 can be expanded by $q^{n}$, $q^{1/3+n}$ and $q^{2/3+n}$ for $n\in\{0,1,2,...\}$, respectively.}.
For example, suppose that the up quark mass matrix elements have the following $T$-charges under $A_4$ modular symmetry:
\begin{align}
  M_u: ~
  \begin{pmatrix}
    2 & 2 & 1 \\
    1 & 1 & 0 \\
    1 & 1 & 0 \\
  \end{pmatrix}. \label{eq:221110110}
\end{align}
Then, the mass matrix can be estimated as
\begin{align}
  M_u \sim
  \begin{pmatrix}
    q^{2/3} & q^{2/3} & q^{1/3} \\
    q^{1/3} & q^{1/3}  & 1 \\
    q^{1/3} & q^{1/3} & 1 \\
  \end{pmatrix},
\end{align}
by the first order approximation in $q$-expansion.
Since we focus on the region $\varepsilon\sim0.15$ to generate large quark mass hierarchies, we can ignore the second order of ${\cal O}(10^{-3})$ compared with the first order.
We note that this charge pattern can be obtained by charge assignments of fields,
\begin{align}
  Q: ~(2,0,0), \quad u_R:~(2,2,0), \quad H_u:~0.
\end{align}
To understand the origin of CP violation, let us see phase factors in this mass matrix,
\begin{align}
  M_u &\sim 
  \begin{pmatrix}
    |q|^{2/3}e^{4\pi i\textrm{Re}\tau/3} & |q|^{2/3}e^{4\pi i\textrm{Re}\tau/3} & |q|^{1/3}e^{2\pi i\textrm{Re}\tau/3} \\
    |q|^{1/3}e^{2\pi i\textrm{Re}\tau/3} & |q|^{1/3}e^{2\pi i\textrm{Re}\tau/3}  & 1 \\
    |q|^{1/3}e^{2\pi i\textrm{Re}\tau/3} & |q|^{1/3}e^{2\pi i\textrm{Re}\tau/3} & 1 \\
  \end{pmatrix} \\
  &=
  \begin{pmatrix}
    |q|^{2/3}e^{2i\alpha} & |q|^{2/3}e^{2i\alpha} & |q|^{1/3}e^{i\alpha} \\
    |q|^{1/3}e^{i\alpha} & |q|^{1/3}e^{i\alpha}  & 1 \\
    |q|^{1/3}e^{i\alpha} & |q|^{1/3}e^{i\alpha} & 1 \\
  \end{pmatrix}, %\quad
%  \alpha = 2\pi\textrm{Re}\tau/3.
\end{align}
where $\alpha = 2\pi\textrm{Re}\tau/3$.
All of these phase factors can be canceled by the following basis transformations for fields,
\begin{align}
  M_u &\rightarrow u_L^\dagger M_u u_R \notag \\
  &\sim u_L^\dagger
  \begin{pmatrix}
    |q|^{2/3}e^{2i\alpha} & |q|^{2/3}e^{2i\alpha} & |q|^{1/3}e^{i\alpha} \\
    |q|^{1/3}e^{i\alpha} & |q|^{1/3}e^{i\alpha}  & 1 \\
    |q|^{1/3}e^{i\alpha} & |q|^{1/3}e^{i\alpha} & 1 \\
  \end{pmatrix} 
  u_R \\
  &=
  \begin{pmatrix}
    e^{-i\alpha} & & \\
    & 1 & \\
    & & 1 \\
  \end{pmatrix}
  \begin{pmatrix}
    |q|^{2/3}e^{2i\alpha} & |q|^{2/3}e^{2i\alpha} & |q|^{1/3}e^{i\alpha} \\
    |q|^{1/3}e^{i\alpha} & |q|^{1/3}e^{i\alpha}  & 1 \\
    |q|^{1/3}e^{i\alpha} & |q|^{1/3}e^{i\alpha} & 1 \\
  \end{pmatrix} 
  \begin{pmatrix}
    e^{-i\alpha} & & \\
    & e^{-i\alpha} & \\
    & & 1 \\
  \end{pmatrix} \\
  &=
  \begin{pmatrix}
    |q|^{2/3} & |q|^{2/3} & |q|^{1/3} \\
    |q|^{1/3} & |q|^{1/3}  & 1 \\
    |q|^{1/3} & |q|^{1/3} & 1 \\
  \end{pmatrix}.
\end{align} 
Here basis transformations $u_L^\dagger$ and $u_R$ are given by
\begin{align}
  &u_L^\dagger =
  \begin{pmatrix}
    e^{-i\phi^1\alpha} & & \\
    & e^{-i\phi^2\alpha} & \\
    & & e^{-i\phi^3\alpha} \\
  \end{pmatrix}, \quad \phi^i = [-(T\textrm{-charge~of~}Q^i)]_{\textrm{mod~3}}, \label{eq:u_L} \\
  &u_R =
  \begin{pmatrix}
    e^{-i\psi^1\alpha} & & \\
    & e^{-i\psi^2\alpha} & \\
    & & e^{-i\psi^3\alpha} \\
  \end{pmatrix}, \quad \psi^i = [-(T\textrm{-charge~of~}u_R^i)]_{\textrm{mod~3}}. \label{eq:u_R}
\end{align}
Here we use the notation $[q]_{\textrm{mod~3}}=r$ when $q=3n+r$ with the maximum integer $n$ such that $r=0,1,2$.
Consequently, all phase factors in the first order approximation of the mass matrix with $T$-charge Eq.~(\ref{eq:221110110}) 
vanish \footnote{Similar behaviors at fixed points were studied in Refs.~\cite{Kobayashi:2019uyt,Kikuchi:2022geu}.}.

On the other hand, there are charge patterns whose phase factors in mass matrices survive after the basis transformations $u_L^\dagger$ and $u_R$ in Eqs.~(\ref{eq:u_L}) and (\ref{eq:u_R}).
Let us consider the case that the up quark mass matrix elements have the following $T$-charges under $A_4$ modular symmetry,
\begin{align}
  M_u:~
  \begin{pmatrix}
    0 & 2 & 1 \\
    2 & 1 & 0 \\
    2 & 1 & 0 \\
  \end{pmatrix},
\end{align}
which is obtained by charge assignments of fields,
\begin{align}
  Q: ~(2,0,0), \quad u_R:~(1,2,0), \quad H_u:~0.
\end{align}
Then, the mass matrix is estimated as
\begin{align}
  M_u &\sim
  \begin{pmatrix}
    1 & q^{2/3} & q^{1/3} \\
    q^{2/3} & q^{1/3} & 1 \\
    q^{2/3} & q^{1/3} & 1 \\
  \end{pmatrix} \\
  &=
  \begin{pmatrix}
    1 & |q|^{2/3}e^{4\pi i\textrm{Re}\tau/3} & |q|^{1/3}e^{2\pi i\textrm{Re}\tau/3} \\
    |q|^{2/3}e^{4\pi i\textrm{Re}\tau/3} & |q|^{1/3}e^{2\pi i\textrm{Re}\tau/3} & 1 \\
    |q|^{2/3}e^{4\pi i\textrm{Re}\tau/3} & |q|^{1/3}e^{2\pi i\textrm{Re}\tau/3} & 1 \\
  \end{pmatrix} \\
  &=
  \begin{pmatrix}
    1 & |q|^{2/3}e^{2i\alpha} & |q|^{1/3}e^{i\alpha} \\
    |q|^{2/3}e^{2i\alpha} & |q|^{1/3}e^{i\alpha} & 1 \\
    |q|^{2/3}e^{2i\alpha} & |q|^{1/3}e^{i\alpha} & 1 \\
  \end{pmatrix}, 
%\quad \alpha = 2\pi \textrm{Re}\tau/3,
\end{align}
by the first order approximation in $q$-expansion, 
where $\alpha = 2\pi \textrm{Re}\tau/3$.
Using the basis transformations in Eqs.~(\ref{eq:u_L}) and (\ref{eq:u_R}), the phase factors in this matrix are partially canceled as follows,
\begin{align}
  M_u &\rightarrow u_L^\dagger M_u u_R \notag \\
  &\sim
  u_L^\dagger
  \begin{pmatrix}
    1 & |q|^{2/3}e^{2i\alpha} & |q|^{1/3}e^{i\alpha} \\
    |q|^{2/3}e^{2i\alpha} & |q|^{1/3}e^{i\alpha} & 1 \\
    |q|^{2/3}e^{2i\alpha} & |q|^{1/3}e^{i\alpha} & 1 \\
  \end{pmatrix}
  u_R \\
  &=
  \begin{pmatrix}
    e^{-i\alpha} & & \\
    & 1 & \\
    & & 1 \\
  \end{pmatrix}
  \begin{pmatrix}
    1 & |q|^{2/3}e^{2i\alpha} & |q|^{1/3}e^{i\alpha} \\
    |q|^{2/3}e^{2i\alpha} & |q|^{1/3}e^{i\alpha} & 1 \\
    |q|^{2/3}e^{2i\alpha} & |q|^{1/3}e^{i\alpha} & 1 \\
  \end{pmatrix}
  \begin{pmatrix}
    e^{-2i\alpha} & & \\
    & e^{-i\alpha} & \\
    & & 1 \\
  \end{pmatrix} \\
  &=
  \begin{pmatrix}
    e^{-3i\alpha} & |q|^{2/3} & |q|^{1/3} \\
    |q|^{2/3} & |q|^{1/3} & 1 \\
    |q|^{2/3} & |q|^{1/3} & 1 \\
  \end{pmatrix}.
\end{align}
Why does the phase factor of (1,1) matrix element remain after the basis transformations ?
Its reason is as follows.
The $T$-charge of the $M_u^{ij}$ element is given by
\begin{align}
  T\textrm{-charge~of~}M_u^{ij} = [-(T\textrm{-charge~of~}Q^i) - (T\textrm{-charge~of~}u_R^j)]_{\textrm{mod~3}}.
\end{align}
%Here we use the notation $[q]_{\textrm{mod~3}}=r$ when $q=3n+r$ with the maximum integer $n$ such that $r=0,1,2$.
Then, the $M^{ij}_u$ element has the phase factor,
\begin{align}
  \textrm{exp}\left[i\alpha(T\textrm{-charge~of~}M_u^{ij})\right] =
  \textrm{exp}\left[i\alpha[-(T\textrm{-charge~of~}Q^i) - (T\textrm{-charge~of~}u_R^j)]_{\textrm{mod~3}}\right],
\end{align}
while the phase cancellation by the basis transformations is given by
\begin{align}
  \textrm{exp}\left[-i\alpha([-(T\textrm{-charge~of~}Q^i)]_{\textrm{mod~3}} + [-(T\textrm{-charge~of~}u_R^j)]_{\textrm{mod~3}})\right].
\end{align}
Thus when
\begin{align}
  [-(T\textrm{-charge~of~}Q^i)]_{\textrm{mod~3}} +[- ( T\textrm{-charge~of~}u_R^j)]_{\textrm{mod~3}} \geq 3, \label{eq:CP_condition_up}
\end{align}
$M_u^{ij}$ gets the phase factor $e^{-3i\alpha}$ after the basis transformations in Eqs.~(\ref{eq:u_L}) and (\ref{eq:u_R}).
The same condition for down quark mass matrix is given by
\begin{align}
  [-(T\textrm{-charge~of~}Q^i)]_{\textrm{mod~3}} +[- ( T\textrm{-charge~of~}d_R^j)]_{\textrm{mod~3}} \geq 3. \label{eq:CP_condition_down}
\end{align}
These conditions suggest that residual charge assignments into fields decide phase factors in mass matrices as well as hierarchical structures.
In other words, CP violation is strongly related to hierarchical quark masses through the residual charges.
This is also true in other models since above analysis only depends on the residual charges.

Now we are ready to discuss CP violation induced by the VEV of the modulus $\tau$.
When either the up sector or the down sector does not satisfy the conditions in Eqs.~(\ref{eq:CP_condition_up}) and (\ref{eq:CP_condition_down}), mass matrices become completely real.
Hence CP violation obviously does not occur in this type of mass matrices even if we freely choose the value of the modulus $\tau$.
In contrast, if at least one element of either up or down quark mass matrices satisfies the conditions in Eqs.~(\ref{eq:CP_condition_up}) or (\ref{eq:CP_condition_down}), mass matrices become complex and CP violation can occur depending on the value of the modulus $\tau$.
As a result, Eqs.~(\ref{eq:CP_condition_up}) and (\ref{eq:CP_condition_down}) are regarded as necessary conditions for CP violation.
We again note that this is the results of the first order approximation in $q$-expansion but the second order is estimated as ${\cal O}(10^{-3})$ compared with the first order and sufficiently negligible.

Next let us consider the vicinity of $\tau=\omega$.
At $\tau=\omega$, the mass matrix elements can be written in terms of $\varepsilon \propto u\equiv \frac{\tau-\omega}{\tau-\omega^2}$ ($u$-expansion).
At $\tau\sim\omega$ ($|u|\ll 1$), it is still good approximation.
In this way, powers of $\varepsilon=u$ in mass matrix elements at $\tau\sim\omega$ are determined by its $ST$-charge at $\tau=\omega$ since $u$ has $ST$-charge 1.
Thus, the same results for CP violation at $\tau\sim i\infty$ can be obtained at $\tau\sim\omega$ by reading $q$ as $u$ and $T$-charge as $ST$-charge.
That is, when at least one element of either up or down quark mass matrices satisfies the following conditions, 
\begin{align}
  &[-(ST\textrm{-charge~of~}Q^i)]_{\textrm{mod~3}} +[ -( ST\textrm{-charge~of~}u_R^j)]_{\textrm{mod~3}} \geq 3, \label{eq:CP_condition_up_ST} \\
  &[-(ST\textrm{-charge~of~}Q^i)]_{\textrm{mod~3}} +[- ( ST\textrm{-charge~of~}d_R^j)]_{\textrm{mod~3}} \geq 3, \label{eq:CP_condition_down_ST}
\end{align}
CP violation can occur depending on the value of the modulus $\tau$.

We can extend these results to the models in $A_4\times A_4\times A_4$ modular symmetry.
Then we need the conditions Eqs.~(\ref{eq:CP_condition_up}) and (\ref{eq:CP_condition_down}) or Eqs.~(\ref{eq:CP_condition_up_ST}) and (\ref{eq:CP_condition_down_ST}) for each $A_4$ modular symmetry.
In appendix \ref{app:viable_models}, we classify the phase factors after the basis transformations in Eqs.~(\ref{eq:u_L}) and (\ref{eq:u_R}), and hierarchical structures of the mass matrices of favorable models summarized in Tables \ref{tab:chi<0.01atinfinite} and \ref{tab:chi<0.01atomega}.
As a result, we find that all favorable models satisfy these necessary conditions.
Nevertheless from the numerical analysis it is also found that all of those models cannot induce sufficiently large CP violation when the modulus $\tau$ lies on the region satisfying hierarchy conditions in Eq.~(\ref{eq: mass_ratio_order}).
This can be checked by the argument at the first order approximation in $\varepsilon$ expansion.
From the mass matrix structures summarized in Table \ref{tab:mass_matrix_structures}, we find that all favorable models have the following structures of the CKM matrix at the first order approximation.
\begin{align}
  V_{\textrm{CKM}} =
&\begin{pmatrix}
1 & 1.5 |\varepsilon|p^* & - |\varepsilon|^{3}p^* \\
- 1.5 |\varepsilon| p & 1 & - 2 |\varepsilon|^{2} \\
- 2 |\varepsilon|^{3} p & 2 |\varepsilon|^{2} & 1 \\
\end{pmatrix} , \quad
\begin{pmatrix}
1 & - 1.5 |\varepsilon| p^* & |\varepsilon|^{3}p^* \\
1.5 |\varepsilon| p & 1 & - 2 |\varepsilon|^{2} \\
2 |\varepsilon|^{3} p & 2 |\varepsilon|^{2} & 1 \\
\end{pmatrix}, \label{eq:CKM_at_first_order}
\end{align}
where $p$ is given by $u/|u|$ for $\tau\sim\omega$ and $(q/|q|)^{1/3}$ for $\tau\sim i\infty$.
This directly leads to that Jarlskog invariant $J_{\textrm{CP}}$ vanishes at the first order approximation as
\begin{align}
  J_{\textrm{CP}} = |\textrm{Im}(V_{\textrm{CKM}}^{us}V_{\textrm{CKM}}^{cb}(V_{\textrm{CKM}}^{ub}V_{\textrm{CKM}}^{cs})^*)| = \textrm{Im}(3p^*p|\varepsilon|^{6}) = 0. \label{eq:first_p_cancellation}
\end{align}
We have checked that the second order contribution to $V_{\textrm{CKM}}$ is ${\cal O}(\varepsilon^2)$ compared to the first order.
Therefore we can expect
\begin{align}
  J_{\textrm{CP}} \lesssim 3 \times |\varepsilon|^8, \label{eq:J_CP_second}
\end{align}
at the second order approximation.
On the other hand, we need $\varepsilon\sim 0.15$ to realize large quark mass hierarchies and Jarlskog invariant is given by $J_{\textrm{CP}}\lesssim 7.7\times 10^{-7}$ which are extremely small compared with the observed value $J_{\textrm{CP}}=2.8\times 10^{-5}$.
In the following subsection, we will confirm this point by a concrete model.

%----------------------------------------------------------------------
%----------------------------------------------------------------------

\subsection{Numerical example of CP violation}
\label{subsec:numerical_CP}

To illustrate CP violation in $A_4\times A_4\times A_4$ modular symmetric models, let us consider the model in type $1^22\textrm{-}1^23^2\textrm{-}3^2$.
In type $1^22\textrm{-}1^23^2\textrm{-}3^2$, quarks have the following $Z_3 \times Z_3 \times Z_3$ charges:
\begin{align}
  &\{Q^1,Q^2,Q^3\}:~\{(a_1,a_2,a_3),(b_1,b_2,b_3),(0,0,0)\}, \\
  &\{u_R^1,u_R^2,u_R^3\}:~\{(2-a_1,2-a_2,2-a_3)_{\textrm{mod~3}},(2-b_1,1-b_2,-b_3)_{\textrm{mod~3}},(0,0,0)\}, \\
  &\{d_R^1,d_R^2,d_R^3\}:~\{(2-a_1,-a_2,2-a_3)_{\textrm{mod~3}},(-b_1,-b_2,2-b_3)_{\textrm{mod~3}},(0,0,0)\},
\end{align}
where $a_i\in\{0,1,2\}$ and $b_i\in\{0,1,2\}$ are $Z_3$-charges of the $i$-th $A_4$ for $Q^1$ and $Q^2$ respectively.

\begin{figure}[H]
  \centering
  \includegraphics[width=8cm]{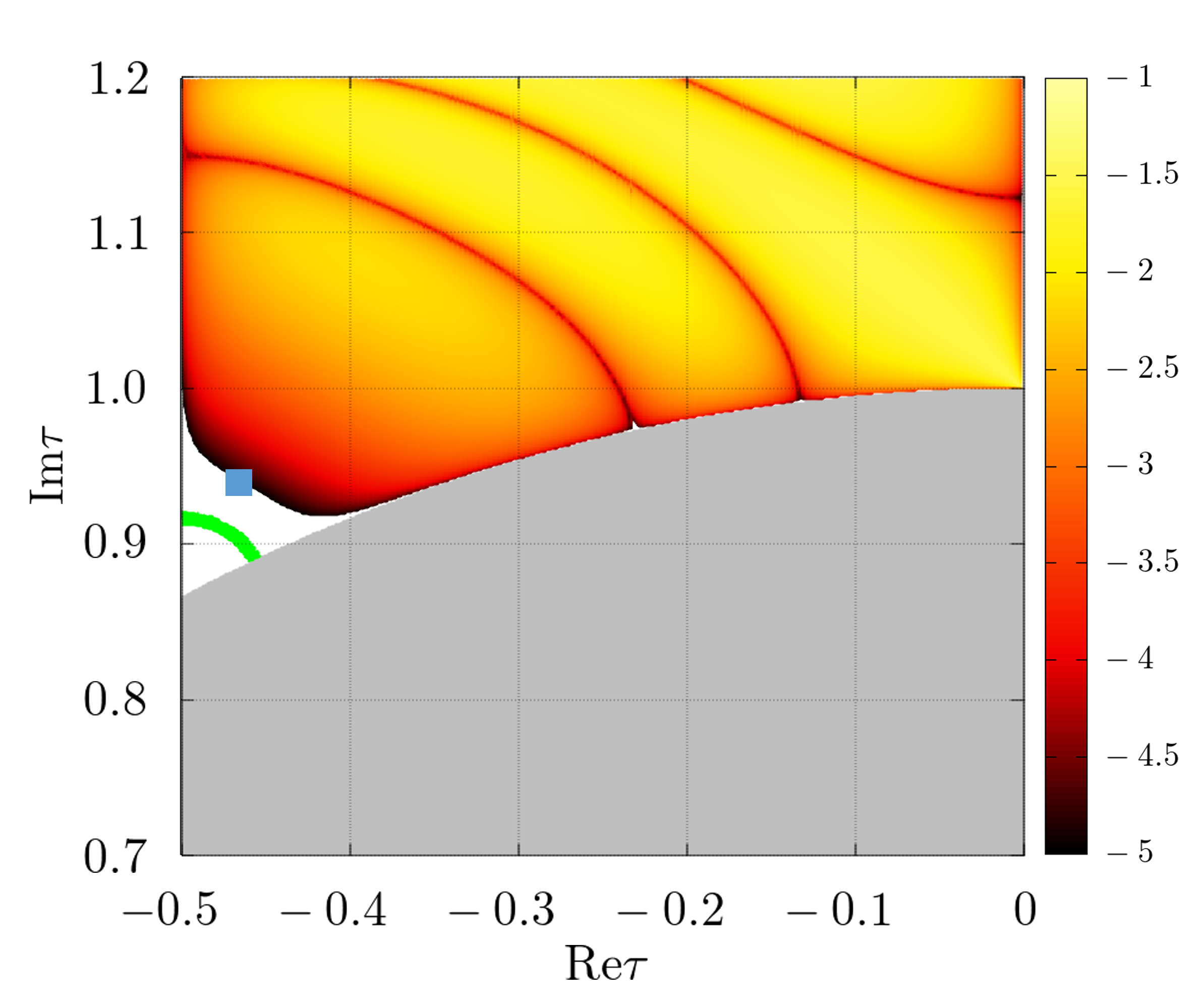}
  \caption{Allowed regions and Jarlskog invariant in the $\tau$ plane around $\tau=\omega$ for the model in type $1^22\textrm{-}1^23^2\textrm{-}3^2$.
  Green is the region satisfying hierarchy conditions in Eq.~(\ref{eq: mass_ratio_order}), and black, red and yellow colors correspond to $\textrm{log}_{10}J_{\textrm{CP}}$.
  White is the region with $\textrm{log}_{10}J_{\textrm{CP}}<-5$.
  Note that Jarlskog invariant $J_{\textrm{CP}}$ has been observed as $2.80\times 10^{-5}$.
  Blue square denotes the point $\tau=\omega+(0.0326+0.0753i)$ on numerical example in Table \ref{tab:J>000001_omega}.
  Gray is out of fundamental region.}
  \label{fig:chi_J_omega}
\end{figure}

First we focus on the vicinity of $\tau=\omega$ and study the following model,
\begin{align}
  &a_1=1,\quad a_2=1,\quad a_3=1,\quad b_1=1,\quad b_2=0,\quad b_3=0, \label{eq:J_model_charge} \\
  &\begin{pmatrix}
    \alpha^{11} & \alpha^{12} & \alpha^{13} \\
    \alpha^{21} & \alpha^{22} & \alpha^{23} \\
    \alpha^{31} & \alpha^{32} & \alpha^{33} \\
  \end{pmatrix}
  =
  \begin{pmatrix}
    1 & 1 & 1 \\
    1 & -1 & -1 \\
    1 & 1 & -1 \\
  \end{pmatrix}, \quad
  \begin{pmatrix}
    \beta^{11} & \beta^{12} & \beta^{13} \\
    \beta^{21} & \beta^{22} & \beta^{23} \\
    \beta^{31} & \beta^{32} & \beta^{33} \\
  \end{pmatrix}
  =
  \begin{pmatrix}
    1 & 1 & 1 \\
   - 1 & 1 & -1 \\
    -1 & -1 & 1 \\
  \end{pmatrix}. \label{eq:J_model_sign}
\end{align}
The mass matrices are given by
\begin{align}
  &M_u = \langle H_u \rangle
  \begin{pmatrix}
    Y^{(8)}_{\bm{1}}Y^{(8)}_{\bm{1}}Y^{(8)}_{\bm{1}} & Y^{(8)}_{\bm{1}}Y^{(8)}_{\bm{1''}}Y^{(8)}_{\bm{1}} & Y^{(8)}_{\bm{1}}Y^{(8)}_{\bm{1}}Y^{(8)}_{\bm{1}} \\
    Y^{(8)}_{\bm{1}}Y^{(8)}_{\bm{1''}}Y^{(8)}_{\bm{1''}} & -Y^{(8)}_{\bm{1}}Y^{(8)}_{\bm{1'}}Y^{(8)}_{\bm{1''}} & -Y^{(8)}_{\bm{1}}Y^{(8)}_{\bm{1''}}Y^{(8)}_{\bm{1''}} \\
    Y^{(8)}_{\bm{1''}}Y^{(8)}_{\bm{1''}}Y^{(8)}_{\bm{1''}} & Y^{(8)}_{\bm{1''}}Y^{(8)}_{\bm{1'}}Y^{(8)}_{\bm{1''}} & -Y^{(8)}_{\bm{1''}}Y^{(8)}_{\bm{1''}}Y^{(8)}_{\bm{1''}} \\
  \end{pmatrix}, \\
  &M_d = \langle H_d \rangle
  \begin{pmatrix}
    Y^{(8)}_{\bm{1}}Y^{(8)}_{\bm{1''}}Y^{(8)}_{\bm{1}} & Y^{(8)}_{\bm{1''}}Y^{(8)}_{\bm{1}}Y^{(8)}_{\bm{1'}} & Y^{(8)}_{\bm{1}}Y^{(8)}_{\bm{1}}Y^{(8)}_{\bm{1}} \\
    -Y^{(8)}_{\bm{1}}Y^{(8)}_{\bm{1'}}Y^{(8)}_{\bm{1''}} & Y^{(8)}_{\bm{1''}}Y^{(8)}_{\bm{1''}}Y^{(8)}_{\bm{1}} & -Y^{(8)}_{\bm{1}}Y^{(8)}_{\bm{1''}}Y^{(8)}_{\bm{1''}} \\
    -Y^{(8)}_{\bm{1''}}Y^{(8)}_{\bm{1'}}Y^{(8)}_{\bm{1''}} & -Y^{(8)}_{\bm{1'}}Y^{(8)}_{\bm{1''}}Y^{(8)}_{\bm{1}} & Y^{(8)}_{\bm{1''}}Y^{(8)}_{\bm{1''}}Y^{(8)}_{\bm{1''}} \\
  \end{pmatrix}.
\end{align}
This is a model counted in Table \ref{tab:chi<0.01atomega} and can satisfy hierarchy conditions in Eq.~(\ref{eq: mass_ratio_order}) at the benchmark point $\tau=\omega+0.051i$ although CP violation does not occur at this value of $\tau$.
To obtain non-vanishing CP phase, we calculate Jarlskog invariant $J_{\textrm{CP}}=\textrm{Im}(V_{\textrm{CKM}}^{us}V_{\textrm{CKM}}^{cb}(V_{\textrm{CKM}}^{ub}V_{\textrm{CKM}}^{cs})^*)$ in the $\tau$ plane around $\tau=\omega$.
The results are shown in Figure \ref{fig:chi_J_omega}.
Clearly, there are no regions satisfying Eq.~(\ref{eq: mass_ratio_order}) and $J_{\textrm{CP}}>10^{-5}$.
Eq.~(\ref{eq: mass_ratio_order}) can be satisfied at $|\tau-\omega|\sim0.05$ ($\varepsilon\sim 0.15$) while $J_{\textrm{CP}}>10^{-5}$ can be realized at $|\tau-\omega|> 0.080$ ($\varepsilon> 0.23$) \footnote{This result is consistent with the estimation in Eq.~(\ref{eq:J_CP_second}).
At $\varepsilon\sim0.23$, it gives $J_{\textrm{CP}}\lesssim 2.3\times 10^{-5}\sim 10^{-5}$.}.
Since $\varepsilon\sim 0.15$ is required to generate large quark mass hierarchies, particularly the up quark mass ratio, it is difficult to obtain both realistic quark mass ratios and Jarlskog invariant simultaneously.
As a numerical example realizing $J_{\textrm{CP}}>10^{-5}$, we show the results at $\tau=\omega+(0.0326+0.0753i)$ in Table \ref{tab:J>000001_omega}.
\begin{table}[H]
\small
  \begin{center}
    \renewcommand{\arraystretch}{1.3}
    \begin{tabular}{c|cccccccc} \hline
      & $\frac{m_u}{m_t}{\times10^{6}}$ & $\frac{m_c}{m_t}{\times10^3}$ & $\frac{m_d}{m_b}{\times10^4}$ & $\frac{m_s}{m_b}{\times10^2}$ & $|V_{\textrm{CKM}}^{us}|$ & $|V_{\textrm{CKM}}^{cb}|$ & $|V_{\textrm{CKM}}^{ub}|$ & $J_{\textrm{CP}}{\times 10^5}$ \\ \hline
      obtained values & 162 & 17.8 & 76.9 & 5.93 & 0.287 & 0.100 & 0.0128 & 1.01 \\ \hline
      GUT scale values & 5.39 & 2.80 & 9.21 & 1.82 & 0.225 & 0.0400 & 0.00353 & 2.80 \\
      $1\sigma$ errors & $\pm 1.68$ & $\pm 0.12$ & $\pm 1.02$ & $\pm 0.10$ & $\pm 0.0007$ & $\pm 0.0008$ & $\pm 0.00013$ & $^{+0.14}_{-0.12}$ \\ \hline
    \end{tabular}
  \end{center}
  \caption{The mass ratios of the quarks and the absolute values of the CKM matrix elements at the benchmark point $\tau=\omega+(0.0326+0.0753i)$.
GUT scale values at $2\times 10^{16}$ GeV with $\tan \beta=5$ \cite{Antusch:2013jca,Bjorkeroth:2015ora} and $1\sigma$ errors are shown.}
\label{tab:J>000001_omega}
\normalsize
\end{table}
In this example, the CKM matrix elements are roughly realized but especially the up quark mass ratio is deviated by ${\cal O}(10)$ compared to observed value.
In other words, it may be possible to describe realistic quark mass ratios as well as the Jarlskog invariant by use of ${\cal O}(10)$ coefficients in Yukawa couplings.

\begin{figure}[H]
  \centering
  \includegraphics[width=15cm]{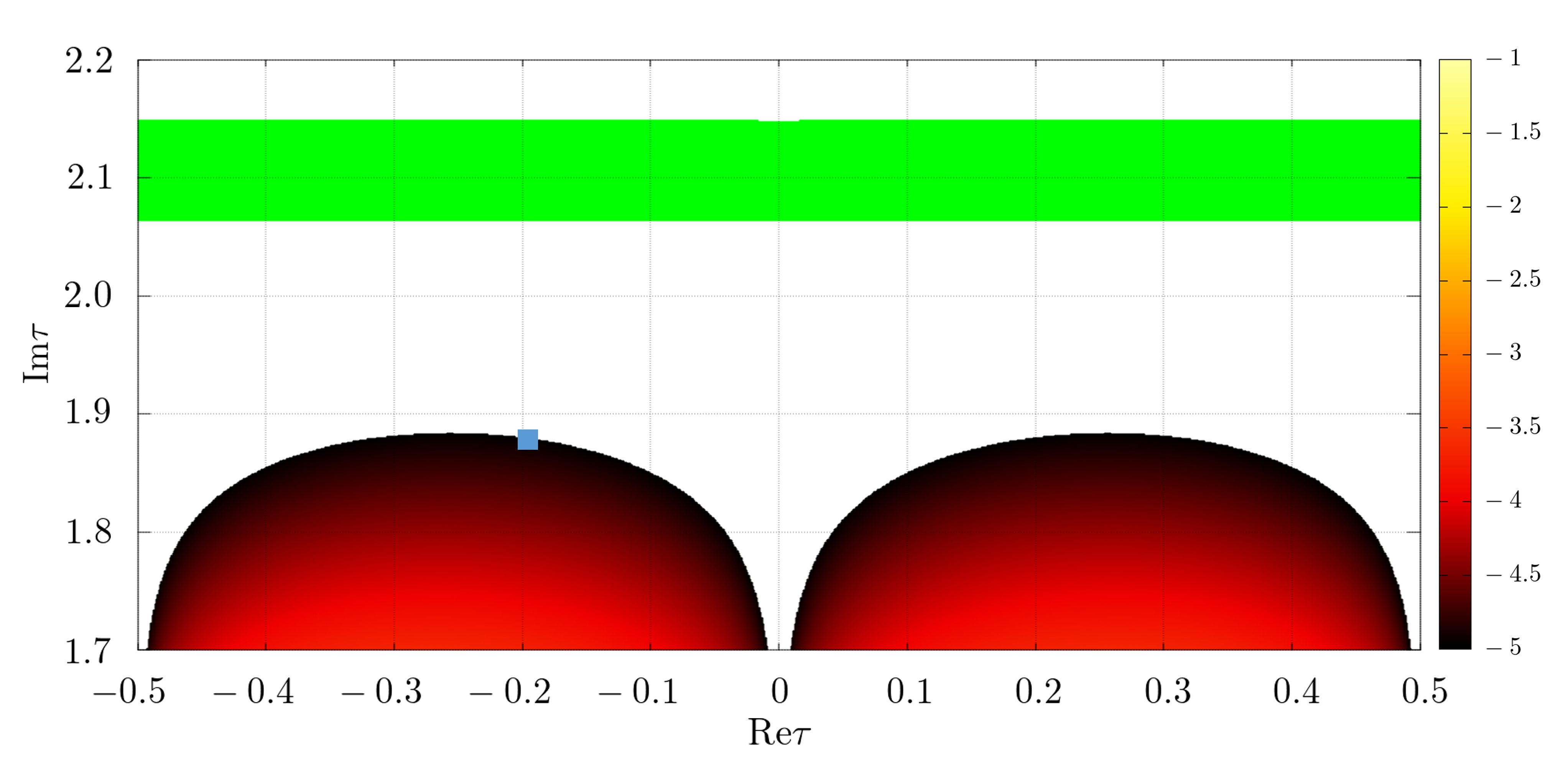}
  \caption{Allowed regions and Jarlskog invariant in the $\tau$ plane around $\tau=i \infty$ for the model in type $1^22\textrm{-}1^23^2\textrm{-}3^2$.
  Green is the region satisfying hierarchy conditions in Eq.~(\ref{eq: mass_ratio_order}), and black, red and yellow colors correspond to $\textrm{log}_{10}J_{\textrm{CP}}$.
  White is the region with $\textrm{log}_{10}J_{\textrm{CP}}<-5$.
  Note that Jarlskog invariant $J_{\textrm{CP}}$ has been observed as $2.80\times 10^{-5}$.
  Blue square denotes the point $\tau=-0.198+1.88i$ on numerical example in Table \ref{tab:J>000001_infinite}.}
  \label{fig:chi_J_infinite}
\end{figure}

Second, let us consider the vicinity of $\tau=i\infty$.
We use the model given by Eqs.~(\ref{eq:J_model_charge}) and (\ref{eq:J_model_sign}) as same as the analysis of $\tau\sim\omega$.
The mass matrices are given by
\begin{align}
  &M_u = \langle H_u \rangle
  \begin{pmatrix}
    Y^{(8)}_{\bm{1}}Y^{(8)}_{\bm{1''}}Y^{(8)}_{\bm{1''}} & Y^{(8)}_{\bm{1''}}Y^{(8)}_{\bm{1}}Y^{(8)}_{\bm{1''}} & Y^{(8)}_{\bm{1''}}Y^{(8)}_{\bm{1''}}Y^{(8)}_{\bm{1''}} \\
    Y^{(8)}_{\bm{1''}}Y^{(8)}_{\bm{1}}Y^{(8)}_{\bm{1}} & -Y^{(8)}_{\bm{1''}}Y^{(8)}_{\bm{1'}}Y^{(8)}_{\bm{1}} & -Y^{(8)}_{\bm{1''}}Y^{(8)}_{\bm{1}}Y^{(8)}_{\bm{1}} \\
    Y^{(8)}_{\bm{1}}Y^{(8)}_{\bm{1}}Y^{(8)}_{\bm{1}} & Y^{(8)}_{\bm{1}}Y^{(8)}_{\bm{1'}}Y^{(8)}_{\bm{1}} & -Y^{(8)}_{\bm{1}}Y^{(8)}_{\bm{1}}Y^{(8)}_{\bm{1}} \\
  \end{pmatrix}, \\
  &M_d = \langle H_d \rangle
  \begin{pmatrix}
    Y^{(8)}_{\bm{1''}}Y^{(8)}_{\bm{1}}Y^{(8)}_{\bm{1''}} & Y^{(8)}_{\bm{1}}Y^{(8)}_{\bm{1''}}Y^{(8)}_{\bm{1'}} & Y^{(8)}_{\bm{1''}}Y^{(8)}_{\bm{1''}}Y^{(8)}_{\bm{1''}} \\
    -Y^{(8)}_{\bm{1''}}Y^{(8)}_{\bm{1'}}Y^{(8)}_{\bm{1}} & Y^{(8)}_{\bm{1}}Y^{(8)}_{\bm{1}}Y^{(8)}_{\bm{1''}} & -Y^{(8)}_{\bm{1''}}Y^{(8)}_{\bm{1}}Y^{(8)}_{\bm{1}} \\
    -Y^{(8)}_{\bm{1}}Y^{(8)}_{\bm{1'}}Y^{(8)}_{\bm{1}} & -Y^{(8)}_{\bm{1'}}Y^{(8)}_{\bm{1}}Y^{(8)}_{\bm{1''}} & Y^{(8)}_{\bm{1}}Y^{(8)}_{\bm{1}}Y^{(8)}_{\bm{1}} \\
  \end{pmatrix}.
\end{align}
This is also a model counted in Table \ref{tab:chi<0.01atinfinite} and can satisfy hierarchy conditions in Eq.~(\ref{eq: mass_ratio_order}) at the benchmark point $\tau=2.1i$ although CP violation does not occur at this value of $\tau$.
To obtain non-vanishing CP phase, we calculate the Jarlskog invariant $J_{\textrm{CP}}$ in the $\tau$ plane around $\tau=2.1i$ ($\sim i\infty$).
The results are shown in Figure \ref{fig:chi_J_infinite}.
Clearly, there are no regions satisfying Eq.~(\ref{eq: mass_ratio_order}) and $J_{\textrm{CP}}>10^{-5}$.
Eq.~(\ref{eq: mass_ratio_order}) can be satisfied at $\textrm{Im}\tau\sim2.1$ ($\varepsilon\sim 0.15$) while $J_{\textrm{CP}}>10^{-5}$ can be realized at $\textrm{Im}\tau< 1.88$ ($\varepsilon> 0.23$).
Hence it is difficult to obtain both realistic quark mass ratios and the Jarlskog invariant simultaneously.
As a numerical example realizing $J_{\textrm{CP}}>10^{-5}$, we show the results at $\tau=-0.198+1.88i$ in Table \ref{tab:J>000001_infinite}.
\begin{table}[H]
\small
  \begin{center}
    \renewcommand{\arraystretch}{1.3}
    \begin{tabular}{c|cccccccc} \hline
      & $\frac{m_u}{m_t}{\times10^{6}}$ & $\frac{m_c}{m_t}{\times10^3}$ & $\frac{m_d}{m_b}{\times10^4}$ & $\frac{m_s}{m_b}{\times10^2}$ & $|V_{\textrm{CKM}}^{us}|$ & $|V_{\textrm{CKM}}^{cb}|$ & $|V_{\textrm{CKM}}^{ub}|$ & $J_{\textrm{CP}}{\times 10^5}$ \\ \hline
      obtained values & 160 & 17.7 & 75.9 & 5.94 & 0.287 & 0.0997 & 0.0127 & 1.00 \\ \hline
      GUT scale values & 5.39 & 2.80 & 9.21 & 1.82 & 0.225 & 0.0400 & 0.00353 & 2.80 \\ 
      $1\sigma$ errors & $\pm 1.68$ & $\pm 0.12$ & $\pm 1.02$ & $\pm 0.10$ & $\pm 0.0007$ & $\pm 0.0008$ & $\pm 0.00013$ & $^{+0.14}_{-0.12}$ \\ \hline
    \end{tabular}
  \end{center}
  \caption{The mass ratios of the quarks and the absolute values of the CKM matrix elements at the benchmark point $\tau=-0.198+1.88i$.
GUT scale values at $2\times 10^{16}$ GeV with $\tan \beta=5$ \cite{Antusch:2013jca,Bjorkeroth:2015ora} and $1\sigma$ errors are shown.}
\label{tab:J>000001_infinite}
\normalsize
\end{table}
Consequently $\textrm{Im}\tau\sim i\infty$ leads almost  same results as $\tau\sim\omega$.
We note that these results are given in the model of Eqs.~(\ref{eq:J_model_charge}) and (\ref{eq:J_model_sign}) but similar results can be obtained in all other models shown in Tables \ref{tab:chi<0.01atinfinite} and \ref{tab:chi<0.01atomega}.
No models can satisfy Eq.~(\ref{eq: mass_ratio_order}) and $J_{\textrm{CP}}>10^{-5}$ simultaneously although it may be possible when we use ${\cal O}(10)$ coefficients in Yukawa couplings.

Before ending this subsection, we also comment on further possibilities realizing quark mass hierarchies, CKM matrix elements and the Jarlskog invariant simultaneously.
As mentioned above, the Jarlskog invariant $J_{\textrm{CP}}\sim10^{-5}$ is obtained at $\varepsilon\sim0.23$ in our models.
This fact may be available for other modular flavor symmetry.
In $A_4\times A_4\times A_4$ modular symmetry, we need $\varepsilon\sim 0.15$ to produce $\varepsilon^6=(0.15)^6=1.14\times 10^{-5}\sim m_u/m_t$.
On the other hand, in $Z_l\times Z_m\times Z_n$ symmetry for $(l-1)+(m-1)+(n-1)\geq 8$, the modular forms of $\varepsilon^8$ exist and at $\varepsilon=0.23$ we can produce $\varepsilon^8=(0.23)^8=7.83\times 10^{-6}\sim m_u/m_t$.
In addition, $\varepsilon=0.23$ is nearly equal to the Cabibbo angle.
It is easy to check that the following mass matrix,
\begin{align}
  M_u =
  \begin{pmatrix}
    {\cal O}(\varepsilon^8) & {\cal O}(\varepsilon^5) & {\cal O}(\varepsilon^3) \\
    {\cal O}(\varepsilon^7) & {\cal O}(\varepsilon^4) & {\cal O}(\varepsilon^2) \\
    {\cal O}(\varepsilon^5) & {\cal O}(\varepsilon^2) & {\cal O}(1) \\
  \end{pmatrix},
\end{align}
is approximately triangularizable as
\begin{align}
  &U_L^\dagger M_u \sim
  \begin{pmatrix}
    {\cal O}(\varepsilon^8) & 0 & 0 \\
    {\cal O}(\varepsilon^7) & {\cal O}(\varepsilon^4) & 0 \\
    {\cal O}(\varepsilon^5) & {\cal O}(\varepsilon^2) & {\cal O}(1) \\
  \end{pmatrix},
\end{align}
where
\begin{align}
  U_L^\dagger &=
  \begin{pmatrix}
    1 & {\cal O}(\varepsilon) & 0 \\
    {\cal O}(\varepsilon) & 1 & 0 \\
    0 & 0 & 1 \\
  \end{pmatrix}
  \begin{pmatrix}
    1 & 0 & {\cal O}(\varepsilon^3) \\
    0 & 1 & 0 \\
    {\cal O}(\varepsilon^3) & 0 & 1 \\
  \end{pmatrix}
  \begin{pmatrix}
    1 & 0 & 0 \\
    0 & 1 & {\cal O}(\varepsilon^2) \\
    0 & {\cal O}(\varepsilon^2) & 1 \\
  \end{pmatrix} \\
  &\sim
  \begin{pmatrix}
    1 & {\cal O}(\varepsilon) & {\cal O}(\varepsilon^3) \\
    {\cal O}(\varepsilon) & 1 & {\cal O}(\varepsilon^2) \\
    {\cal O}(\varepsilon^3) & {\cal O}(\varepsilon^2) & 1 \\
  \end{pmatrix}.
\end{align}
Since eigenvalues of a matrix are equal to diagonal elements of its triangular matrix, mass ratios of above mass matrix are given by $(\varepsilon^8,\varepsilon^4,1)$.
Then, choosing $|\varepsilon|=0.230$ ($\sim$Cabibbo angle), we can obtain
\begin{align}
  &(m_u,m_c,m_t)/m_t \sim (|\varepsilon|^8,|\varepsilon|^4,1) = (7.83\times 10^{-6}, 2.80\times 10^{-3},1), \\
  &|U_L^\dagger| 
  \sim
  \begin{pmatrix}
    1 & |\varepsilon| & |\varepsilon|^3 \\
    |\varepsilon| & 1 & |\varepsilon|^2 \\
    |\varepsilon|^3 & |\varepsilon|^2 & 1 \\
  \end{pmatrix} 
  \sim
  \begin{pmatrix}
    1 & 0.230 & 0.0122 \\
    0.230 & 1 & 0.0529 \\
    0.0122 & 0.0529 & 1 \\
  \end{pmatrix}.
\end{align}
When down quark mass matrix is in diagonalized base as 
\begin{align}
  |(M_d^{11},M_d^{22},M_d^{33})/M_d^{33}| = (m_d,m_s,m_b)/m_b = (|\varepsilon|^5,|\varepsilon|^3,1)=(6.44\times 10^{-4},1.22\times 10^{-2},1),
\end{align}
The CKM matrix is given by $V_{\textrm{CKM}}=U_L^\dagger$ and these results are good realization of quark flavors. 
Thus the modular symmetry which breaks into $Z_l\times Z_m\times Z_n$ symmetry with $(l-1)+(m-1)+(n-1)\geq 8$ at the modular fixed point has the possibility realizing the quark flavor structure including the Jarlskog invariant.

%----------------------------------------------------------------------
%----------------------------------------------------------------------

\subsection{Non-universal moduli}

So far, we have studied the moduli values satisfying $\tau_1=\tau_2=\tau_3=\tau$.
Here we study the models with non-universal moduli as a possibility realizing quark flavors including CP violation.
As we have mentioned in section \ref{subsec:necessary}, the first order approximation in $\varepsilon$ expansion suggests that sufficient CP violation does not occur in our favorable models when $\varepsilon\sim 0.15$.
This is because the phase factor $p$ in mass matrices cannot contribute to Jarlskog invariant $J_{\textrm{CP}}$ at the first order approximation as shown in Eq.~(\ref{eq:first_p_cancellation}).
That is, $J_{\textrm{CP}}$ becomes extremely small when mass matrices possess only one kind of phase factor since it is canceled at the first order approximation.
In other words, when several kind of phase factors appear in mass matrices, they may not be canceled and $J_{\textrm{CP}}$ would have non-vanishing contribution at the first order approximation.
Such phase factors can be obtained when three moduli $\tau_1$, $\tau_2$ and $\tau_3$ take different VEVs.
In this subsection, we consider non-universal moduli and study CP violation by a concrete model.

For simplicity, let us consider following moduli:
\begin{align}
  \tau_1=\tau_2 \equiv \tau \neq \tau_3, \quad |\tau_1-\omega| = |\tau_2-\omega| = |\tau_3-\omega|.
\end{align}
Then we consider the model in type $123\textrm{-}1^22^2\textrm{-}1^2$.
In type $123\textrm{-}1^22^2\textrm{-}1^2$, quarks have the following $Z_3\times Z_3\times Z_3$ charges:
\begin{align}
&\{Q^1,Q^2,Q^3\}:~\{(a_1,a_2,a_3),(b_1,b_2,b_3),(0,0,0)\}, \\
&\{u_R^1,u_R^2,u_R^3\}:~\{(1-a_1,1-a_2,1-a_3)_{\textrm{mod~3}},(2-b_1,2-b_2,2-b_3)_{\textrm{mod~3}},(0,0,0)\}, \\
&\{d_R^1,d_R^2,d_R^3\}:~\{(1-a_1,1-a_2,-a_3)_{\textrm{mod~3}},(1-b_1,-b_2,-b_3)_{\textrm{mod~3}},(0,0,0)\},
\end{align}
where $a_i\in\{0,1,2\}$ and $b_i\in\{0,1,2\}$ are $Z_3$-charges of the $i$-th $A_4$ for $Q^1$ and $Q^2$ respectively.
We focus on the vicinity of $\tau=\omega$ and study the following model,
\begin{align}
  &a_1 = 1, \quad a_2 = 1, \quad a_3 = 1, \quad b_1 = 2, \quad b_2 = 2, \quad b_3 = 0, \\
  &\begin{pmatrix}
\alpha^{11} & \alpha^{12} & \alpha^{13} \\
\alpha^{21} & \alpha^{22} & \alpha^{23} \\
\alpha^{31} & \alpha^{32} & \alpha^{33} \\
\end{pmatrix}
=
\begin{pmatrix}
1 & 1 & 1 \\
1 & 1 & -1 \\
1 & -1 & -1 \\
\end{pmatrix}, \quad\begin{pmatrix}
\beta^{11} & \beta^{12} & \beta^{13} \\
\beta^{21} & \beta^{22} & \beta^{23} \\
\beta^{31} & \beta^{32} & \beta^{33} \\
\end{pmatrix}
=
\begin{pmatrix}
1 & 1 & 1 \\
-1 & -1 & 1 \\
1 & 1 & -1 \\
\end{pmatrix}.
\end{align}
The mass matrices are given by
\begin{align}
&\frac{M_u}{ \langle H_u \rangle} =  \begin{pmatrix}
Y^{(8)}_{\bm{1''}}(\tau)Y^{(8)}_{\bm{1''}}(\tau)Y^{(8)}_{\bm{1''}}(\tau_3) & Y^{(8)}_{\bm{1''}}(\tau)Y^{(8)}_{\bm{1''}}(\tau)Y^{(8)}_{\bm{1}}(\tau_3) & Y^{(8)}_{\bm{1''}}(\tau)Y^{(8)}_{\bm{1''}}(\tau)Y^{(8)}_{\bm{1''}}(\tau_3) \\
Y^{(8)}_{\bm{1'}}(\tau)Y^{(8)}_{\bm{1'}}(\tau)Y^{(8)}_{\bm{1}}(\tau_3) & Y^{(8)}_{\bm{1'}}(\tau)Y^{(8)}_{\bm{1'}}(\tau)Y^{(8)}_{\bm{1'}}(\tau_3) & -Y^{(8)}_{\bm{1'}}(\tau)Y^{(8)}_{\bm{1'}}(\tau)Y^{(8)}_{\bm{1}}(\tau_3) \\
Y^{(8)}_{\bm{1}}(\tau)Y^{(8)}_{\bm{1}}(\tau)Y^{(8)}_{\bm{1}}(\tau_3) & -Y^{(8)}_{\bm{1}}(\tau)Y^{(8)}_{\bm{1}}(\tau)Y^{(8)}_{\bm{1'}}(\tau_3) & -Y^{(8)}_{\bm{1}}(\tau)Y^{(8)}_{\bm{1}}(\tau)Y^{(8)}_{\bm{1}}(\tau_3) \\
\end{pmatrix}, \\
&\frac{M_d }{ \langle H_d \rangle}= \begin{pmatrix}
Y^{(8)}_{\bm{1''}}(\tau)Y^{(8)}_{\bm{1''}}(\tau)Y^{(8)}_{\bm{1}}(\tau_3) & Y^{(8)}_{\bm{1}}(\tau)Y^{(8)}_{\bm{1'}}(\tau)Y^{(8)}_{\bm{1''}}(\tau_3) & Y^{(8)}_{\bm{1''}}(\tau)Y^{(8)}_{\bm{1''}}(\tau)Y^{(8)}_{\bm{1''}}(\tau_3) \\
-Y^{(8)}_{\bm{1'}}(\tau)Y^{(8)}_{\bm{1'}}(\tau)Y^{(8)}_{\bm{1'}}(\tau_3) & -Y^{(8)}_{\bm{1''}}(\tau)Y^{(8)}_{\bm{1}}(\tau)Y^{(8)}_{\bm{1}}(\tau_3) & Y^{(8)}_{\bm{1'}}(\tau)Y^{(8)}_{\bm{1'}}(\tau)Y^{(8)}_{\bm{1}}(\tau_3) \\
Y^{(8)}_{\bm{1}}(\tau)Y^{(8)}_{\bm{1}}(\tau)Y^{(8)}_{\bm{1'}}(\tau_3) & Y^{(8)}_{\bm{1'}}(\tau)Y^{(8)}_{\bm{1''}}(\tau)Y^{(8)}_{\bm{1}}(\tau_3) & -Y^{(8)}_{\bm{1}}(\tau)Y^{(8)}_{\bm{1}}(\tau)Y^{(8)}_{\bm{1}}(\tau_3) \\
\end{pmatrix}.
\end{align}
This is a model counted in Table \ref{tab:chi<0.01atomega} and can satisfy hierarchy conditions in Eq.~(\ref{eq: mass_ratio_order}) at the benchmark point $\tau=\tau_3=\omega+0.051i$ although CP violation does not occur at this value of $\tau$.
After the basis transformation in Eqs.~(\ref{eq:u_L}) and (\ref{eq:u_R}), these mass matrices are estimated as
\begin{align}
  M_u \sim
  \begin{pmatrix}
    |\varepsilon|^6 & |\varepsilon|^4p_3^{-1} & |\varepsilon|^6 \\
    |\varepsilon|^2 & |\varepsilon|^3 & -|\varepsilon|^2 \\
    1 & -|\varepsilon| & -1 \\
  \end{pmatrix}, \quad
  M_d \sim
  \begin{pmatrix}
    |\varepsilon|^4p_3^{-1} & |\varepsilon|^3p^{-2} & |\varepsilon|^6 \\
    -|\varepsilon|^3 & -|\varepsilon|^2p^{-1} & |\varepsilon|^2 \\
    |\varepsilon| & |\varepsilon|^3 & -1 \\
  \end{pmatrix},
\end{align}
at the first order approximation.
In above, $p$ is given by $u/|u|$ for $u=(\tau-\omega)/(\tau-\omega^2)$ and $p_3$ is given by $u_3/|u_3|$ for $u_3=(\tau_3-\omega)/(\tau_3-\omega^2)$.
From these mass matrices, we can find the following structures of the CKM matrix,
\begin{align}
  V_{\textrm{CKM}} =
  \begin{pmatrix}
1 & -|\varepsilon|\left(p^* + 0.5p_3^*\right) & |\varepsilon|^{3}p_3^* \\
|\varepsilon| \left(p + 0.5 p_3\right) & 1 & - 2 |\varepsilon|^{2} \\
2 |\varepsilon|^{3} p & 2 |\varepsilon|^{2} & 1 \\
  \end{pmatrix},
\end{align}
at the first order approximation.
This directly leads to Jarlskog invariant, 
\begin{align}
  J_{\textrm{CP}} &= |\textrm{Im}(- 2|\varepsilon|^{6} \left(- p^* - 0.5p_3^*\right) p_3)| = 2|\varepsilon|^6\cdot |\textrm{Im}(p^*p_3)|.
\end{align}
Thus, when $p\neq p_3$, hence $\tau\neq\tau_3$, we can obtain non-vanishing Jarlskog invariant at the first order approximation.
As a numerical example realizing realistic Jarlskog invariant, we choose
\begin{align}
  \tau = \omega+0.055i, \quad \tau_3 = \omega+0.055e^{2\pi i/5},
\end{align}
and show the results in Table \ref{tab:non-universal_result}.
\begin{table}[H]
\small
  \begin{center}
    \renewcommand{\arraystretch}{1.3}
    \begin{tabular}{c|cccccccc} \hline
      & $\frac{m_u}{m_t}{\times10^{6}}$ & $\frac{m_c}{m_t}{\times10^3}$ & $\frac{m_d}{m_b}{\times10^4}$ & $\frac{m_s}{m_b}{\times10^2}$ & $|V_{\textrm{CKM}}^{us}|$ & $|V_{\textrm{CKM}}^{cb}|$ & $|V_{\textrm{CKM}}^{ub}|$ & $J_{\textrm{CP}}{\times 10^5}$ \\ \hline
      obtained values & 16.0 & 5.63 & 6.16 & 2.52 & 0.214 & 0.0498 & 0.00411 & 2.53 \\ \hline
      GUT scale values & 5.39 & 2.80 & 9.21 & 1.82 & 0.225 & 0.0400 & 0.00353 & 2.80 \\ 
      $1\sigma$ errors & $\pm 1.68$ & $\pm 0.12$ & $\pm 1.02$ & $\pm 0.10$ & $\pm 0.0007$ & $\pm 0.0008$ & $\pm 0.00013$ & $^{+0.14}_{-0.12}$ \\ \hline
    \end{tabular}
  \end{center}
  \caption{The mass ratios of the quarks and the absolute values of the CKM matrix elements at $\tau=\omega+0.055i$ and $\tau_3=\omega+0.055e^{2\pi i/5}$.
GUT scale values at $2\times 10^{16}$ GeV with $\tan \beta=5$ \cite{Antusch:2013jca,Bjorkeroth:2015ora} and $1\sigma$ errors are shown.}
\label{tab:non-universal_result}
\normalsize
\end{table}
This result satisfies hierarchy conditions in Eq.~(\ref{eq: mass_ratio_order}).
Consequently, we can simultaneously obtain realistic quark mass ratios, absolute values of CKM matrix elements and Jarlskog invariant through non-universal moduli.

%----------------------------------------------------------------------
%----------------------------------------------------------------------
%----------------------------------------------------------------------

\section{Conclusion}
\label{sec:4}

We have discussed the possibilities of explaining quark flavor structures, in particular large quark mass hierarchies, without fine-tuning.
In modular symmetric flavor models, mass matrices are written in terms of the modular forms.
The values of the modular forms become hierarchical as close to the modular fixed points depending on the residual $Z_n$ charges.
In more detail, the modular forms with $Z_n$ residual charge $r$ can be estimated as $\varepsilon^r$ where $\varepsilon$ is the deviation of the modulus $\tau$ from the modular fixed points.
Along in this way we study large quark mass hierarchies as well as CKM matrix elements in $A_4\times A_4\times A_4$ modular symmetry.
We have focused two fixed points, $\tau=i\infty$ and $\omega$ where $A_4\times A_4\times A_4$ breaks into $Z_3\times Z_3\times Z_3$.
Then we can obtain the modular forms whose orders are $1$, $\varepsilon$, ... , $\varepsilon^6$.

The modular forms of level 3 and weight 8 contain three singlets denoted by $Y^{(8)}_{\bm{1}}(\tau)$, $Y^{(8)}_{\bm{1'}}(\tau)$ and $Y^{(8)}_{\bm{1''}}(\tau)$.
At $\tau\sim i\infty$, they are estimated as $1$, $\varepsilon$ and $\varepsilon^2$ since their $T$-charges are 0, 1 and 2.
Similarly $\tau\sim \omega$, they are estimated as $\varepsilon^2$, $\varepsilon$ and $1$ since their $ST$-charges are 2, 1 and 0.
Using these modular forms, we have classified charge assignments (types) leading to the up quark mass matrix with $\textrm{diag}(M_u)=({\cal O}(\varepsilon^6),{\cal O}(\varepsilon^3),{\cal O}(1))$ and down quark mass matrix with $\textrm{diag}(M_d)=({\cal O}(\varepsilon^4),{\cal O}(\varepsilon^2),{\cal O}(1))$ which are plausible to realize quark masses.
In addition, we have fixed coupling constants $\alpha$ and $\beta$ in Yukawa couplings to $\pm 1$ to avoid fine-tuning of them.
We have enumerated the models for each choice of the signs in $\alpha$ and $\beta$ for each types, and investigated the models satisfying hierarchy conditions in Eq.~(\ref{eq: mass_ratio_order}).
Consequently, we have obtained 1,584 number of passed models for both two benchmark points $\tau=2.1i$ ($\sim i\infty$) and $\omega+0.051i$ ($\sim\omega$) as shown in Tables \ref{tab:chi<0.01atinfinite} and \ref{tab:chi<0.01atomega}.
Actually our numerical examples by the models satisfying Eq.~(\ref{eq: mass_ratio_order}) present realistic quark mass ratios and absolute values of CKM matrix elements as shown in section \ref{subsec:numerical}.

We also study CP violation induced by the VEV of the modulus $\tau$.
To understand the origin of CP violation, we have studied the necessary conditions for CP violation.
They suggest that phase factors and hierarchical structures of mass matrices in the vicinity of the modular fixed points are  determined by the residual charge assignments into fields.
In other words, hierarchical quark masses and CP violation are related each other through the residual charges.

It was found that favorable models in Tables \ref{tab:chi<0.01atinfinite} and \ref{tab:chi<0.01atomega} satisfy the necessary conditions for CP violation.
However it was also found from the numerical analysis that they cannot induce sufficient CP violation in the regions satisfying hierarchy conditions in Eq.~(\ref{eq: mass_ratio_order}).
This weak CP violation may be caused by the size of the deviation of $\tau$, $\varepsilon$.
In the region satisfying Eq.~(\ref{eq: mass_ratio_order}), we find $\varepsilon\sim 0.15$ while $J_{\textrm{CP}}>10^{-5}$ is realized in $\varepsilon>0.23$.
Although we give numerical examples in $\varepsilon\sim 0.23$, $J_{\textrm{CP}}\sim 10^{-5}$ and $m_u/m_t\sim 10^{-4}$ have been obtained.
To obtain realistic values of quark flavors including the Jarlskog invariant in our models, we would need tuning by ${\cal O}(10)$ constants in Yukawa couplings.

We have commented on the further possibilities describing quark flavors.
To realize the up quark mass ratio by the modular forms of $\varepsilon^6$ in $A_4\times A_4\times A_4$ modular symmetry, we need $\varepsilon\sim0.15$.
When we introduce the residual $Z_n$ symmetry with $n\geq 9$, we can obtain the modular forms of $\varepsilon^8$ and can relax the size of $\varepsilon$ to $0.23$.
Moreover, this value is nearly equal to Cabibbo angle and therefore there are the possibilities explaining quark mass hierarchies, mixing angles and CP violation simultaneously in the $Z_l\times Z_m\times Z_n$ residual symmetry with $(l-1)+(m-1)+(n-1)\geq 8$.
We will study this in near future.

We have focus the case that the moduli values satisfy $\tau_1=\tau_2=\tau_3=\tau$ for simplicity.
In general, these moduli values $\tau_i$ can be different from each other.
We may have more rich structure in variation of types and numerical results.
%We would study it elsewhere.
Actually, in the end of section \ref{sec:CP} we have studied the model at $\tau_i\sim\omega$ with non-universal moduli $\tau_1=\tau_2= \tau\neq\tau_3$, $|\tau_1-\omega|=|\tau_2-\omega|=|\tau_3-\omega|$, and obtained realistic quark flavor observations including Jarlskog invariant.
Then Jarlskog invariant originates from the difference between $(\tau-\omega)/|(\tau-\omega)|$ and $(\tau_3-\omega)/|(\tau_3-\omega)|$.
Note that quark mass hierarchies originate from the deviation from the modular fixed point $|\tau-\omega|=|\tau_3-\omega|$ as same as the results in $\tau_1=\tau_2=\tau_3=\tau$.
In this way, the modulus value is important in our models.
Thus, the moduli stabilization is the key issue \footnote{See for moduli stabilization in moduli flavor models 
Refs.~\cite{Kobayashi:2019xvz,Ishiguro:2020tmo,Abe:2020vmv,Novichkov:2022wvg,Ishiguro:2022pde}.}.
We leave it for future study.

%-------- acknowledgement -------%
\vspace{1.5 cm}
\noindent
{\large\bf Acknowledgement}\\

This work was supported by JSPS KAKENHI Grant Numbers JP22J10172 (SK) and JP20J20388 (HU), 
and 
JST SPRING Grant Number JPMJSP2119(KN).

%----------------------------------------------------------------------
%----------------------------------------------------------------------
%----------------------------------------------------------------------

\appendix
\section*{Appendix}

%----------------------------------------------------------------------
%----------------------------------------------------------------------
%----------------------------------------------------------------------

\section{Group theoretical aspects of $A_4$}
\label{app:A4group}

Here, we give a review on group theoretical aspects of  $A_4$.
The generators of $A_4$ are denoted by $S$ and $T$, and 
they satisfy the following algebraic relations:
\begin{align}
  S^2 = (ST)^3 = T^3 = 1.
\end{align}
In $A_4$ group, there are four irreducible representations, three singlets $\bm{1}$, $\bm{1'}$ and $\bm{1''}$ and one triplet $3$.
Each irreducible representation is given by
\begin{align}
  &\bm{1}\quad \rho(S)=1, ~\rho(T)=1, \\
  &\bm{1'} \quad \rho(S)=1,~\rho(T)=\omega, \\
  &\bm{1''} \quad \rho(S)=1, ~\rho(T)=\omega^2, \\
  &\bm{3} \quad 
  \rho(S) = \frac{1}{3}
  \begin{pmatrix}
    -1 & 2 & 2 \\
    2 & -1 & 2 \\
    2 & 2 & -1 \\
  \end{pmatrix},\quad
  \rho(T) =
  \begin{pmatrix}
    1 & 0 & 0 \\
    0 & \omega & 0 \\
    0 & 0 & \omega^2 \\
  \end{pmatrix},
\end{align}
in the $T$-diagonal basis.
Their multiplication rules are shown in Table \ref{tab:MultiRuleinA4}.
\begin{table}[H]
\begin{center}
\renewcommand{\arraystretch}{1}
\begin{tabular}{c|c} \hline
  Tensor product  & $T$-diagonal basis \\ \hline
  $\bm{1''} \otimes \bm{1''} = \bm{1'}$ & \multirow{3}{*}{$a^1b^1$} \\
  $~~~~~~~~~~\bm{1'} \otimes \bm{1'} = \bm{1''}$ ~~$(a^1 b^1)$ & \\
  $\bm{1''} \otimes \bm{1'} = \bm{1}$ & \\ \hline
  \multirow{2}{*}{$\bm{1''} \otimes \bm{3} = \bm{3}$ ~~$(a^1 b^i)$} & \multirow{2}{*}{$\left(\begin{smallmatrix} a^1b^3\\ a^1b^1\\ a^1b^2\\\end{smallmatrix}\right)$} \\
  & \\ \hline
  \multirow{2}{*}{$\bm{1'} \otimes \bm{3} = \bm{3}$~~$(a^1 b^i)$} & \multirow{2}{*}{$\left(\begin{smallmatrix} a^1b^2\\ a^1b^3\\ a^1b^1\\ \end{smallmatrix}\right)$} \\
  & \\ \hline
  \multirow{5}{*}{$\bm{3}\otimes \bm{3}=\bm{1}\oplus \bm{1''} \oplus \bm{1'} \oplus \bm{3} \oplus \bm{3}$} & $\begin{smallmatrix}(a^1b^1+a^2b^3+a^3b^2)\end{smallmatrix}$ \\
  & $\oplus\begin{smallmatrix}(a^1b^2+a^2b^1+a^3b^3)\end{smallmatrix}$ \\
  & $\oplus\begin{smallmatrix}(a^1b^3+a^2b^2+a^3b^1)\end{smallmatrix}$ \\
 \multirow{2}{*}{$(a^ib^j)$}  & \multirow{2}{*}{$\oplus\frac{1}{3}\left(\begin{smallmatrix} 2a^1b^1-a^2b^3-a^3b^2\\ -a^1b^2-a^2b^1+2a^3b^3\\ -a^1b^3+2a^2b^2-a^3b^1\\ \end{smallmatrix}\right)$} \\
  & \\
  & \multirow{2}{*}{$\oplus\frac{1}{2}\left(\begin{smallmatrix} a^2b^3-a^3b^2\\ a^1b^2-a^2b^1\\ -a^1b^3+a^3b^1\\ \end{smallmatrix}\right)$} \\
  & \\ \hline
\end{tabular}
\end{center}
\caption{Multiplication rule in irreducible representations of $A_4$.}
\label{tab:MultiRuleinA4}
\end{table}

%----------------------------------------------------------------------
%----------------------------------------------------------------------
%----------------------------------------------------------------------

\section{Modular forms of $A_4$}
\label{app:A4modularforms}

Here we give a review on the modular forms of $\Gamma_3 \simeq A_4$.
The modular forms of  even weights can be constructed from the Dedekind eta function $\eta(\tau)$ and its derivative,
\begin{align}
  &\eta(\tau) = q^{1/24} \prod_{n=1}^{\infty} (1-q^n), \quad q = e^{2\pi i\tau}, \\
  &\eta'(\tau) \equiv \frac{d}{d\tau} \eta(\tau).
\end{align}
Using $\eta$ and $\eta'$, the modular forms of weight 2 belonging to $A_4$ triplet $\bm{3}$ can be written down as 
\cite{Feruglio:2017spp}
\begin{align}
  Y^{(2)}_{\bm{3}}(\tau) = 
  \begin{pmatrix}
    Y_1 \\ Y_2 \\ Y_3 \\
  \end{pmatrix},
\end{align}
where
\begin{align}
  &Y_1(\tau) = \frac{i}{2\pi} \left(\frac{\eta'(\tau/3)}{\eta(\tau/3)} + \frac{\eta'((\tau+1)/3)}{\eta((\tau+1)/3)} + \frac{\eta'((\tau+2)/3)}{\eta((\tau+2)/3)} - \frac{27\eta'(3\tau)}{\eta(3\tau)}\right), \\
  &Y_2(\tau) = \frac{-i}{\pi} \left(\frac{\eta'(\tau/3)}{\eta(\tau/3)} + \omega^2\frac{\eta'((\tau+1)/3)}{\eta((\tau+1)/3)} + \omega \frac{\eta'((\tau+2)/3)}{\eta((\tau+2)/3)}\right), \\
  &Y_3(\tau) = \frac{-i}{\pi} \left(\frac{\eta'(\tau/3)}{\eta(\tau/3)} + \omega\frac{\eta'((\tau+1)/3)}{\eta((\tau+1)/3)} + \omega^2\frac{\eta'((\tau+2)/3)}{\eta((\tau+2)/3)}\right).
\end{align}
They have the following $q$-expansions:
\begin{align}
  Y^{(2)}_{\bm{3}}(\tau) = 
  \begin{pmatrix}
    Y_1 \\ Y_2 \\ Y_3 \\
  \end{pmatrix}
  =
  \begin{pmatrix}
    1+12q+36q^2+12q^3+\cdots \\
    -6q^{1/3}(1+7q+8q^2+\cdots) \\
    -18q^{2/3}(1+2q+5q^2+\cdots) \\
  \end{pmatrix}.
\end{align}
Higher modular forms can be obtained by tensor products of $Y^{(2)}_{\bm{3}}(\tau)$.
Here we show the modular forms up to weight 8.
The linearly independent three modular forms of weight 4 are given by
\begin{align}
  &Y_{\bm{1}}^{(4)}(\tau) = Y^2_1+2Y_2Y_3, \quad Y_{\bm{1'}}^{(4)}(\tau) = Y^2_3+2Y_1Y_2, \notag \\
  &Y_{\bm{3}}^{(4)}(\tau) 
  =
  \begin{pmatrix}
    Y^2_1-Y_2Y_3 \\ Y^2_3-Y_1Y_2 \\ Y^2_2-Y_1Y_3 \\
  \end{pmatrix}.
\end{align}
The linearly independent three modular forms of weight 6 are given by
\begin{align}
  &Y^{(6)}_{\bm{1}}(\tau) = Y^3_1+Y^3_2+Y^3_3-3Y_1Y_2Y_3, \notag \\
  &Y_{\bm{3}}^{(6)}(\tau) = (Y^2_1+2Y_2Y_3)
  \begin{pmatrix}
    Y_1 \\ Y_2 \\ Y_3 \\
  \end{pmatrix}, \quad
  Y_{\bm{3'}}^{(6)}(\tau) = (Y^2_3+2Y_1Y_2)
  \begin{pmatrix}
    Y_3 \\ Y_1 \\ Y_2 \\
  \end{pmatrix}.
\end{align}
The linearly independent five modular forms of weight 8 are given by
\begin{align}
  &Y_{\bm{1}}^{(8)}(\tau) = (Y^2_1+2Y_2Y_3)^2, \quad Y_{\bm{1'}}^{(8)}(\tau) = (Y^2_1+2Y_2Y_3)(Y^2_3+2Y_1Y_2), \quad Y_{\bm{1''}}^{(8)}(\tau) = (Y^2_3+2Y_1Y_2)^2, \notag \\
  &Y_{\bm{3}}^{(8)}(\tau) = (Y^2_1+2Y_2Y_3)
  \begin{pmatrix}
    Y^2_1-Y_2Y_3 \\ Y^2_3-Y_1Y_2 \\ Y^2_2-Y_1Y_3 \\
  \end{pmatrix}, \quad
  Y_{\bm{3'}}^{(8)}(\tau) = (Y^2_3+2Y_1Y_2)
  \begin{pmatrix}
    Y^2_2-Y_1Y_3 \\
    Y^2_1-Y_2Y_3 \\
    Y^2_3-Y_1Y_2 \\
  \end{pmatrix}.
\end{align}

%----------------------------------------------------------------------
%----------------------------------------------------------------------
%----------------------------------------------------------------------

\section{Mass matrix structures in favorable models}
\label{app:viable_models}

Here we show the mass matrix structures in favorable models summarized in Tables \ref{tab:chi<0.01atinfinite} and \ref{tab:chi<0.01atomega}.
We express the structures of up and down quark mass matrices by phase factors after the basis transformations Eqs.~(\ref{eq:u_L}) and (\ref{eq:u_R}) and powers of $\varepsilon\sim 0.15$.
Then mass matrix structures satisfying hierarchy conditions in Eq.~(\ref{eq: mass_ratio_order}) at $\tau=2.1i$ and $\omega+0.051i$ are shown in Table \ref{tab:mass_matrix_structures}.
Note that we show different structures which are not related by unitary transformations for fields.
In total we find 128 number of different structures.
As we mentioned in section \ref{subsec:numerical_CP}, we would realize realistic quark flavor observations including the Jarlskog invariant in these mass matrix structures by using ${\cal O}(10)$ constants.

\begin{table}[H]
  \centering
  \caption{Phase factors and hierarchical structures of up and down quark mass matrices after the basis transformations Eqs.~(\ref{eq:u_L}) and (\ref{eq:u_R}) in favorable models in Tables \ref{tab:chi<0.01atinfinite} and \ref{tab:chi<0.01atomega}.
  First row denotes the structure of up quark mass matrix and other rows denote ones of down quark, up to $\langle H_u \rangle$ and $\langle H_d \rangle$.
  $p$ is given by $u/|u|$ for $\tau\sim\omega$ and $(q/|q|)^{1/3}$ for $\tau\sim i\infty$.
  We show different structures which are not related by unitary transformations for fields.
  In total we find 128 number of different structures.}
  \label{tab:mass_matrix_structures}
  \renewcommand{\arraystretch}{1.8}
  \begin{tabular}{p{9em}p{9em}p{9em}p{9em}} \hline
\multicolumn{4}{c}{$M_u=\begin{psmallmatrix}
|\varepsilon|^6 & |\varepsilon|^4p^{-1} & |\varepsilon|^6 \\
|\varepsilon|^2 & -|\varepsilon|^3 & -|\varepsilon|^2 \\
1 & |\varepsilon| & -1 \\
\end{psmallmatrix}$} \\ \hline
$\begin{psmallmatrix}
|\varepsilon|^4p^{-1} & |\varepsilon|^3p^{-2} & |\varepsilon|^6 \\
|\varepsilon|^3 & |\varepsilon|^2p^{-1} & |\varepsilon|^2 \\
|\varepsilon| & -|\varepsilon|^3 & -1 \\
\end{psmallmatrix}$
&
$\begin{psmallmatrix}
|\varepsilon|^4p^{-1} & |\varepsilon|^3p^{-2} & |\varepsilon|^6 \\
-|\varepsilon|^3 & |\varepsilon|^2p^{-1} & |\varepsilon|^2 \\
|\varepsilon| & -|\varepsilon|^3 & -1 \\
\end{psmallmatrix}$
&
$\begin{psmallmatrix}
|\varepsilon|^4p^{-1} & |\varepsilon|^3p^{-2} & |\varepsilon|^6 \\
|\varepsilon|^3 & |\varepsilon|^2p^{-1} & |\varepsilon|^2 \\
-|\varepsilon| & -|\varepsilon|^3 & -1 \\
\end{psmallmatrix}$
&
$\begin{psmallmatrix}
|\varepsilon|^4p^{-1} & |\varepsilon|^3p^{-2} & |\varepsilon|^6 \\
-|\varepsilon|^3 & |\varepsilon|^2p^{-1} & |\varepsilon|^2 \\
-|\varepsilon| & -|\varepsilon|^3 & -1 \\
\end{psmallmatrix}$
\\
$\begin{psmallmatrix}
|\varepsilon|^4p^{-1} & |\varepsilon|^3p^{-2} & |\varepsilon|^6 \\
|\varepsilon|^3 & |\varepsilon|^2p^{-1} & |\varepsilon|^2 \\
|\varepsilon| & |\varepsilon|^3 & -1 \\
\end{psmallmatrix}$
&
$\begin{psmallmatrix}
|\varepsilon|^4p^{-1} & |\varepsilon|^3p^{-2} & |\varepsilon|^6 \\
-|\varepsilon|^3 & |\varepsilon|^2p^{-1} & |\varepsilon|^2 \\
|\varepsilon| & |\varepsilon|^3 & -1 \\
\end{psmallmatrix}$
&
$\begin{psmallmatrix}
|\varepsilon|^4p^{-1} & |\varepsilon|^3p^{-2} & |\varepsilon|^6 \\
|\varepsilon|^3 & |\varepsilon|^2p^{-1} & |\varepsilon|^2 \\
-|\varepsilon| & |\varepsilon|^3 & -1 \\
\end{psmallmatrix}$
&
$\begin{psmallmatrix}
|\varepsilon|^4p^{-1} & |\varepsilon|^3p^{-2} & |\varepsilon|^6 \\
-|\varepsilon|^3 & |\varepsilon|^2p^{-1} & |\varepsilon|^2 \\
-|\varepsilon| & |\varepsilon|^3 & -1 \\
\end{psmallmatrix}$
\\
$\begin{psmallmatrix}
|\varepsilon|^4p^{-1} & |\varepsilon|^3p^{-2} & |\varepsilon|^6 \\
|\varepsilon|^3 & |\varepsilon|^2p^{-1} & -|\varepsilon|^2 \\
|\varepsilon| & -|\varepsilon|^3 & 1 \\
\end{psmallmatrix}$
&
$\begin{psmallmatrix}
|\varepsilon|^4p^{-1} & |\varepsilon|^3p^{-2} & |\varepsilon|^6 \\
-|\varepsilon|^3 & |\varepsilon|^2p^{-1} & -|\varepsilon|^2 \\
|\varepsilon| & -|\varepsilon|^3 & 1 \\
\end{psmallmatrix}$
&
$\begin{psmallmatrix}
|\varepsilon|^4p^{-1} & |\varepsilon|^3p^{-2} & |\varepsilon|^6 \\
|\varepsilon|^3 & |\varepsilon|^2p^{-1} & -|\varepsilon|^2 \\
-|\varepsilon| & -|\varepsilon|^3 & 1 \\
\end{psmallmatrix}$
&
$\begin{psmallmatrix}
|\varepsilon|^4p^{-1} & |\varepsilon|^3p^{-2} & |\varepsilon|^6 \\
-|\varepsilon|^3 & |\varepsilon|^2p^{-1} & -|\varepsilon|^2 \\
-|\varepsilon| & -|\varepsilon|^3 & 1 \\
\end{psmallmatrix}$
\\
$\begin{psmallmatrix}
|\varepsilon|^4p^{-1} & |\varepsilon|^3p^{-2} & |\varepsilon|^6 \\
|\varepsilon|^3 & |\varepsilon|^2p^{-1} & -|\varepsilon|^2 \\
|\varepsilon| & |\varepsilon|^3 & 1 \\
\end{psmallmatrix}$
&
$\begin{psmallmatrix}
|\varepsilon|^4p^{-1} & |\varepsilon|^3p^{-2} & |\varepsilon|^6 \\
-|\varepsilon|^3 & |\varepsilon|^2p^{-1} & -|\varepsilon|^2 \\
|\varepsilon| & |\varepsilon|^3 & 1 \\
\end{psmallmatrix}$
&
$\begin{psmallmatrix}
|\varepsilon|^4p^{-1} & |\varepsilon|^3p^{-2} & |\varepsilon|^6 \\
|\varepsilon|^3 & |\varepsilon|^2p^{-1} & -|\varepsilon|^2 \\
-|\varepsilon| & |\varepsilon|^3 & 1 \\
\end{psmallmatrix}$
&
$\begin{psmallmatrix}
|\varepsilon|^4p^{-1} & |\varepsilon|^3p^{-2} & |\varepsilon|^6 \\
-|\varepsilon|^3 & |\varepsilon|^2p^{-1} & -|\varepsilon|^2 \\
-|\varepsilon| & |\varepsilon|^3 & 1 \\
\end{psmallmatrix}$
\\
$\begin{psmallmatrix}
|\varepsilon|^4p^{-1} & |\varepsilon|^3p^{-3} & |\varepsilon|^6 \\
|\varepsilon|^3 & |\varepsilon|^2p^{-2} & |\varepsilon|^2 \\
|\varepsilon| & |\varepsilon|^6 & -1 \\
\end{psmallmatrix}$
&
$\begin{psmallmatrix}
|\varepsilon|^4p^{-1} & |\varepsilon|^3p^{-3} & |\varepsilon|^6 \\
-|\varepsilon|^3 & |\varepsilon|^2p^{-2} & |\varepsilon|^2 \\
|\varepsilon| & |\varepsilon|^6 & -1 \\
\end{psmallmatrix}$
&
$\begin{psmallmatrix}
|\varepsilon|^4p^{-1} & |\varepsilon|^3p^{-3} & |\varepsilon|^6 \\
|\varepsilon|^3 & |\varepsilon|^2p^{-2} & |\varepsilon|^2 \\
-|\varepsilon| & |\varepsilon|^6 & -1 \\
\end{psmallmatrix}$
&
$\begin{psmallmatrix}
|\varepsilon|^4p^{-1} & |\varepsilon|^3p^{-3} & |\varepsilon|^6 \\
-|\varepsilon|^3 & |\varepsilon|^2p^{-2} & |\varepsilon|^2 \\
-|\varepsilon| & |\varepsilon|^6 & -1 \\
\end{psmallmatrix}$
\\
$\begin{psmallmatrix}
|\varepsilon|^4p^{-1} & |\varepsilon|^3p^{-3} & |\varepsilon|^6 \\
|\varepsilon|^3 & |\varepsilon|^2p^{-2} & |\varepsilon|^2 \\
|\varepsilon| & -|\varepsilon|^6 & -1 \\
\end{psmallmatrix}$
&
$\begin{psmallmatrix}
|\varepsilon|^4p^{-1} & |\varepsilon|^3p^{-3} & |\varepsilon|^6 \\
-|\varepsilon|^3 & |\varepsilon|^2p^{-2} & |\varepsilon|^2 \\
|\varepsilon| & -|\varepsilon|^6 & -1 \\
\end{psmallmatrix}$
&
$\begin{psmallmatrix}
|\varepsilon|^4p^{-1} & |\varepsilon|^3p^{-3} & |\varepsilon|^6 \\
|\varepsilon|^3 & |\varepsilon|^2p^{-2} & |\varepsilon|^2 \\
-|\varepsilon| & -|\varepsilon|^6 & -1 \\
\end{psmallmatrix}$
&
$\begin{psmallmatrix}
|\varepsilon|^4p^{-1} & |\varepsilon|^3p^{-3} & |\varepsilon|^6 \\
-|\varepsilon|^3 & |\varepsilon|^2p^{-2} & |\varepsilon|^2 \\
-|\varepsilon| & -|\varepsilon|^6 & -1 \\
\end{psmallmatrix}$
\\
$\begin{psmallmatrix}
|\varepsilon|^4p^{-1} & |\varepsilon|^3p^{-3} & |\varepsilon|^6 \\
|\varepsilon|^3 & |\varepsilon|^2p^{-2} & -|\varepsilon|^2 \\
|\varepsilon| & |\varepsilon|^6 & 1 \\
\end{psmallmatrix}$
&
$\begin{psmallmatrix}
|\varepsilon|^4p^{-1} & |\varepsilon|^3p^{-3} & |\varepsilon|^6 \\
-|\varepsilon|^3 & |\varepsilon|^2p^{-2} & -|\varepsilon|^2 \\
|\varepsilon| & |\varepsilon|^6 & 1 \\
\end{psmallmatrix}$
&
$\begin{psmallmatrix}
|\varepsilon|^4p^{-1} & |\varepsilon|^3p^{-3} & |\varepsilon|^6 \\
|\varepsilon|^3 & |\varepsilon|^2p^{-2} & -|\varepsilon|^2 \\
-|\varepsilon| & |\varepsilon|^6 & 1 \\
\end{psmallmatrix}$
&
$\begin{psmallmatrix}
|\varepsilon|^4p^{-1} & |\varepsilon|^3p^{-3} & |\varepsilon|^6 \\
-|\varepsilon|^3 & |\varepsilon|^2p^{-2} & -|\varepsilon|^2 \\
-|\varepsilon| & |\varepsilon|^6 & 1 \\
\end{psmallmatrix}$
\\
$\begin{psmallmatrix}
|\varepsilon|^4p^{-1} & |\varepsilon|^3p^{-3} & |\varepsilon|^6 \\
|\varepsilon|^3 & |\varepsilon|^2p^{-2} & -|\varepsilon|^2 \\
|\varepsilon| & -|\varepsilon|^6 & 1 \\
\end{psmallmatrix}$
&
$\begin{psmallmatrix}
|\varepsilon|^4p^{-1} & |\varepsilon|^3p^{-3} & |\varepsilon|^6 \\
-|\varepsilon|^3 & |\varepsilon|^2p^{-2} & -|\varepsilon|^2 \\
|\varepsilon| & -|\varepsilon|^6 & 1 \\
\end{psmallmatrix}$
&
$\begin{psmallmatrix}
|\varepsilon|^4p^{-1} & |\varepsilon|^3p^{-3} & |\varepsilon|^6 \\
|\varepsilon|^3 & |\varepsilon|^2p^{-2} & -|\varepsilon|^2 \\
-|\varepsilon| & -|\varepsilon|^6 & 1 \\
\end{psmallmatrix}$
&
$\begin{psmallmatrix}
|\varepsilon|^4p^{-1} & |\varepsilon|^3p^{-3} & |\varepsilon|^6 \\
-|\varepsilon|^3 & |\varepsilon|^2p^{-2} & -|\varepsilon|^2 \\
-|\varepsilon| & -|\varepsilon|^6 & 1 \\
\end{psmallmatrix}$
\\
$\begin{psmallmatrix}
|\varepsilon|^4p^{-2} & |\varepsilon|^3p^{-2} & |\varepsilon|^6 \\
-|\varepsilon|^3p^{-1} & |\varepsilon|^2p^{-1} & |\varepsilon|^2 \\
-|\varepsilon|^4 & -|\varepsilon|^3 & -1 \\
\end{psmallmatrix}$
&
$\begin{psmallmatrix}
|\varepsilon|^4p^{-2} & |\varepsilon|^3p^{-2} & |\varepsilon|^6 \\
-|\varepsilon|^3p^{-1} & |\varepsilon|^2p^{-1} & |\varepsilon|^2 \\
|\varepsilon|^4 & -|\varepsilon|^3 & -1 \\
\end{psmallmatrix}$
&
$\begin{psmallmatrix}
|\varepsilon|^4p^{-2} & |\varepsilon|^3p^{-2} & |\varepsilon|^6 \\
-|\varepsilon|^3p^{-1} & |\varepsilon|^2p^{-1} & |\varepsilon|^2 \\
-|\varepsilon|^4 & |\varepsilon|^3 & -1 \\
\end{psmallmatrix}$
&
$\begin{psmallmatrix}
|\varepsilon|^4p^{-2} & |\varepsilon|^3p^{-2} & |\varepsilon|^6 \\
-|\varepsilon|^3p^{-1} & |\varepsilon|^2p^{-1} & |\varepsilon|^2 \\
|\varepsilon|^4 & |\varepsilon|^3 & -1 \\
\end{psmallmatrix}$
\\
$\begin{psmallmatrix}
|\varepsilon|^4p^{-2} & |\varepsilon|^3p^{-2} & |\varepsilon|^6 \\
-|\varepsilon|^3p^{-1} & |\varepsilon|^2p^{-1} & -|\varepsilon|^2 \\
-|\varepsilon|^4 & -|\varepsilon|^3 & 1 \\
\end{psmallmatrix}$
&
$\begin{psmallmatrix}
|\varepsilon|^4p^{-2} & |\varepsilon|^3p^{-2} & |\varepsilon|^6 \\
-|\varepsilon|^3p^{-1} & |\varepsilon|^2p^{-1} & -|\varepsilon|^2 \\
|\varepsilon|^4 & -|\varepsilon|^3 & 1 \\
\end{psmallmatrix}$
&
$\begin{psmallmatrix}
|\varepsilon|^4p^{-2} & |\varepsilon|^3p^{-2} & |\varepsilon|^6 \\
-|\varepsilon|^3p^{-1} & |\varepsilon|^2p^{-1} & -|\varepsilon|^2 \\
-|\varepsilon|^4 & |\varepsilon|^3 & 1 \\
\end{psmallmatrix}$
&
$\begin{psmallmatrix}
|\varepsilon|^4p^{-2} & |\varepsilon|^3p^{-2} & |\varepsilon|^6 \\
-|\varepsilon|^3p^{-1} & |\varepsilon|^2p^{-1} & -|\varepsilon|^2 \\
|\varepsilon|^4 & |\varepsilon|^3 & 1 \\
\end{psmallmatrix}$
\\
$\begin{psmallmatrix}
|\varepsilon|^4p^{-2} & |\varepsilon|^3p^{-3} & |\varepsilon|^6 \\
-|\varepsilon|^3p^{-1} & |\varepsilon|^2p^{-2} & |\varepsilon|^2 \\
-|\varepsilon|^4 & |\varepsilon|^6 & -1 \\
\end{psmallmatrix}$
&
$\begin{psmallmatrix}
|\varepsilon|^4p^{-2} & |\varepsilon|^3p^{-3} & |\varepsilon|^6 \\
-|\varepsilon|^3p^{-1} & |\varepsilon|^2p^{-2} & |\varepsilon|^2 \\
|\varepsilon|^4 & |\varepsilon|^6 & -1 \\
\end{psmallmatrix}$
&
$\begin{psmallmatrix}
|\varepsilon|^4p^{-2} & |\varepsilon|^3p^{-3} & |\varepsilon|^6 \\
-|\varepsilon|^3p^{-1} & |\varepsilon|^2p^{-2} & |\varepsilon|^2 \\
-|\varepsilon|^4 & -|\varepsilon|^6 & -1 \\
\end{psmallmatrix}$
&
$\begin{psmallmatrix}
|\varepsilon|^4p^{-2} & |\varepsilon|^3p^{-3} & |\varepsilon|^6 \\
-|\varepsilon|^3p^{-1} & |\varepsilon|^2p^{-2} & |\varepsilon|^2 \\
|\varepsilon|^4 & -|\varepsilon|^6 & -1 \\
\end{psmallmatrix}$
\\
$\begin{psmallmatrix}
|\varepsilon|^4p^{-2} & |\varepsilon|^3p^{-3} & |\varepsilon|^6 \\
-|\varepsilon|^3p^{-1} & |\varepsilon|^2p^{-2} & -|\varepsilon|^2 \\
-|\varepsilon|^4 & |\varepsilon|^6 & 1 \\
\end{psmallmatrix}$
&
$\begin{psmallmatrix}
|\varepsilon|^4p^{-2} & |\varepsilon|^3p^{-3} & |\varepsilon|^6 \\
-|\varepsilon|^3p^{-1} & |\varepsilon|^2p^{-2} & -|\varepsilon|^2 \\
|\varepsilon|^4 & |\varepsilon|^6 & 1 \\
\end{psmallmatrix}$
&
$\begin{psmallmatrix}
|\varepsilon|^4p^{-2} & |\varepsilon|^3p^{-3} & |\varepsilon|^6 \\
-|\varepsilon|^3p^{-1} & |\varepsilon|^2p^{-2} & -|\varepsilon|^2 \\
-|\varepsilon|^4 & -|\varepsilon|^6 & 1 \\
\end{psmallmatrix}$
&
$\begin{psmallmatrix}
|\varepsilon|^4p^{-2} & |\varepsilon|^3p^{-3} & |\varepsilon|^6 \\
-|\varepsilon|^3p^{-1} & |\varepsilon|^2p^{-2} & -|\varepsilon|^2 \\
|\varepsilon|^4 & -|\varepsilon|^6 & 1 \\
\end{psmallmatrix}$
\\
$\begin{psmallmatrix}
|\varepsilon|^4p^{-2} & |\varepsilon|^3p^{-2} & |\varepsilon|^6 \\
-|\varepsilon|^6 & |\varepsilon|^2p^{-1} & |\varepsilon|^2 \\
-|\varepsilon|^4 & -|\varepsilon|^3 & -1 \\
\end{psmallmatrix}$
&
$\begin{psmallmatrix}
|\varepsilon|^4p^{-2} & |\varepsilon|^3p^{-2} & |\varepsilon|^6 \\
|\varepsilon|^6 & |\varepsilon|^2p^{-1} & |\varepsilon|^2 \\
-|\varepsilon|^4 & -|\varepsilon|^3 & -1 \\
\end{psmallmatrix}$
&
$\begin{psmallmatrix}
|\varepsilon|^4p^{-2} & |\varepsilon|^3p^{-2} & |\varepsilon|^6 \\
-|\varepsilon|^6 & |\varepsilon|^2p^{-1} & |\varepsilon|^2 \\
|\varepsilon|^4 & -|\varepsilon|^3 & -1 \\
\end{psmallmatrix}$
&
$\begin{psmallmatrix}
|\varepsilon|^4p^{-2} & |\varepsilon|^3p^{-2} & |\varepsilon|^6 \\
|\varepsilon|^6 & |\varepsilon|^2p^{-1} & |\varepsilon|^2 \\
|\varepsilon|^4 & -|\varepsilon|^3 & -1 \\
\end{psmallmatrix}$
\\
$\begin{psmallmatrix}
|\varepsilon|^4p^{-2} & |\varepsilon|^3p^{-2} & |\varepsilon|^6 \\
-|\varepsilon|^6 & |\varepsilon|^2p^{-1} & |\varepsilon|^2 \\
-|\varepsilon|^4 & |\varepsilon|^3 & -1 \\
\end{psmallmatrix}$
&
$\begin{psmallmatrix}
|\varepsilon|^4p^{-2} & |\varepsilon|^3p^{-2} & |\varepsilon|^6 \\
|\varepsilon|^6 & |\varepsilon|^2p^{-1} & |\varepsilon|^2 \\
-|\varepsilon|^4 & |\varepsilon|^3 & -1 \\
\end{psmallmatrix}$
&
$\begin{psmallmatrix}
|\varepsilon|^4p^{-2} & |\varepsilon|^3p^{-2} & |\varepsilon|^6 \\
-|\varepsilon|^6 & |\varepsilon|^2p^{-1} & |\varepsilon|^2 \\
|\varepsilon|^4 & |\varepsilon|^3 & -1 \\
\end{psmallmatrix}$
&
$\begin{psmallmatrix}
|\varepsilon|^4p^{-2} & |\varepsilon|^3p^{-2} & |\varepsilon|^6 \\
|\varepsilon|^6 & |\varepsilon|^2p^{-1} & |\varepsilon|^2 \\
|\varepsilon|^4 & |\varepsilon|^3 & -1 \\
\end{psmallmatrix}$
\\
$\begin{psmallmatrix}
|\varepsilon|^4p^{-2} & |\varepsilon|^3p^{-2} & |\varepsilon|^6 \\
-|\varepsilon|^6 & |\varepsilon|^2p^{-1} & -|\varepsilon|^2 \\
-|\varepsilon|^4 & -|\varepsilon|^3 & 1 \\
\end{psmallmatrix}$
&
$\begin{psmallmatrix}
|\varepsilon|^4p^{-2} & |\varepsilon|^3p^{-2} & |\varepsilon|^6 \\
|\varepsilon|^6 & |\varepsilon|^2p^{-1} & -|\varepsilon|^2 \\
-|\varepsilon|^4 & -|\varepsilon|^3 & 1 \\
\end{psmallmatrix}$
&
$\begin{psmallmatrix}
|\varepsilon|^4p^{-2} & |\varepsilon|^3p^{-2} & |\varepsilon|^6 \\
-|\varepsilon|^6 & |\varepsilon|^2p^{-1} & -|\varepsilon|^2 \\
|\varepsilon|^4 & -|\varepsilon|^3 & 1 \\
\end{psmallmatrix}$
&
$\begin{psmallmatrix}
|\varepsilon|^4p^{-2} & |\varepsilon|^3p^{-2} & |\varepsilon|^6 \\
|\varepsilon|^6 & |\varepsilon|^2p^{-1} & -|\varepsilon|^2 \\
|\varepsilon|^4 & -|\varepsilon|^3 & 1 \\
\end{psmallmatrix}$
\\
$\begin{psmallmatrix}
|\varepsilon|^4p^{-2} & |\varepsilon|^3p^{-2} & |\varepsilon|^6 \\
-|\varepsilon|^6 & |\varepsilon|^2p^{-1} & -|\varepsilon|^2 \\
-|\varepsilon|^4 & |\varepsilon|^3 & 1 \\
\end{psmallmatrix}$
&
$\begin{psmallmatrix}
|\varepsilon|^4p^{-2} & |\varepsilon|^3p^{-2} & |\varepsilon|^6 \\
|\varepsilon|^6 & |\varepsilon|^2p^{-1} & -|\varepsilon|^2 \\
-|\varepsilon|^4 & |\varepsilon|^3 & 1 \\
\end{psmallmatrix}$
&
$\begin{psmallmatrix}
|\varepsilon|^4p^{-2} & |\varepsilon|^3p^{-2} & |\varepsilon|^6 \\
-|\varepsilon|^6 & |\varepsilon|^2p^{-1} & -|\varepsilon|^2 \\
|\varepsilon|^4 & |\varepsilon|^3 & 1 \\
\end{psmallmatrix}$
&
$\begin{psmallmatrix}
|\varepsilon|^4p^{-2} & |\varepsilon|^3p^{-2} & |\varepsilon|^6 \\
|\varepsilon|^6 & |\varepsilon|^2p^{-1} & -|\varepsilon|^2 \\
|\varepsilon|^4 & |\varepsilon|^3 & 1 \\
\end{psmallmatrix}$
\\ \hline
  \end{tabular}
\end{table}
\begin{table}[H]
  \centering
  \renewcommand{\arraystretch}{1.8}
  \begin{tabular}{p{9em}p{9em}p{9em}p{9em}} \hline
\multicolumn{4}{c}{$M_u=\begin{psmallmatrix}
|\varepsilon|^6 & |\varepsilon|^4p^{-1} & |\varepsilon|^6 \\
|\varepsilon|^2 & |\varepsilon|^3 & -|\varepsilon|^2 \\
1 & -|\varepsilon| & -1 \\
\end{psmallmatrix}$} \\ \hline
$\begin{psmallmatrix}
|\varepsilon|^4p^{-1} & |\varepsilon|^3p^{-2} & |\varepsilon|^6 \\
|\varepsilon|^3 & -|\varepsilon|^2p^{-1} & |\varepsilon|^2 \\
|\varepsilon| & -|\varepsilon|^3 & -1 \\
\end{psmallmatrix}$
&
$\begin{psmallmatrix}
|\varepsilon|^4p^{-1} & |\varepsilon|^3p^{-2} & |\varepsilon|^6 \\
-|\varepsilon|^3 & -|\varepsilon|^2p^{-1} & |\varepsilon|^2 \\
|\varepsilon| & -|\varepsilon|^3 & -1 \\
\end{psmallmatrix}$
&
$\begin{psmallmatrix}
|\varepsilon|^4p^{-1} & |\varepsilon|^3p^{-2} & |\varepsilon|^6 \\
|\varepsilon|^3 & -|\varepsilon|^2p^{-1} & |\varepsilon|^2 \\
-|\varepsilon| & -|\varepsilon|^3 & -1 \\
\end{psmallmatrix}$
&
$\begin{psmallmatrix}
|\varepsilon|^4p^{-1} & |\varepsilon|^3p^{-2} & |\varepsilon|^6 \\
-|\varepsilon|^3 & -|\varepsilon|^2p^{-1} & |\varepsilon|^2 \\
-|\varepsilon| & -|\varepsilon|^3 & -1 \\
\end{psmallmatrix}$
\\
$\begin{psmallmatrix}
|\varepsilon|^4p^{-1} & |\varepsilon|^3p^{-2} & |\varepsilon|^6 \\
|\varepsilon|^3 & -|\varepsilon|^2p^{-1} & |\varepsilon|^2 \\
|\varepsilon| & |\varepsilon|^3 & -1 \\
\end{psmallmatrix}$
&
$\begin{psmallmatrix}
|\varepsilon|^4p^{-1} & |\varepsilon|^3p^{-2} & |\varepsilon|^6 \\
-|\varepsilon|^3 & -|\varepsilon|^2p^{-1} & |\varepsilon|^2 \\
|\varepsilon| & |\varepsilon|^3 & -1 \\
\end{psmallmatrix}$
&
$\begin{psmallmatrix}
|\varepsilon|^4p^{-1} & |\varepsilon|^3p^{-2} & |\varepsilon|^6 \\
|\varepsilon|^3 & -|\varepsilon|^2p^{-1} & |\varepsilon|^2 \\
-|\varepsilon| & |\varepsilon|^3 & -1 \\
\end{psmallmatrix}$
&
$\begin{psmallmatrix}
|\varepsilon|^4p^{-1} & |\varepsilon|^3p^{-2} & |\varepsilon|^6 \\
-|\varepsilon|^3 & -|\varepsilon|^2p^{-1} & |\varepsilon|^2 \\
-|\varepsilon| & |\varepsilon|^3 & -1 \\
\end{psmallmatrix}$
\\
$\begin{psmallmatrix}
|\varepsilon|^4p^{-1} & |\varepsilon|^3p^{-2} & |\varepsilon|^6 \\
|\varepsilon|^3 & -|\varepsilon|^2p^{-1} & -|\varepsilon|^2 \\
|\varepsilon| & -|\varepsilon|^3 & 1 \\
\end{psmallmatrix}$
&
$\begin{psmallmatrix}
|\varepsilon|^4p^{-1} & |\varepsilon|^3p^{-2} & |\varepsilon|^6 \\
-|\varepsilon|^3 & -|\varepsilon|^2p^{-1} & -|\varepsilon|^2 \\
|\varepsilon| & -|\varepsilon|^3 & 1 \\
\end{psmallmatrix}$
&
$\begin{psmallmatrix}
|\varepsilon|^4p^{-1} & |\varepsilon|^3p^{-2} & |\varepsilon|^6 \\
|\varepsilon|^3 & -|\varepsilon|^2p^{-1} & -|\varepsilon|^2 \\
-|\varepsilon| & -|\varepsilon|^3 & 1 \\
\end{psmallmatrix}$
&
$\begin{psmallmatrix}
|\varepsilon|^4p^{-1} & |\varepsilon|^3p^{-2} & |\varepsilon|^6 \\
-|\varepsilon|^3 & -|\varepsilon|^2p^{-1} & -|\varepsilon|^2 \\
-|\varepsilon| & -|\varepsilon|^3 & 1 \\
\end{psmallmatrix}$
\\
$\begin{psmallmatrix}
|\varepsilon|^4p^{-1} & |\varepsilon|^3p^{-2} & |\varepsilon|^6 \\
|\varepsilon|^3 & -|\varepsilon|^2p^{-1} & -|\varepsilon|^2 \\
|\varepsilon| & |\varepsilon|^3 & 1 \\
\end{psmallmatrix}$
&
$\begin{psmallmatrix}
|\varepsilon|^4p^{-1} & |\varepsilon|^3p^{-2} & |\varepsilon|^6 \\
-|\varepsilon|^3 & -|\varepsilon|^2p^{-1} & -|\varepsilon|^2 \\
|\varepsilon| & |\varepsilon|^3 & 1 \\
\end{psmallmatrix}$
&
$\begin{psmallmatrix}
|\varepsilon|^4p^{-1} & |\varepsilon|^3p^{-2} & |\varepsilon|^6 \\
|\varepsilon|^3 & -|\varepsilon|^2p^{-1} & -|\varepsilon|^2 \\
-|\varepsilon| & |\varepsilon|^3 & 1 \\
\end{psmallmatrix}$
&
$\begin{psmallmatrix}
|\varepsilon|^4p^{-1} & |\varepsilon|^3p^{-2} & |\varepsilon|^6 \\
-|\varepsilon|^3 & -|\varepsilon|^2p^{-1} & -|\varepsilon|^2 \\
-|\varepsilon| & |\varepsilon|^3 & 1 \\
\end{psmallmatrix}$
\\
$\begin{psmallmatrix}
|\varepsilon|^4p^{-1} & |\varepsilon|^3p^{-3} & |\varepsilon|^6 \\
|\varepsilon|^3 & -|\varepsilon|^2p^{-2} & |\varepsilon|^2 \\
|\varepsilon| & |\varepsilon|^6 & -1 \\
\end{psmallmatrix}$
&
$\begin{psmallmatrix}
|\varepsilon|^4p^{-1} & |\varepsilon|^3p^{-3} & |\varepsilon|^6 \\
-|\varepsilon|^3 & -|\varepsilon|^2p^{-2} & |\varepsilon|^2 \\
|\varepsilon| & |\varepsilon|^6 & -1 \\
\end{psmallmatrix}$
&
$\begin{psmallmatrix}
|\varepsilon|^4p^{-1} & |\varepsilon|^3p^{-3} & |\varepsilon|^6 \\
|\varepsilon|^3 & -|\varepsilon|^2p^{-2} & |\varepsilon|^2 \\
-|\varepsilon| & |\varepsilon|^6 & -1 \\
\end{psmallmatrix}$
&
$\begin{psmallmatrix}
|\varepsilon|^4p^{-1} & |\varepsilon|^3p^{-3} & |\varepsilon|^6 \\
-|\varepsilon|^3 & -|\varepsilon|^2p^{-2} & |\varepsilon|^2 \\
-|\varepsilon| & |\varepsilon|^6 & -1 \\
\end{psmallmatrix}$
\\
$\begin{psmallmatrix}
|\varepsilon|^4p^{-1} & |\varepsilon|^3p^{-3} & |\varepsilon|^6 \\
|\varepsilon|^3 & -|\varepsilon|^2p^{-2} & |\varepsilon|^2 \\
|\varepsilon| & -|\varepsilon|^6 & -1 \\
\end{psmallmatrix}$
&
$\begin{psmallmatrix}
|\varepsilon|^4p^{-1} & |\varepsilon|^3p^{-3} & |\varepsilon|^6 \\
-|\varepsilon|^3 & -|\varepsilon|^2p^{-2} & |\varepsilon|^2 \\
|\varepsilon| & -|\varepsilon|^6 & -1 \\
\end{psmallmatrix}$
&
$\begin{psmallmatrix}
|\varepsilon|^4p^{-1} & |\varepsilon|^3p^{-3} & |\varepsilon|^6 \\
|\varepsilon|^3 & -|\varepsilon|^2p^{-2} & |\varepsilon|^2 \\
-|\varepsilon| & -|\varepsilon|^6 & -1 \\
\end{psmallmatrix}$
&
$\begin{psmallmatrix}
|\varepsilon|^4p^{-1} & |\varepsilon|^3p^{-3} & |\varepsilon|^6 \\
-|\varepsilon|^3 & -|\varepsilon|^2p^{-2} & |\varepsilon|^2 \\
-|\varepsilon| & -|\varepsilon|^6 & -1 \\
\end{psmallmatrix}$
\\
$\begin{psmallmatrix}
|\varepsilon|^4p^{-1} & |\varepsilon|^3p^{-3} & |\varepsilon|^6 \\
|\varepsilon|^3 & -|\varepsilon|^2p^{-2} & -|\varepsilon|^2 \\
|\varepsilon| & |\varepsilon|^6 & 1 \\
\end{psmallmatrix}$
&
$\begin{psmallmatrix}
|\varepsilon|^4p^{-1} & |\varepsilon|^3p^{-3} & |\varepsilon|^6 \\
-|\varepsilon|^3 & -|\varepsilon|^2p^{-2} & -|\varepsilon|^2 \\
|\varepsilon| & |\varepsilon|^6 & 1 \\
\end{psmallmatrix}$
&
$\begin{psmallmatrix}
|\varepsilon|^4p^{-1} & |\varepsilon|^3p^{-3} & |\varepsilon|^6 \\
|\varepsilon|^3 & -|\varepsilon|^2p^{-2} & -|\varepsilon|^2 \\
-|\varepsilon| & |\varepsilon|^6 & 1 \\
\end{psmallmatrix}$
&
$\begin{psmallmatrix}
|\varepsilon|^4p^{-1} & |\varepsilon|^3p^{-3} & |\varepsilon|^6 \\
-|\varepsilon|^3 & -|\varepsilon|^2p^{-2} & -|\varepsilon|^2 \\
-|\varepsilon| & |\varepsilon|^6 & 1 \\
\end{psmallmatrix}$
\\
$\begin{psmallmatrix}
|\varepsilon|^4p^{-1} & |\varepsilon|^3p^{-3} & |\varepsilon|^6 \\
|\varepsilon|^3 & -|\varepsilon|^2p^{-2} & -|\varepsilon|^2 \\
|\varepsilon| & -|\varepsilon|^6 & 1 \\
\end{psmallmatrix}$
&
$\begin{psmallmatrix}
|\varepsilon|^4p^{-1} & |\varepsilon|^3p^{-3} & |\varepsilon|^6 \\
-|\varepsilon|^3 & -|\varepsilon|^2p^{-2} & -|\varepsilon|^2 \\
|\varepsilon| & -|\varepsilon|^6 & 1 \\
\end{psmallmatrix}$
&
$\begin{psmallmatrix}
|\varepsilon|^4p^{-1} & |\varepsilon|^3p^{-3} & |\varepsilon|^6 \\
|\varepsilon|^3 & -|\varepsilon|^2p^{-2} & -|\varepsilon|^2 \\
-|\varepsilon| & -|\varepsilon|^6 & 1 \\
\end{psmallmatrix}$
&
$\begin{psmallmatrix}
|\varepsilon|^4p^{-1} & |\varepsilon|^3p^{-3} & |\varepsilon|^6 \\
-|\varepsilon|^3 & -|\varepsilon|^2p^{-2} & -|\varepsilon|^2 \\
-|\varepsilon| & -|\varepsilon|^6 & 1 \\
\end{psmallmatrix}$
\\
$\begin{psmallmatrix}
|\varepsilon|^4p^{-2} & |\varepsilon|^3p^{-2} & |\varepsilon|^6 \\
|\varepsilon|^3p^{-1} & -|\varepsilon|^2p^{-1} & |\varepsilon|^2 \\
-|\varepsilon|^4 & -|\varepsilon|^3 & -1 \\
\end{psmallmatrix}$
&
$\begin{psmallmatrix}
|\varepsilon|^4p^{-2} & |\varepsilon|^3p^{-2} & |\varepsilon|^6 \\
|\varepsilon|^3p^{-1} & -|\varepsilon|^2p^{-1} & |\varepsilon|^2 \\
|\varepsilon|^4 & -|\varepsilon|^3 & -1 \\
\end{psmallmatrix}$
&
$\begin{psmallmatrix}
|\varepsilon|^4p^{-2} & |\varepsilon|^3p^{-2} & |\varepsilon|^6 \\
|\varepsilon|^3p^{-1} & -|\varepsilon|^2p^{-1} & |\varepsilon|^2 \\
-|\varepsilon|^4 & |\varepsilon|^3 & -1 \\
\end{psmallmatrix}$
&
$\begin{psmallmatrix}
|\varepsilon|^4p^{-2} & |\varepsilon|^3p^{-2} & |\varepsilon|^6 \\
|\varepsilon|^3p^{-1} & -|\varepsilon|^2p^{-1} & |\varepsilon|^2 \\
|\varepsilon|^4 & |\varepsilon|^3 & -1 \\
\end{psmallmatrix}$
\\
$\begin{psmallmatrix}
|\varepsilon|^4p^{-2} & |\varepsilon|^3p^{-2} & |\varepsilon|^6 \\
|\varepsilon|^3p^{-1} & -|\varepsilon|^2p^{-1} & -|\varepsilon|^2 \\
-|\varepsilon|^4 & -|\varepsilon|^3 & 1 \\
\end{psmallmatrix}$
&
$\begin{psmallmatrix}
|\varepsilon|^4p^{-2} & |\varepsilon|^3p^{-2} & |\varepsilon|^6 \\
|\varepsilon|^3p^{-1} & -|\varepsilon|^2p^{-1} & -|\varepsilon|^2 \\
|\varepsilon|^4 & -|\varepsilon|^3 & 1 \\
\end{psmallmatrix}$
&
$\begin{psmallmatrix}
|\varepsilon|^4p^{-2} & |\varepsilon|^3p^{-2} & |\varepsilon|^6 \\
|\varepsilon|^3p^{-1} & -|\varepsilon|^2p^{-1} & -|\varepsilon|^2 \\
-|\varepsilon|^4 & |\varepsilon|^3 & 1 \\
\end{psmallmatrix}$
&
$\begin{psmallmatrix}
|\varepsilon|^4p^{-2} & |\varepsilon|^3p^{-2} & |\varepsilon|^6 \\
|\varepsilon|^3p^{-1} & -|\varepsilon|^2p^{-1} & -|\varepsilon|^2 \\
|\varepsilon|^4 & |\varepsilon|^3 & 1 \\
\end{psmallmatrix}$
\\
$\begin{psmallmatrix}
|\varepsilon|^4p^{-2} & |\varepsilon|^3p^{-3} & |\varepsilon|^6 \\
|\varepsilon|^3p^{-1} & -|\varepsilon|^2p^{-2} & |\varepsilon|^2 \\
-|\varepsilon|^4 & |\varepsilon|^6 & -1 \\
\end{psmallmatrix}$
&
$\begin{psmallmatrix}
|\varepsilon|^4p^{-2} & |\varepsilon|^3p^{-3} & |\varepsilon|^6 \\
|\varepsilon|^3p^{-1} & -|\varepsilon|^2p^{-2} & |\varepsilon|^2 \\
|\varepsilon|^4 & |\varepsilon|^6 & -1 \\
\end{psmallmatrix}$
&
$\begin{psmallmatrix}
|\varepsilon|^4p^{-2} & |\varepsilon|^3p^{-3} & |\varepsilon|^6 \\
|\varepsilon|^3p^{-1} & -|\varepsilon|^2p^{-2} & |\varepsilon|^2 \\
-|\varepsilon|^4 & -|\varepsilon|^6 & -1 \\
\end{psmallmatrix}$
&
$\begin{psmallmatrix}
|\varepsilon|^4p^{-2} & |\varepsilon|^3p^{-3} & |\varepsilon|^6 \\
|\varepsilon|^3p^{-1} & -|\varepsilon|^2p^{-2} & |\varepsilon|^2 \\
|\varepsilon|^4 & -|\varepsilon|^6 & -1 \\
\end{psmallmatrix}$
\\
$\begin{psmallmatrix}
|\varepsilon|^4p^{-2} & |\varepsilon|^3p^{-3} & |\varepsilon|^6 \\
|\varepsilon|^3p^{-1} & -|\varepsilon|^2p^{-2} & -|\varepsilon|^2 \\
-|\varepsilon|^4 & |\varepsilon|^6 & 1 \\
\end{psmallmatrix}$
&
$\begin{psmallmatrix}
|\varepsilon|^4p^{-2} & |\varepsilon|^3p^{-3} & |\varepsilon|^6 \\
|\varepsilon|^3p^{-1} & -|\varepsilon|^2p^{-2} & -|\varepsilon|^2 \\
|\varepsilon|^4 & |\varepsilon|^6 & 1 \\
\end{psmallmatrix}$
&
$\begin{psmallmatrix}
|\varepsilon|^4p^{-2} & |\varepsilon|^3p^{-3} & |\varepsilon|^6 \\
|\varepsilon|^3p^{-1} & -|\varepsilon|^2p^{-2} & -|\varepsilon|^2 \\
-|\varepsilon|^4 & -|\varepsilon|^6 & 1 \\
\end{psmallmatrix}$
&
$\begin{psmallmatrix}
|\varepsilon|^4p^{-2} & |\varepsilon|^3p^{-3} & |\varepsilon|^6 \\
|\varepsilon|^3p^{-1} & -|\varepsilon|^2p^{-2} & -|\varepsilon|^2 \\
|\varepsilon|^4 & -|\varepsilon|^6 & 1 \\
\end{psmallmatrix}$
\\
$\begin{psmallmatrix}
|\varepsilon|^4p^{-2} & |\varepsilon|^3p^{-2} & |\varepsilon|^6 \\
-|\varepsilon|^6 & -|\varepsilon|^2p^{-1} & |\varepsilon|^2 \\
-|\varepsilon|^4 & -|\varepsilon|^3 & -1 \\
\end{psmallmatrix}$
&
$\begin{psmallmatrix}
|\varepsilon|^4p^{-2} & |\varepsilon|^3p^{-2} & |\varepsilon|^6 \\
|\varepsilon|^6 & -|\varepsilon|^2p^{-1} & |\varepsilon|^2 \\
-|\varepsilon|^4 & -|\varepsilon|^3 & -1 \\
\end{psmallmatrix}$
&
$\begin{psmallmatrix}
|\varepsilon|^4p^{-2} & |\varepsilon|^3p^{-2} & |\varepsilon|^6 \\
-|\varepsilon|^6 & -|\varepsilon|^2p^{-1} & |\varepsilon|^2 \\
|\varepsilon|^4 & -|\varepsilon|^3 & -1 \\
\end{psmallmatrix}$
&
$\begin{psmallmatrix}
|\varepsilon|^4p^{-2} & |\varepsilon|^3p^{-2} & |\varepsilon|^6 \\
|\varepsilon|^6 & -|\varepsilon|^2p^{-1} & |\varepsilon|^2 \\
|\varepsilon|^4 & -|\varepsilon|^3 & -1 \\
\end{psmallmatrix}$
\\
$\begin{psmallmatrix}
|\varepsilon|^4p^{-2} & |\varepsilon|^3p^{-2} & |\varepsilon|^6 \\
-|\varepsilon|^6 & -|\varepsilon|^2p^{-1} & |\varepsilon|^2 \\
-|\varepsilon|^4 & |\varepsilon|^3 & -1 \\
\end{psmallmatrix}$
&
$\begin{psmallmatrix}
|\varepsilon|^4p^{-2} & |\varepsilon|^3p^{-2} & |\varepsilon|^6 \\
|\varepsilon|^6 & -|\varepsilon|^2p^{-1} & |\varepsilon|^2 \\
-|\varepsilon|^4 & |\varepsilon|^3 & -1 \\
\end{psmallmatrix}$
&
$\begin{psmallmatrix}
|\varepsilon|^4p^{-2} & |\varepsilon|^3p^{-2} & |\varepsilon|^6 \\
-|\varepsilon|^6 & -|\varepsilon|^2p^{-1} & |\varepsilon|^2 \\
|\varepsilon|^4 & |\varepsilon|^3 & -1 \\
\end{psmallmatrix}$
&
$\begin{psmallmatrix}
|\varepsilon|^4p^{-2} & |\varepsilon|^3p^{-2} & |\varepsilon|^6 \\
|\varepsilon|^6 & -|\varepsilon|^2p^{-1} & |\varepsilon|^2 \\
|\varepsilon|^4 & |\varepsilon|^3 & -1 \\
\end{psmallmatrix}$
\\
$\begin{psmallmatrix}
|\varepsilon|^4p^{-2} & |\varepsilon|^3p^{-2} & |\varepsilon|^6 \\
-|\varepsilon|^6 & -|\varepsilon|^2p^{-1} & -|\varepsilon|^2 \\
-|\varepsilon|^4 & -|\varepsilon|^3 & 1 \\
\end{psmallmatrix}$
&
$\begin{psmallmatrix}
|\varepsilon|^4p^{-2} & |\varepsilon|^3p^{-2} & |\varepsilon|^6 \\
|\varepsilon|^6 & -|\varepsilon|^2p^{-1} & -|\varepsilon|^2 \\
-|\varepsilon|^4 & -|\varepsilon|^3 & 1 \\
\end{psmallmatrix}$
&
$\begin{psmallmatrix}
|\varepsilon|^4p^{-2} & |\varepsilon|^3p^{-2} & |\varepsilon|^6 \\
-|\varepsilon|^6 & -|\varepsilon|^2p^{-1} & -|\varepsilon|^2 \\
|\varepsilon|^4 & -|\varepsilon|^3 & 1 \\
\end{psmallmatrix}$
&
$\begin{psmallmatrix}
|\varepsilon|^4p^{-2} & |\varepsilon|^3p^{-2} & |\varepsilon|^6 \\
|\varepsilon|^6 & -|\varepsilon|^2p^{-1} & -|\varepsilon|^2 \\
|\varepsilon|^4 & -|\varepsilon|^3 & 1 \\
\end{psmallmatrix}$
\\
$\begin{psmallmatrix}
|\varepsilon|^4p^{-2} & |\varepsilon|^3p^{-2} & |\varepsilon|^6 \\
-|\varepsilon|^6 & -|\varepsilon|^2p^{-1} & -|\varepsilon|^2 \\
-|\varepsilon|^4 & |\varepsilon|^3 & 1 \\
\end{psmallmatrix}$
&
$\begin{psmallmatrix}
|\varepsilon|^4p^{-2} & |\varepsilon|^3p^{-2} & |\varepsilon|^6 \\
|\varepsilon|^6 & -|\varepsilon|^2p^{-1} & -|\varepsilon|^2 \\
-|\varepsilon|^4 & |\varepsilon|^3 & 1 \\
\end{psmallmatrix}$
&
$\begin{psmallmatrix}
|\varepsilon|^4p^{-2} & |\varepsilon|^3p^{-2} & |\varepsilon|^6 \\
-|\varepsilon|^6 & -|\varepsilon|^2p^{-1} & -|\varepsilon|^2 \\
|\varepsilon|^4 & |\varepsilon|^3 & 1 \\
\end{psmallmatrix}$
&
$\begin{psmallmatrix}
|\varepsilon|^4p^{-2} & |\varepsilon|^3p^{-2} & |\varepsilon|^6 \\
|\varepsilon|^6 & -|\varepsilon|^2p^{-1} & -|\varepsilon|^2 \\
|\varepsilon|^4 & |\varepsilon|^3 & 1 \\
\end{psmallmatrix}$
\\ \hline
  \end{tabular}
\end{table}

%-----------------------------------------------------
%-----------------------------------------------------
%-----------------------------------------------------

\end{document}